\documentclass[apj,twocolumn]{openjournal}
\usepackage{array} % for table spacing
\usepackage{multirow}
\usepackage{amsmath}
\usepackage[colorlinks=true, citecolor=blue]{hyperref}
\usepackage{makecell}
\usepackage{microtype}
\usepackage[normalem]{ulem} % For \sout{}

\usepackage[usenames,dvipsnames,table]{xcolor}

\newcommand{\PC}[1]{\textcolor{cyan}{#1}}

\newcommand{\soutPC}{\bgroup\markoverwith{\textcolor{cyan}{\rule[0.5ex]{2pt}{1pt}}}\ULon}
\definecolor{nbgreen}{rgb}{0.15,0.68,0.38}

\newcommand{\mach}{\mathcal{M}}

\newcommand{\orcidauthor}[3]{\author{#2\href{http://orcid.org/#1}{\includegraphics[height=0.9em]{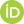}}$^{#3}$}}

\begin{document} 

\title{The LISA Astrophysics ``Disc-IMRI'' Code Comparison Project: \\Intermediate-Mass-Ratio Binaries in AGN-Like Discs}

%Last-name alphabetical within each tier, with coordinators also listing emails for correspondence 
%Current scheme: Coordinators -> Code captains -> Tier 1 -> Tier 2

%% Coordinators
\orcidauthor{0000-0001-9880-8929}{Andrea~Derdzinski}{1,2,*} %Fisk, Vanderbilt
\thanks{$^*$\href{mailto:aderdzinski@fisk.edu}{aderdzinski@fisk.edu}}
\orcidauthor{0000-0001-6157-6722}{Alexander~J.~Dittmann}{3,{\hbox{\small\textdagger}},{\hbox{\small\textexclamdown}}} %IAS
\thanks{\textsuperscript{\textdagger}\href{mailto:dittmann@ias.edu}{dittmann@ias.edu}}
\orcidauthor{0000-0002-8400-0969}{Alessia~Franchini}{4{\hbox{\small\textdaggerdbl}}}  % Pontremoli
\thanks{\textsuperscript{\textdaggerdbl}\href{mailto:alessia.franchini@unimi.it}{alessia.franchini@unimi.it}}
\orcidauthor{0000-0001-6106-7821}{Alessandro~Lupi}{5,6,7,{\hbox{\small\textsection}}} % Como, INFN, INAF Bolagna
\thanks{\textsuperscript{\textsection}\href{mailto:alessandro.lupi@uninsubria.it}{alessandro.lupi@uninsubria.it}}
\thanks{\textsuperscript{\textexclamdown} NASA Einstein Fellow}
%% Code Captains
\orcidauthor{0000-0003-3094-9674}{No\'{e}~Brucy}{8,9} %ENS de Lyon, Heidelberg
\orcidauthor{0000-0002-1786-963X}{Pedro~R.~Capelo}{10} % Zurich
\orcidauthor{0000-0002-9626-2210}{Fr\'{e}d\'{e}ric~S.~Masset}{11} % Universidad Nacional
\orcidauthor{0000-0002-3072-1496}{Rapha\"{e}l~Mignon-Risse}{12,13,14} %Valladolid, Trondheim, LAM Marseille
\orcidauthor{0000-0001-6721-4909}{Michael~Rizzo~Smith}{2} % Vanderbilt
\orcidauthor{0000-0001-9026-0380}{Edwin~Santiago-Leandro}{15} % INAOE, Mexico
\orcidauthor{0000-0001-5997-7148}{Martina~Toscani}{6,16} % Occhialini, INFN
\orcidauthor{0000-0002-6093-891X}{David~A.~Velasco-Romero}{3} %IAS
\orcidauthor{0000-0001-7039-4592}{Robert~Wissing}{17}  %Oslo
% Tier 1
\orcidauthor{0000-0002-9032-9103}{Mudit~Garg}{10} % Zurich
\orcidauthor{0000-0002-7078-2074}{Lucio~Mayer}{10} % Zurich
\orcidauthor{0000-0003-1200-5071}{Roberto~Serafinelli}{18,19} % Diego Portales,  INAF Rome
% Tier 2
\orcidauthor{0009-0006-4981-1092}{Lazaros~Souvaitzis}{20} % MPIA
%% Tier 3 
\orcidauthor{0000-0002-1271-6247}{Daniel~J.~D'Orazio}{21,22,23} %STScI, JHU, NBIA
\orcidauthor{0000-0001-6905-1840}{Jonathan~Menu}{24,25} % Leuven

%% Coordinator affiliations
\affiliation{$^{1}$Department of Life and Physical Sciences, Fisk University, 1000 17th Avenue N., Nashville, TN 37208, USA}
\affiliation{$^{2}$Department of Physics \& Astronomy, Vanderbilt University, 2301 Vanderbilt Place, Nashville, TN 37235, USA}
\affiliation{$^{3}$School of Natural Sciences, Institute for Advanced Study, 1 Einstein Drive, Princeton, NJ 08540, USA}
\affiliation{$^{4}$Dipartimento di Fisica ``A. Pontremoli'', Università degli Studi di Milano, Via Giovanni Celoria 16, 20134 Milano, Italy}
\affiliation{$^{5}$Como Lake Center for Astrophysics, DiSAT, Universit\`a degli Studi dell'Insubria,  via Valleggio 11, I-22100, Como, Italy}
\affiliation{$^{6}$INFN, Sezione di Milano-Bicocca, Piazza della Scienza 3, I-20126 Milano, Italy}
\affiliation{$^{7}$INAF, Osservatorio Astronomico di Bologna, Via Gobetti 93/3, I-40129 Bologna, Italy}
%% Code Captain
\affiliation{$^{8}$ENS de Lyon, CRAL UMR5574, Universite Claude Bernard Lyon 1, CNRS, Lyon 69007, France}
\affiliation{$^{9}$Universität Heidelberg, Zentrum für Astronomie, Institut für Theoretische Astrophysik, Albert-Ueberle-Str. 2, 69120 Heidelberg, Germany}
\affiliation{$^{10}$Department of Astrophysics, University of Zurich, Winterthurerstrasse 190, CH-8057 Z{\"u}rich, Switzerland}
\affiliation{$^{11}$Instituto de Ciencias F\'{i}sicas, Universidad Nacional Aut\'{o}noma de México, Av. Universidad s/n, 62210 Cuernavaca, Mor., Mexico}
\affiliation{$^{12}$Department of Theoretical Physics, Atomic and Optics, Campus Miguel Delibes, University of Valladolid, Paseo Bel\'{e}n, 7, 47011, Valladolid, Spain}
\affiliation{$^{13}$Department of Physics, Norwegian University of Science and Technology, NO-7491 Trondheim, Norway}
\affiliation{$^{14}$Aix-Marseille Université, CNRS, CNES, LAM, Marseille, France}
\affiliation{$^{15}$ Instituto Nacional de Astrofísica Óptica y Electrónica, Luis Enrique Erro 1, Tonantzintla CP 72840, Puebla, México}
\affiliation{$^{16}$Dipartimento di Fisica ``G. Occhialini'', Universit\`{a} degli Studi di Milano-Bicocca, Piazza della Scienza 3, I-20126 Milano, Italy}
\affiliation{$^{17}$Institute of Theoretical Astrophysics, University of Oslo, Postboks 1029, 0315 Oslo, Norway} 
% T1 affiliations
\affiliation{$^{18}$Instituto de Estudios Astrof\'{i}sicos, Facultad de Ingenier\'{i}a y Ciencias, Universidad Diego Portales, Avenida Ej\'{e}rcito Libertador 441, Santiago, Chile}
\affiliation{$^{19}$INAF - Osservatorio Astronomico di Roma, Via Frascati 33, 00078, Monte Porzio Catone (Roma), Italy}
%% T2/3 affiliations
\affiliation{$^{20}$Max Planck Institute for Astrophysics, Karl-Schwarzschild-Str. 1, 85748, Garching, Germany}
\affiliation{$^{21}$ Space Telescope Science Institute, 3700 San Martin Drive, Baltimore, MD 21218, USA}
\affiliation{$^{22}$ Department of Physics and Astronomy, Johns Hopkins University, 3400 North Charles Street, Baltimore, Maryland 21218, USA}
\affiliation{$^{23}$ Niels Bohr International Academy, Niels Bohr Institute, Blegdamsvej 17, 2100 Copenhagen, Denmark}
\affiliation{$^{24}$Institute of Theoretical Physics, KU Leuven, Celestijnenlaan 200D, 3001 Leuven, Belgium}
\affiliation{$^{25}$Leuven Gravity Institute, KU Leuven, Celestijnenlaan 200D, 3001 Leuven, Belgium}

\begin{abstract}

Upcoming space-based gravitational wave detectors such as LISA, the Laser Interferometer Space Antenna, will be sensitive to extreme- and intermediate-mass-ratio inspirals (EMRIs and IMRIs). These binaries are comprised of a supermassive black hole and a stellar-mass object or intermediate-mass black hole. Their detection will probe the structure of galactic nuclei and enable tests of general relativity. As these events will be observed over thousands of orbital cycles, they will be extremely sensitive to both the underlying spacetime and astrophysical environment, demanding exquisite theoretical models on both fronts to avoid biased or even erroneous results. In particular, many (E/)IMRIs are expected to occur within accretion discs around supermassive black holes, and the nonlinearities present when modeling these systems require numerical simulations. In preparation for future modeling of LISA sources, we have conducted a comparison between eight different hydrodynamical codes and applied them to the problem of a $q=10^{-4}$ mass ratio binary interacting with an accretion disc. Thicker discs appear more lenient, and all codes at sufficiently high resolutions are in good agreement with each other and analytical predictions. For thinner discs, beyond the reach of analytical models, we find substantial disagreement between 2D and 3D simulations and between different codes, including both the magnitude and sign of the torque. 
With time and energy efficiency in mind, codes that leverage moving meshes or grid-based Lagrangian remapping seem preferable, as do codes that can leverage graphical processing units and other energy-efficient hardware.
\end{abstract}

\section{Introduction}\label{sec:introduction}

There is significant and growing interest in how binaries evolve within accretion discs. Binary-disc interactions are indeed fundamental in protostellar binaries, protoplanetary systems, and supermassive black hole (SMBH) binaries \citep[e.g.,][]{2012ARA&A..50..211K,LaiDongMunoz2023}. A binary embedded in a gaseous disc will exchange  energy and angular momentum with the surrounding material, leading to orbital contraction or expansion depending on the physical properties of the system \citep{Goldreich1980}. Linear theory, both fully analytic \citep{Ward:1986, Artymowicz_1993, Ward:1997, Goldreich2003} and semi-analytic \citep[e.g.,][]{Korycansky1993,Tanaka2002,MirandaRafikov_2020,TanakaOkada2024,FairbairnRafikov_2025}, has enabled the study of low-mass secondaries, while numerical simulations have allowed confirmed linear theory and extend such studies to higher mass-ratio binaries \citep[e.g.][]{Paardekooper2023,2025MNRAS.543..565F}. These studies have revealed that the result of migration depends sensitively on the disc structure and thermodynamics \citep{Paardekooper:2008,Baruteau:2014}, as well as the binary mass ratio and its orbital properties. However, large regions of the binary-disc parameter space remain unexplored, particularly in the highly viscous, radiatively efficient disc conditions expected in active galactic nuclei (AGN) \citep[e.g.,][]{2002apa..book.....F}. Furthermore, linear theory becomes largely inapplicable at more equal mass ratios, when the Hill radius of the secondary object becomes larger than the scale height of the disc, making simulations necessary to study higher-mass-ratio binaries within thin discs.

The upcoming space-based gravitational wave (GW) interferometer, the Laser Interferometer Space Antenna (LISA), has energized research on how AGN accretion discs affect the orbits of embedded black holes, as those disc-binary interactions may influence the GW signals from these systems observed by LISA \citep[e.g.,][]{Yunes2011,2011PhRvD..84b4032K,Garg2022,2022PhRvL.129x1103C,2023MNRAS.521.4645Z,2024PhRvD.110j3005Z,2025PhRvD.111h4006D,2025PhRvD.111j4079C,2025PhRvD.112f3005Z,2025arXiv251002433D}. LISA is expected to detect a few to a few thousands extreme-mass-ratio inspirals (EMRIs) \citep{babak_science_2017,Pan2021a,Derdzinski2023} and up to hundreds of intermediate-mass-ratio inspirals (IMRIs) \citep{arca_sedda_merging_2021}, per year. In this study we focus on intermediate mass ratio binaries prior to GW inspiral. Formation of these IMRI precursors is supported by detections of intermediate mass black hole mergers with ground-based detectors (the LIGO-VIRGO-Kagra network), which provides evidence for formation channels that facilitate hierarchical mergers \citep{2025arXiv250818082T,2025arXiv250923897L,2022MNRAS.517.5827F,2022MNRAS.514.3886M}. Understanding this initial phase of the binary evolution, i.e. prior to significant GW-driven evolution, will enable us to predict if such systems will reach small separations to become GW sources, and, if so, how prevalent this channel may be for future GW detectors. 

Understanding binary interaction with AGN discs is important within the GW landscape, since the presence of a gas disc influences the dynamics and merger rates of BHs. Furthermore, the class of X-ray transients known as quasi-periodic eruptions, i.e. periodic flares in the soft X-ray band coming from the center of nearby galactic nuclei \citep{Miniutti2019,Giustini2020,Arcodia2021,Chakraborty2021,Quintin2023,Nicholl2024,HernandezGarcia2025}, if interpreted as counterparts of EMRIs \citep{Xian2021,Linial2023b,Franchini2023,Tagawa2023}, suggests that there might be plenty of stellar-mass objects in galactic nuclei that emit GWs in the LISA band. The periodicity of the detected sources, however, may suggest that relatively few would be observable by LISA owing to its sensitivity to lower frequencies \citep{2025arXiv250510488S}. Current predictions for rates and characteristics (see, e.g., \citealt{Pan2021a,Derdzinski2023}) carry underlying assumptions on migration theories developed primarily in low-viscosity, protoplanetary disc-like scenarios. Such assumptions are especially suspect when considering IMRIs, for which the assumption of linear perturber-disc interaction can severely break down. The applicability of planet-disc interaction theories to AGN discs remains uncertain, motivating a closer numerical examination of these systems.

Accurate modeling of binary-disc interactions will be crucial for correctly interpreting these GW signals. GWs offer unique observational windows into BH environments \citep{LISAastrophysics2023}. In particular, their signals will provide precise measurements of the central SMBH mass and spin and illuminating tests of general relativity in the strong-field regime \citep[e.g.,][]{2024arXiv240108085C}. However, even weak gas torques can appreciable alter binary orbital motion, potentially biasing parameter inference or interfering with precise tests of general relativity. Conversely, measurable gas-induced effects could offer a rare observational probe of inner AGN disc conditions, such as turbulence or thermal structure.

Previous numerical studies of binary-disc interactions span a wide range of methods and assumptions, making it difficult to disentangle physical effects from numerical artifacts. To address this, we perform a controlled comparison study through a suite of hydrodynamical simulations of the same fiducial disc model with an embedded perturber, using eight hydrodynamical codes employing various numerical techniques. Our goal is not to capture the full complexity of realistic AGN environments, but rather to assess code consistency in a simplified but astrophysically motivated setup. A few code comparison studies exist in the literature that explore similar systems. The most similar study was \cite{deValBorro2006}, which focused on the low viscosity, planet migration scenario. Other works explore planet migration in discs with self-gravity \citep{Fletcher2019}, planets in 3D radiatively inefficient discs \citep{Ziampras2023}, or focus on the equal-mass SMBH binary case \citep{Duffell2024}.

Our fiducial system represents an intermediate mass ratio binary with mass ratio $q\equiv M_2/M_1 = 10^{-4}$ embedded in a locally isothermal, highly viscous disc ($\alpha=0.1$) inspired by radiatively efficient AGN discs. Our study focuses on two disc regimes: a commonly studied, relatively thick disc, and a thinner disc that produces, for the secondary mass adopted, non-linear gas structures and lies beyond the reach of analytical predictions. Our goal is to isolate the effects of different numerical techniques, identify where codes agree or diverge, and use these insights to better evaluate complex models and guide future research toward more realistic predictions.

The paper outline is as follows. In Section~\ref{sec:disk-second-inter}, we present the theoretical work on linear perturber-disc interactions. We define our simulation setup and the various codes in Section~\ref{sec:numerics}. In Section~\ref{section:results}, we break down our results to focus on gas morphology and torques on the secondary. In Section~\ref{sec:discussion}, we dive deeper into past code comparison studies and discuss the relevance of this work for GW or electromagnetic signatures. 

\section{Linear Perturber-Disc Interactions} 
\label{sec:disk-second-inter} 

There has been considerable theoretical work on the gravitational interaction between discs and embedded secondaries. We summarize here the main results for a satellite on a circular orbit coplanar with the disc,\footnote{While the interaction can lead to a drift of the semi-major axis --~known as migration~--, we neglect the osculating eccentricity due to this drift and qualify the orbit as circular, even though, strictly speaking, it is locally a very tightly wound linear spiral rather than a circle.} and we focus on the case of a locally isothermal disc, in which the temperature is exclusively a function of the distance to the central primary. In this case, the disc response takes the form of a one-armed spiral wake \citep{2002MNRAS.330..950O}, generated by waves launched at the Lindblad resonances of the satellite with the disc and an additional perturbation in the co-orbital region. For linear perturber-disc interactions,\footnote{When we use the term ``linear'' throughout this work, we refer to the wave-launching process, where linearity demands that $R_H\ll H$, or that the disc scale height is larger than the Hill radius of the secondary, meaning that the wave-launching process occurs in a region of the disk where the background flow is dominated by the primary object rather than the secondary. A different kind of nonlinearity follows from the eventual steepening and shocking of the waves as they propagate through the disk \citep[e.g.][]{2001ApJ...552..793G}, but this process is not important for understanding the orbital evolution of the secondary.} angular momentum between the disc and secondary \textit{must} be mediated by resonances, usually divided between the Lindblad and corotation contributions, with a characteristic torque scale given by 

\begin{equation}
  \label{eq:1}
\Gamma_0\equiv\Sigma\Omega^2a^4q^2h^{-2},
\end{equation}

\noindent where $\Sigma$ is the disc's surface density at the secondary's orbit, $\Omega$ the orbital frequency of the latter, $a$ its semi-major axis, and $h\, {\equiv} \, H/r$ is the aspect ratio of the disc ($H$ being the pressure scale height) \citep{Goldreich1980}. 
Notably, the magnitude of the torque on the binary due to the disc scales $\propto q^2$, as does the rate of change of the binary separation due to GW emission, so disc-induced migration will play a comparatively important role in modifying GW-driven inspirals as long as linear theory is applicable. 

Both the corotation and Lindblad torque contributions depend sensitively on the structure of the accretion disc; often these torques are approximated in terms of local surface density ($\Sigma$) and temperature ($\Theta$) gradients,

\begin{equation}\label{eq:2}
    \mathcal{A}=-\frac{{\rm d}\log\Sigma}{{\rm d}\log r}\mbox{~~and~~}\mathcal{B}=-\frac{{\rm d}\log \Theta}{{\rm d}\log r},
\end{equation}

\noindent such that the Lindblad and corotation torques on the secondary are estimated using $\Gamma_{\rm L}=f_{\rm L}(\mathcal{A}, \mathcal{B})\Gamma_0$ and $\Gamma_{\rm C}=f_{\rm C}(\mathcal{A},\mathcal{B})\Gamma_0$ respectively.\footnote{However, we must caution that even within linear theory, these simple fitting functions can be inadequate to accurately describe the dependence of torques on disc properties, \citep[e.g.,][]{FairbairnRafikov_2025}.} We define the functions $f_{\rm L}$ and $f_{\rm C}$ below as they depend on the geometry, i.e. 2D or 3D, of the system.

The validity of linear torque estimates \textit{requires} that the feedback of the wake on the disc is negligible, which is satisfied when $R_{\rm H}=a(q/3)^{1/3} \ll H$, where $R_{\rm H}$ is the \citet{hill1878} radius of the secondary, or equivalently $q\ll h^3$ \citep{GoodmanRafikov:2001}. Within this low-mass-ratio limit, numerous approaches are taken to approximate the response of the disc to gravitational perturbations. The full set of linearized fluid equations are still sufficiently complex to demand numerical solution without further simplification \citep[e.g.,][]{Korycansky1993,FairbairnRafikov_2025}. In a point approximation, omitting nonlocal details of the discs response, it is possible to analytically evaluate the torque contributions from each resonance individually \citep[e.g.,][]{1993ApJ...419..166A,Ward:1997}.  Still, a common middle ground is also taken by numerically solving the linearized fluid equations in a modified shearing box approximation \citep[e.g.,][]{TanakaI:2002,TanakaOkada2024}, still making some simplifying assumptions but capturing physics such as the finite widths of resonances. Others have fit $f_{\rm L}$ and $f_{\rm C}$ based on fully nonlinear hydrodynamics simulations in the linear regime \citep[e.g.,][]{2011MNRAS.410..293P}.

Especially in two-dimensional calculations, one must ``soften'' the gravitational potential on a length to avoid divergences, with the added benefit of slightly better emulating three-dimensional effects \citep[see, however,][]{2025arXiv250904282C}. In the present study we have approximated the gravitational potential of the secondary to be

\begin{equation}\label{eq:4}
    \Phi_2=-\frac{GM_2}{\sqrt{|\mathbf{r}-\mathbf{r_2}|^2+\epsilon^2}},
\end{equation}

\noindent where $G$ is the gravitational constant and we set the softening length to $\epsilon = 0.6H(r_2)$. This value is assumed in the following expressions.
By numerically solving the linearized fluid equations in a modified shearing box setting, \citet{TanakaOkada2024} derived the formulas

\begin{align}\label{eq:linear3D}
    f_{\rm L}^{\rm T24} &= -2.382 + 0.094\mathcal{A} - 1.297\mathcal{B} ,\\
    f_{\rm C}^{\rm T24} &= 0.945 - 0.63\mathcal{A} + 0.858\mathcal{B} ,\\ 
\end{align}

\noindent for three-dimensional interactions. By fitting the results of nonlinear 2D hydrodynamics simulations, \citet{2010MNRAS.401.1950P} derived the expressions

\begin{align}\label{eq:linear2D}
    f_{\rm L}^\mathrm{P11}&=-1.87+0.075\mathcal{A}-1.27\mathcal{B} ,\\
    f_{\rm C}^\mathrm{P11} &=0.63-0.42\mathcal{A}+0.81\mathcal{B},
\end{align}
which we have evaluated specifically for our softening length $\epsilon$ expressed above. 

Corotation torques are especially affected by weakly nonlinear effects, in particular the influence of viscosity on fluid horseshoe trajectories and gas depletion near the orbit of the secondary. The former effect was cleanly illustrated by \citet{2009MNRAS.394.2283P}, who showed that the corotation torque deviates from the linear value after a time marginally larger than the orbital time, and progressively switches to the ``horseshoe drag'' regime. The horseshoe drag regime is named for the characteristic U-shaped streamlines in the corotating frame, familiar from the study of the circular restricted three-body problem. Gas parcels within this regime alternate between trajectories inside and outside the orbit of the secondary, executing U-turns near the perturber. The transition from linear to horseshoe drag occurs on a timescale comparable to the horseshoe U-turn execution time. If the disc viscosity is very large, as in the discs considered here \citep{HosseinNouri2024}, the horseshoe drag regime itself is never attained and the corotation torque remains perpetually in the linear regime \citep{2010ApJ...723.1393M,2010MNRAS.401.1950P}. \emph{This is a key difference between protoplanets embedded in protoplanetary discs and E/IMRIs}. Although the latter case corresponds to simpler calculations, the regime of very high viscosity remains relatively unexplored in numerical simulations.

Once the mass ratio of the secondary approaches $q\sim h^3$ (in the thin-disc case we consider below, $q\approx3.7h^3$), the waves launched by the secondary begin to alter the structure of the disc, reducing the surface density in the co-orbital region \citep[e.g.,][]{1979MNRAS.186..799L,Duffell2013,2024MNRAS.534.1394C}. \citet{Kanagawa2018} propose that, in terms of the gap-depth parameter $K\equiv q^2h^{-5}\alpha^{-1}$ (where $\alpha$ sets the kinematic viscosity according to $\nu=\alpha H^2\Omega^{-1}$), the total torque acting on the secondary can be approximated as

\begin{equation}\label{eq:kanagawa}
T = \frac{\Gamma_{\rm L} + \Gamma_{\rm C}\exp{(-K/20)}}{1+0.04K}.
\end{equation}

As we will show later, this approximation falls short in the IMRI regime, underscoring the importance of simulations when studying these systems.\footnote{Our thin-disc case has $K\approx4.1$, leading to negligible modifications to the standard linear estimate.}

%--------------------------------------------------------------------
\section{The Numerical Problem}\label{sec:numerics}

In this section , we describe the disc models we take as the initial conditions of our simulations and introduce the codes we use to simulate embedded satellites within each disc. While the binaries in our model have a fixed orbital separation and no GW-driven evolution, we refer to the system as an `IMRI' for brevity. 
We assume that binary separation is sufficiently large for GW emission to have a negligible effect on the orbit over the course of each simulation. We summarize the salient details of each code in Table~\ref{tab:sims}, and provide additional information in Appendix~\ref{app:codes}.

\subsection{Initial Conditions}\label{sec:ICs}

We model an intermediate-mass-ratio binary system with mass ratio $q=10^{-4}$ and a separation $a=1$ placed on a circular orbit, embedded in a viscous, geometrically thin, fluid disc. The total mass of the binary is given by $M=M_1+M_2$. The disc is parameterized by a constant aspect ratio $H/r$ (or, equivalently, the azimuthal Mach number $\mathcal{M} = (H/r)^{-1}$ computed using the Keplerian velocity). We approximate dissipation within the disc using the \citet{SS73} prescription for the kinematic viscosity, setting $\nu = \alpha c_{\rm s} H$, where $c_{\rm s}(r)$ is the sound speed and we set $\alpha=0.1$. This value of $\alpha$ is of the order expected to stem from 
magnetohydrodynamic turbulence in AGN discs \citep[e.g.,][]{HosseinNouri2024}.  

We assume a locally isothermal equation of state, such that

\begin{equation}
c_{\rm s}^2=\frac{GM}{r\mathcal{M}^2},
\end{equation}

\noindent and thus $\nu=\alpha\mathcal{M}^{-2}r^{2}\Omega\propto r^{1/2}$. Accordingly, we set the initial surface density profile of the disc according to $\Sigma(t=0) = \Sigma_0 (r/a)^{-1/2}$, where $a=1$ is the semi-major axis of the binary orbit and $\Sigma_0$ is the corresponding surface density, and set the initial radial velocity of the fluid to $v_r=-(3/2)\alpha c_{\rm s}/\mathcal{M}$, to result in an approximately constant mass flux throughout, at least initially. The initial azimuthal velocity of the disc is set to

\begin{equation}
v_{\phi}^2 = \frac{G M}{r} - \frac{3}{2} \frac{1}{\mathcal{M}^2}\frac{G M}{r},
\end{equation}

\noindent accounting for pressure gradients within the disc.

These initial conditions were implemented slightly differently in the particle-based and grid-based codes. The grid-based codes treated the disc in the test-fluid limit, making the surface density scale-free, while the particle-based codes required explicitly specifying a disc mass ($M_{\rm d}=10^{-6}\,M$, to reduce its dynamical impact). This choice precludes significant changes in $a$ over the course of each simulation, effectively holding fixed the orbital separation. Since we report results in terms of the surface density at the location of the secondary, the difference in scale does not affect our results. All of the particle codes were used in three dimensions, specifying the density to match the locally isothermal temperature profile of the disc:

\begin{equation}
    \rho(r,z) = \Sigma_0 \left(\frac{r}{a}\right)^{-1/2} \exp{\left(\frac{-z^2}{2H^2}\right)}.
    \label{eq:density}
\end{equation}

The particle codes sampled particle positions to match this density profile using Monte Carlo methods \citep{Price2018}.

Each simulation covered a domain of at least $0.5<r/a<3$, although some codes simulated more of the disc beyond that range. We provide those, and other code-specific details in Section~\ref{sec:codes}.

We first simulate an IMRI in a thicker disc with $h=0.1$. Although less representative of a standard AGN disc, this thicker disc satisfies $q\lesssim h^3$ and thus the estimates discussed in Section~\ref{sec:disk-second-inter} can provide some analytical basis for comparison. We also studied a thinner disc, with $h=0.03$, more representative of a typical AGN disc but in a regime where linear theory is inapplicable.  

\subsection{Codes}\label{sec:codes}

We employed a variety of codes to model the disc-IMRI interaction problems described above. The underlying fluid equations were solved in both the Eulerian formulation with grid-based methods, in the Lagrangian formulation using particle-based methods, and in a semi-Lagrangian moving-mesh formulation. These equations were solved in both Cartesian and cylindrical polar coordinates. Moreover, the Eulerian codes leveraged both finite-volume and finite-difference techniques. We summarize the salient features of these codes below, and provide additional details in Appendix~\ref{app:codes}. 

\begin{table*}
    \centering
    \setlength{\extrarowheight}{5pt}
\begin{tabular}{|c|c|c|c|c|}
\hline
Code & Type / Geometry & $H/r$ & \makecell{Standard resolution setting \\ $N_{\rm particles}$ or $N_{\rm cells}$ } & Standard resolution at $r=a$ ($\Delta r$) \\
\hline

\multirow{2}{*}{\texttt{ATHENA++}} 
    & Eulerian, finite volume & 0.1  & $N_r=512$, $N_\phi=1792$ & $ 0.0035a$  \\
    \cline{3-5}
    & Cylindrical polar                  & 0.03 & $N_r=512$, $N_\phi=1792$ & $0.0035a$ \\
\hline

\multirow{2}{*}{\texttt{DISCO}}
    & Moving-mesh, finite volume & 0.1 & $N_r=600$, $N_\phi=2107$  & $0.003a$ \\
    \cline{3-5}
    &  Cylindrical polar            &     0.03                  & $N_r=600$, $N_\phi=2107$  & $0.003a$ \\
\hline

\multirow{2}{*}{\texttt{DISCO v2}} 
    & Moving mesh, finite volume & 0.1 & $N_r=576$, $N_\phi=2017$ & $0.0035a$  \\
    \cline{3-5}
    &  Cylindrical polar            & 0.03                      &   $N_r=576$, $N_\phi=2017$ & $0.0035a$  \\
\hline

\multirow{2}{*}{\texttt{FARGO3D}} 
    & Eulerian, finite difference & \multirow{1}{*}{0.1}  & $N_r=448$, $N_\phi=1562$ & $0.004a$ \\
    \cline{4-5}

    \cline{3-5}
    & Cylindrical polar                & \multirow{1}{*}{0.03} & $N_r=448$, $N_\phi=1562$ & $0.004a$ \\
    %\cline{4-5}

\hline

\multirow{2}{*}{\texttt{GASOLINE}} 
    & Lagrangian, smoothed-particle & 0.1  & $10^7$ &  $0.01a$ \\
    \cline{3-5}
    & Cartesian & 0.03 & $10^7$ & $0.01a$ \\
\hline

\multirow{3}{*}{\texttt{GIZMO}} 
    &  Lagrangian, meshless finite-mass                              & 0.1 & $10^7$             & $0.01a$ \\
    \cline{3-5}
    &     Cartesian                        &           0.03            & $10^7$ & $0.01a$ \\
    \cline{3-5}
    &                             &           0.03            & $10^7$ + splitting & $0.003a$ \\
\hline

\multirow{2}{*}{\texttt{PHANTOM}} 
    & Lagrangian, smoothed-particle  & 0.1 & $10^7$ & $0.01a$ \\
    \cline{3-5}
    & Cartesian &0.03 & $10^7$ & $0.01a$ \\
    \hline

\multirow{2}{*}{\texttt{RAMSES}} 
    & Eulerian, finite volume & 0.1  & $4.5 \times 10^6$ & $2\times10^{-3} a$ \\
    \cline{3-5}
    &  Cartesian            & 0.03 & $4.5 \times 10^6$ & $2\times10^{-3} a$ \\
\hline
\end{tabular}
    \vspace{2mm}
    \caption{List of codes used in this study. }
    \label{tab:sims}
\end{table*}

\subsubsection{\texttt{ATHENA++}}

\texttt{ATHENA++} is a finite-volume code, a rewrite of the \texttt{Athena} code \citep{2008ApJS..178..137S} using C++, that supports Cartesian, cylindrical, and spherical geometries \citep{2020ApJS..249....4S}. \texttt{ATHENA++} also features block-based adaptive mesh refinement (AMR), static mesh refinement, and dynamic task scheduling. In this work, \texttt{ATHENA++} has been used to solve the equations of viscous hydrodynamics in a two-dimensional logically-Cartesian cylindrical geometry using second-order methods, without orbital advection. The domain extended radially from $r_{\rm min} = 0.5$ to $r_{\rm max} = 3$ and was resolved using $N_r=512$ cells, spaced log-uniformly, holding $\Delta r/r$ constant. The cell aspect ratio was held near unity by using $N_\phi=1792$ cells in each annulus.

\subsubsection{\texttt{DISCO}}

\texttt{DISCO} is a moving-mesh finite-volume code that is specifically tailored to modeling accretion discs \citep{Duffell2016}. The code utilizes a cylindrical mesh that can move azimuthally along with the gas, which significantly reduces advection errors even for highly supersonic flows. For this work, we adopt a two-dimensional cylindrical grid with logarithmically spaced zones from $r_{\rm min} = 0.5$ to $r_{\rm max} = 3$. The primary BH is excised from the simulation domain and modeled as a point-mass potential at $r=0$. An HLLC Riemann solver is adopted for solving for the flux between adjacent cell interfaces. The azimuthal velocity of the mesh was set by the volume-averaged value within each annulus. Boundary conditions are fixed at the initial conditions described in  Section~\ref{sec:ICs}. 

The original version of \texttt{DISCO} ignored a number of terms in the velocity shear tensor used in its viscosity implementation to allow for rearranging terms between fluxes and source terms and thereby simplifying the viscosity implementation on a moving mesh, as described in the appendix of \citet{Duffell2016}. We also carried out a set of simulations using an updated version of \texttt{DISCO}, which used the full velocity shear tensor as described in Appendix~A of \citet{2021ApJ...921...71D} but was otherwise identical to the setup described above; we will refer to this code as \texttt{DISCO v2} through the remainder of the paper. 

\subsubsection{\texttt{FARGO3D}}\label{desc:fargo3d}

\texttt{FARGO3D} is a publicly available staggered mesh code \citep{2016ApJS..223...11B}. Although based on finite differences, fluxes of mass and momentum are computed at the interfaces between adjacent cells so that mass and angular momentum are conserved to platform accuracy. It includes orbital advection \citep{2000A&AS..141..165M}, which eliminates the timestep restriction from azimuthal advection, often resulting in much larger allowed timesteps. Time consuming routines in \texttt{FARGO3D} can be automatically translated to CUDA (provided they follow a well-documented template format), hence \texttt{FARGO3D} can run on NVIDIA's graphical processing units (GPUs).
Despite what its name suggests, \texttt{FARGO3D} is not limited to three-dimensional setups and can also run one- or two-dimensional problems. Here we have used it in a two-dimensional, cylindrical configuration, with $N_r$ radial bins in geometric sequence with same limits as the other mesh codes used in the present study ($0.5$ to $3$), and $N_\phi$ equally spaced azimuthal sectors. Unless stated otherwise $N_r=448$ and $N_\phi=1562$. 

For comparison with the particle-based codes, we ran two simulations in 3D as well. The thick-disc simulation used $N_r=340$, $N_\phi=1187$, and $N_\theta=114$; and the thin-disc simulation used $N_r=448$, $N_\phi=1562$, and $N_\theta=45$. All simulations are performed in the frame corotating with the secondary and centered on the primary. 

\subsubsection{\texttt{GASOLINE}}

\textsc{Gasoline2} \citep[][referred to as simply \texttt{GASOLINE} throughout this work]{Wadsley_et_al_2017} is a parallel $N$-body smoothed-particle hydrodynamic (SPH) and magneto-hydrodynamic \citep[][]{Wissing_Shen_2020} code, based on the $N$-body KD-tree code \textsc{pkdgrav} \citep[][]{Stadel_2001} and with a modern implementation of SPH that incorporates several improvements with respect to previous versions of the code \citep[][]{Wadsley_et_al_2004}. These include the geometric density average force expression \citep[][]{Monaghan_1992,Ritchie_Thomas_2001,Keller_et_al_2014}, which minimizes numerical surface tension effects; gradient-based shock detection \citep[][]{Wadsley_et_al_2017}, which limits artificial viscosity; upgraded kernels \citep[][]{Wendland_1995,Dehnen_Aly_2012,Keller_et_al_2014}, which avoid pairing/clumping instability and allow for larger neighbor numbers; and time-dependent local viscosity limiters \citep[][]{Morris_Monaghan_1997,Cullen_Dehnen_2010}. For the specific purposes of this project, we recently added physical viscosity terms, using both the first-derivative and second-derivative method \citep[][]{Price2018}, and a locally isothermal equation of state. Unlike the other particle-based codes used in this work (see below), \texttt{GASOLINE} does not allow one to ignore the self-gravity of the fluid. In addition to potentially changing the physical evolution of the system, computing the self-gravity of the disc typically adds considerable expense (see Table~\ref{tab:CO2}). To reduce the physical impact of disc self-gravity in this study, the mass of the fluid disc was set to $10^{-6}$ times that of the binary system. Thus, each of the $10^7$ particles in this simulation had a mass of $10^{-9}$ times that of the secondary. Owing to their high computational cost, the \texttt{GASOLINE} simulations were halted after 33 and 50 binary orbits for the alignment and thin disc run, respectively.

\subsubsection{\texttt{GIZMO}}

\texttt{GIZMO}  is a multi-methods and multi-physics code for hydrodynamics simulations with both SPH, meshless finite-mass (MFM), and meshless finite-volume (MFV) solvers implemented \citep{Hopkins2015}.  The version of the code used in this study is the same as the one in \cite{Franchini2022,Franchini2023}. The code has been run in MFM mode, i.e. the fully Lagrangian scheme in which the faces shared by neighbor particles are forced to move at the velocity of the contact wave, in order to ensure a vanishing mass flux between particles. Compared to \cite{Franchini2022}, adaptive particle splitting refinement is not employed in the alignment run, but only in the thinner disc case, to investigate the effect of resolution on the torques. The method is similar to the one used in \cite{Franchini2022}, although we split particles in an annulus close to the secondary. For both runs in this study, we employ 10 million particles. The initial disc extended from $r_{\rm in}=0.5$ to $r_{\rm out}=10$ to reduce the effects of the finite disc size and mimic a constant inflow of mass from larger radii. We neglect the self-gravity of the disc. The binary components are modeled as sink particles \citep{Bate1995} and we only allow accretion of gas particles onto the primary. Accretion is done by simply removing particles that enter the accretion radius of the primary, set to $0.2$ for both runs. Our analysis only considers the region between $r=0.5$ and $r=3$.

\subsubsection{\texttt{PHANTOM}}

\texttt{PHANTOM} is a fast, parallel, modular and low-memory SPH and magnetohydrodynamics code \citep{Price2018}. For the present work, we initialized a disc with inner and outer radii of $r_{\rm in}=0.5$ and $r_{\rm out}=10$ , respectively. For comparison with the other codes, during the analysis the disc is truncated at $r=3$. The binary is modeled using a gravitational potential (instead of using sink particles) that describes a binary with total mass $M=1$ and mass ratio $q=10^{-4}$. Accretion of gas particles was suppressed, and the self-gravity of the disc was neglected.

\subsubsection{\texttt{RAMSES}}

\texttt{RAMSES} is an adaptive mesh refinement (AMR) finite-volume code for (radiative magneto-) hydrodynamics simulations \citep{Teyssier2002}. The equations of hydrodynamics are solved on a multi-dimensional Cartesian grid using a second order Godunov scheme. \texttt{RAMSES} has been used for a wide range of applications, from protostar \citep[e.g.,][]{ahmad_2025} and pre-/protostellar disc formation/evolution \citep[e.g.,][]{lebreuilly_2021} to parsec-scale \citep[e.g.,][]{lescaudron_dynamical_2023} and cosmological simulations \citep[e.g.,][]{dubois_2021}. While \texttt{RAMSES} has been previously used for isolated disc simulations \citep[e.g.,][]{brucy_2021}, to our knowledge, this is the first published application of \texttt{RAMSES} using an explicit viscosity.

For the present application, a 2D geometry was chosen and the flux between adjacent cells was computed with a HLLC Riemann solver. We do not use sink particles, and instead design a specific treatment for central cells in order to mimic steady-state inner boundary conditions. The primary BH is modeled using a cubic spline kernel potential which reduces to a Newtonian potential for $r\, {>} \, 0.5$ \citep{2001NewA....6...79S}. The $\alpha$-viscosity treatment is not native and we used a new implementation from \citet[][see Appendix ~\ref{app:ramses}]{Edwin_paper}.

\subsection{Resolution criteria}

All simulations are initialized with a minimum resolution criterion set in the region of the secondary (see Table~\ref{tab:sims}). For grid codes, all runs satisfy a spatial resolution within the radius of the secondary that is a fraction of the Hill sphere: $\Delta x/x \lesssim R_{\rm H}/(6 r)$, where $x$ is a distance coordinate.

Particle codes use an effectively equivalent spatial resolution which can be computed starting from the smoothing length. For these codes, we take the average of the distribution of the smoothing lengths $h_{\rm sml}$ in the region of interest (i.e. within the secondary Hill radius) and then we define the inter-particle spacing as $\Delta x/x = h_{\rm sml}/h_{\rm fact}$, where $h_{\rm fact}=1.2$ for $58$ neighbors (see Section~2.1.3 in \citealt{Price2018}). Since we model a finite-mass disc without external particle injection from the outer edge, the mass of the disk and effective resolution near the secondary tend to decrease with time because of accretion onto the primary object and viscous spreading of the accretion disk outward. The resolution criterion is met in the alignment case by modeling the disc with $10^7$ particles. In the thin disc case, the \texttt{GIZMO} code uses hyper-Lagrangian refinement to achieve an inter-particle spacing $\Delta x/x = 0.003$ around the secondary.

\section{Results}\label{section:results}

First, in Section~\ref{sec:thickDisc}, we review simulations of a $q=10^{-4}$ secondary in an $H/r=0.1$, $\alpha=0.1$ disc, which we refer to as the ``thick disc'' case or ``alignment run'' since these parameters are amenable to linear theory, as described in Section~\ref{sec:disk-second-inter}. Following that, we move to simulations of thinner ($H/r=0.03$) discs in Section~\ref{sec:thinDisc}, which are more representative of AGN discs but to which linear theory is not applicable owing to mild nonlinearities. Afterwards, we investigate convergence and the differences between 2D and 3D calculations in Section~\ref{sec:convergence}, and torque variability in Section~\ref{sec:torquevar}.

\subsection{Alignment run -- thick discs}\label{sec:thickDisc}

\subsubsection{Gas morphology}

\begin{figure*}
    \centering
    \includegraphics[width=.95\textwidth]{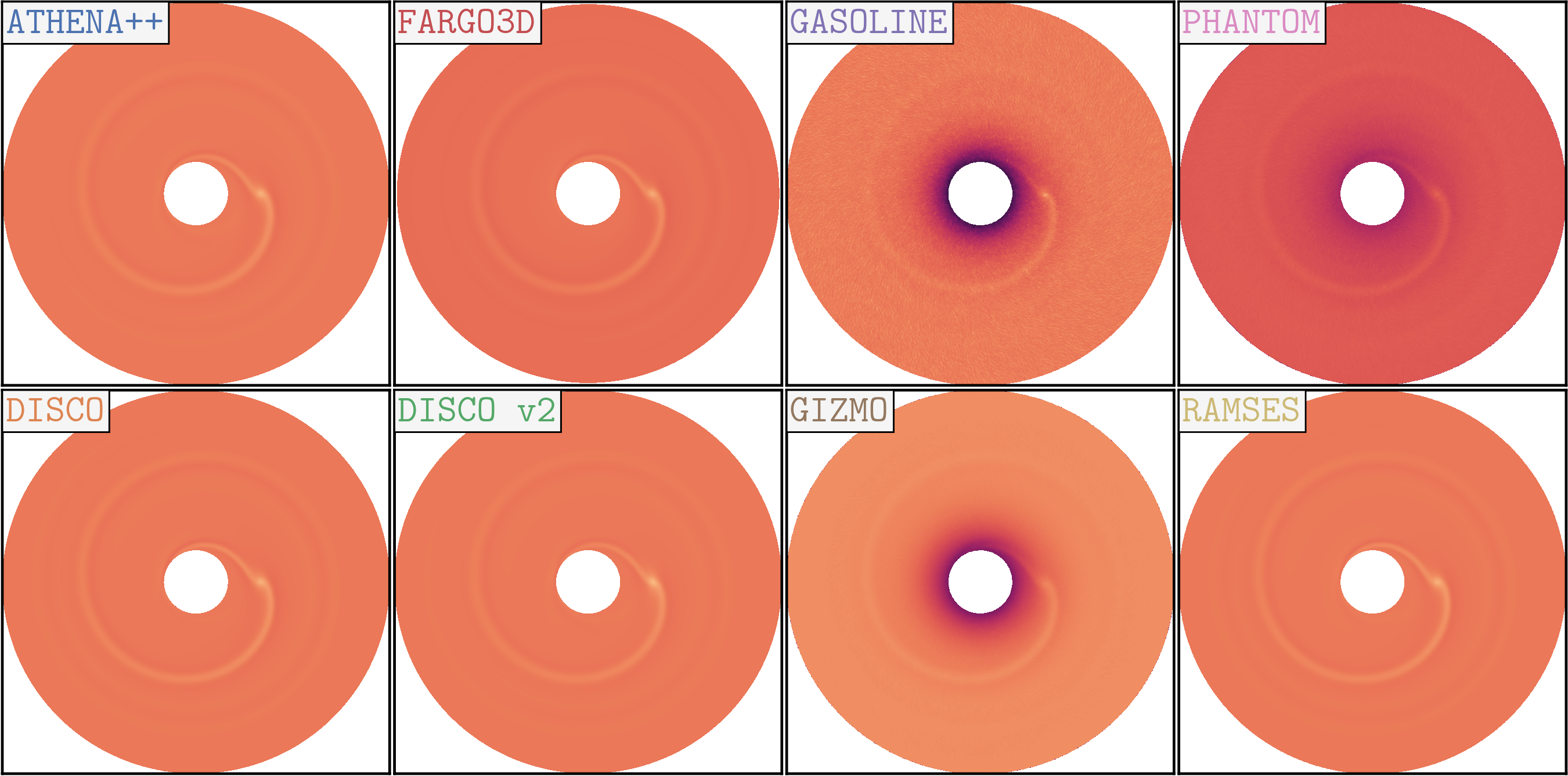}
    \includegraphics[width=.95\textwidth]{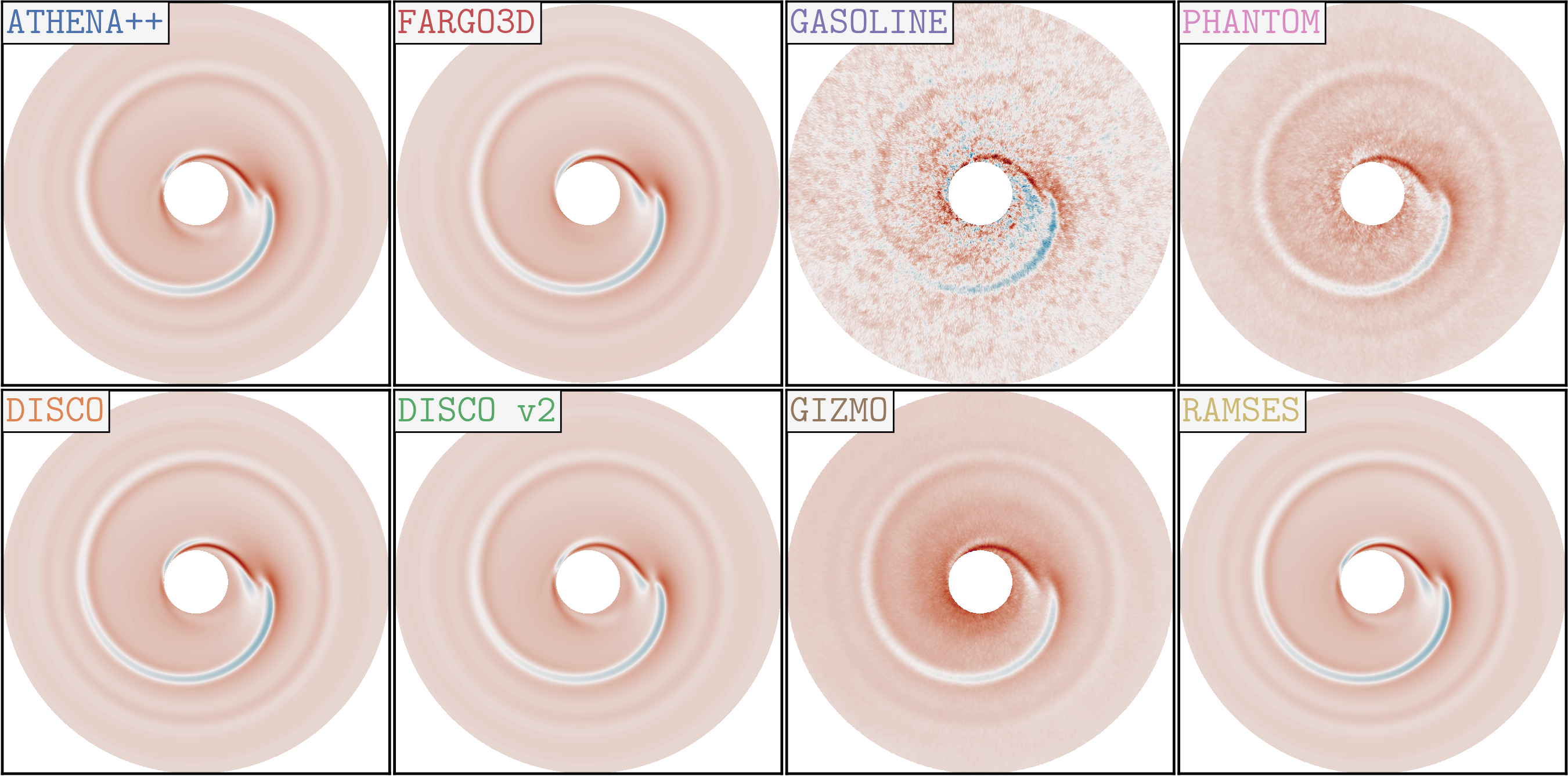}
    \caption{Surface density map (top two rows) and radial velocity map (bottom two rows) for the alignment run ($H/r=0.1$) after 100 orbits (33 for \texttt{GASOLINE}. Specifically, the top panels plot $\log_{10}{(\Sigma/\Sigma(t=0)}$ over the range $[-0.2,0.1]$; the bottom panels plot $v_r$ on a scale of $[-0.01,0.01]$, with red indicating negative velocities. Thus, typical deviations in the surface density due to the forcing from the secondary are on the order of $\sim$10\%, and deviations in the velocity profile of the disc are of order $\sim$1\%. The grid-based codes maintained surface density profiles with the same slope as the initial condition, whereas the particle-based code developed slightly shallower profiles. Both maps illustrate the trailing spiral arm launched by the perturber, which damps away before reaching the outer boundary but interacts with the inner boundary at $r = 0.5a$. A complimentary view in $(r,\phi)$ is provided in Appendix~\ref{app:extra} in Fig.\ref{fig:surfacedensitymap_hr01_unrolled}.}
    \label{fig:surfacedensitymap_vr_hr01}
\end{figure*}

As the secondary interacts with the disc, launching waves near Lindblad resonances, it excites a coherent spiral arm in the disc \citep{2002MNRAS.330..950O}, visualized in both the surface density perturbations ($\log_{10}{[\Sigma/\Sigma(t=0)]}$) and radial velocity in Figure~\ref{fig:surfacedensitymap_vr_hr01} after $100$ orbits for all codes, with the exception of \texttt{GASOLINE} for which we only reached $33$ orbits. The size of the maps is $6a\times 6a$.
The background velocity field is that of slow viscous inflow, over which the wave launched by the pertruber is superimposed. While many of the simulations maintained background surface density profiles with the same slope ($\propto r^{-1/2}$) as the initial condition, the particle-based codes developed central underdensities. These differences are likely because of boundary conditions, as the grid-based codes fixed interior and exterior values to the initial conditions, while the particle codes simulated viscously spreading discs with a finite mass, allowing material to enter the $0.5a$ boundary and reach the sink radius of the primary.

Across all codes, the loosely wound structure of the spiral arm is quite consistent, although in the \texttt{GASOLINE} simulation it is almost washed out by particle noise. 
Often-minor deviations occur near the inner edge of the domain: for example, the interaction between the spiral arm and the inner boundary in \texttt{FARGO3D} clearly differs from \texttt{ATHENA++}, \texttt{DISCO v2}, and \texttt{RAMSES}, as seen in the lower left quadrant. This is somewhat interesting, because \texttt{DISCO v2} used a moving mesh, \texttt{ATHENA++} a fixed mesh without operator splitting for azimuthal advection, and \texttt{RAMSES} used a cartesian domain with a central damping zone rather than an inner boundary, suggesting that this difference follows from the primary-centered coordinate system used by \texttt{FARGO3D} (as opposed to the barycentric coordinate system used in every other case). These contribute to some of the minor torque discrepancies seen between \texttt{FARGO3D} and the other grid codes in the inner regions of the disc in Section~\ref{sec:r_density}. We also see that all  the grid codes display an outward radial velocity wave in the upper left quadrant of the inner boundary, which is absent in \texttt{PHANTOM} and \texttt{GIZMO} and not easily distinguishable from plotting artifacts in \texttt{GASOLINE}. This suggests that this feature is potentially a consequence of the fixed inner boundary condition used in the grid codes.

\subsubsection{Torques}\label{sec:torque}

\begin{figure}
    \centering
    \includegraphics[width=.475\textwidth]{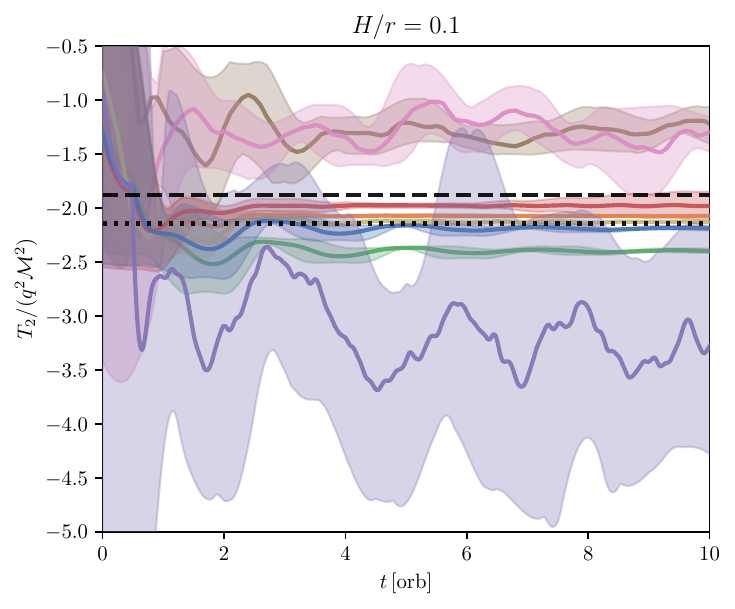}\\
    \includegraphics[width=.48\textwidth]{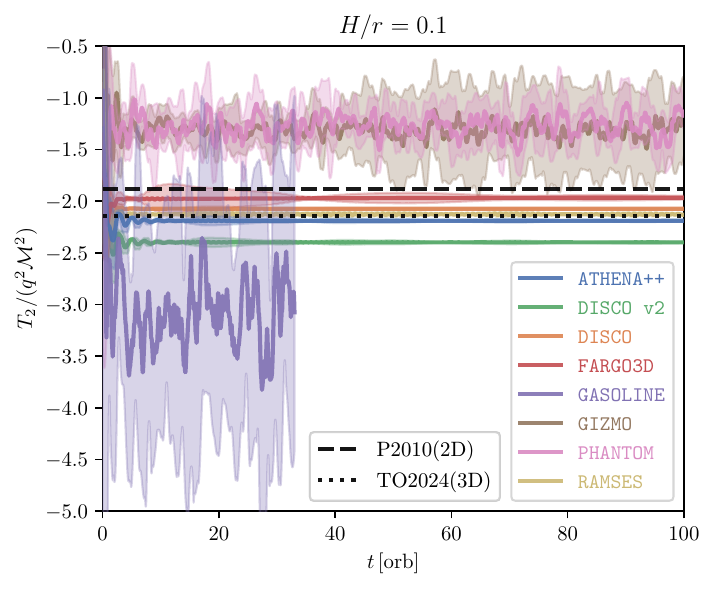}
    \caption{
    Torque on the secondary versus time in orbits for the alignment run ($H/r=0.1$), over the first 10 orbits (top panel) and the entire 100 orbit (33 for \texttt{GASOLINE}) series (bottom panel). 
    The black dashed and dotted lines show analytical predictions from Equation~\eqref{eq:linear2D} \citep[][-1.875]{2010MNRAS.401.1950P} for the 2D case and Equation~\eqref{eq:linear3D} \citep[][-2.144]{TanakaOkada2024} for the 3D case, respectively. The torque is averaged over a 1-orbit window, and the envelope on each lines delineates the
    the variability seen in the unsmoothed torque data.} 
    \label{fig:torque_h01_M2}
\end{figure}

We computed the gravitational torque exerted by the non-axisymmetric perturbations in the gaseous disc onto the binary. The torque on each binary component is given by

\begin{equation}
    T_i = \int {\rm d}A\,\Sigma\mathbf{r}\times(\nabla\Phi_i)=\int {\rm d}A\,\Sigma\partial_\phi\Phi_i,
\end{equation}

\noindent where $i=\{1,2\}$ correspond to the primary and secondary, respectively, and the sign of the torque is chosen to quantify the change in angular momentum of the binary rather than that of the disc. The torque on each binary member was computed by summing over all cells or particles within the disc inside the range $0.5<r/a<3.0$.
The torque was then normalized with the surface density at the location of the secondary in order for all codes to have the same magnitude, regardless of the finite vs infinite disc mass choice, and divided by $q^2\mathcal{M}^2$ (see Equation (\ref{eq:1})).

Figure~\ref{fig:torque_h01_M2} shows the time evolution of the gas torque on the secondary. 
In each case the torque on the binary reached a quasi-steady state after couple orbital periods, or more importantly a few disc sound-crossing times. This is natural in the linear regime studied in this test, since the disc structure is not significantly modified (see Figure~\ref{fig:surfacedensitymap_vr_hr01}) thus limiting the importance of viscosity, and so the structure of the perturbed disc settles in after a few sound crossing times, rather than a viscous time. The top panel shows the first 10 orbits of the simulations, illustrating the rapid convergence of the torque within this window. The dashed and dotted lines in Figure~\ref{fig:torque_h01_M2} denote the linear estimates for the 2D \citep{2010MNRAS.401.1950P} and the 3D torque \citep{TanakaOkada2024}, respectively. All codes find a negative torque, and the time-averaged torque is in each case in the general vicinity of fitting formula predictions. In the case of \texttt{DISCO}, this shows that the density deviations are sufficiently small (see Figure~\ref{fig:surfacedensitymap_vr_hr01}) for the assumptions underlying our simple viscosity implementation to remain acceptable. The torques measured by the (3D) particle codes depart somewhat from the commensurate prediction from \cite{TanakaOkada2024}, which did not include any softening of the potential: \texttt{GIZMO} and \texttt{PHANTOM} found values in agreement with each other but smaller in magnitude than the value from \cite{TanakaOkada2024}, while \texttt{GASOLINE} found a significantly more negative value. In order to pinpoint the origin of the discrepancy between the \texttt{GIZMO} value (and presumably \texttt{PHANTOM}) and the linear prediction, we ran another \texttt{GIZMO} simulation using a smaller softening for the secondary ($\epsilon=0.018a$), finding excellent agreement with the analytical estimate, as shown in Section~\ref{sec:convergence}.

The bottom panel of Figure~\ref{fig:torque_h01_M2} shows the torque time series averaged over a 1-orbit rolling window, with envelopes encompassing the rolling minimum and maximum of the variability. The particle-based codes show pronounced variability in the torque, which is generally expected, given their discretization scheme. 
Indeed, the use of a large but finite number of particles has a Poisson noise associated with their distribution, which is reflected in the instantaneous torque measurement. The torques measured in the \texttt{GASOLINE} simulations deviate significantly from those of the other particle-based codes, suggesting that disc self-gravity strongly affects the disc asymmetry and torque, even for low disk masses.

\subsubsection{Radial density profiles}\label{sec:r_density}

\begin{figure}
    \includegraphics[width=.49\textwidth]{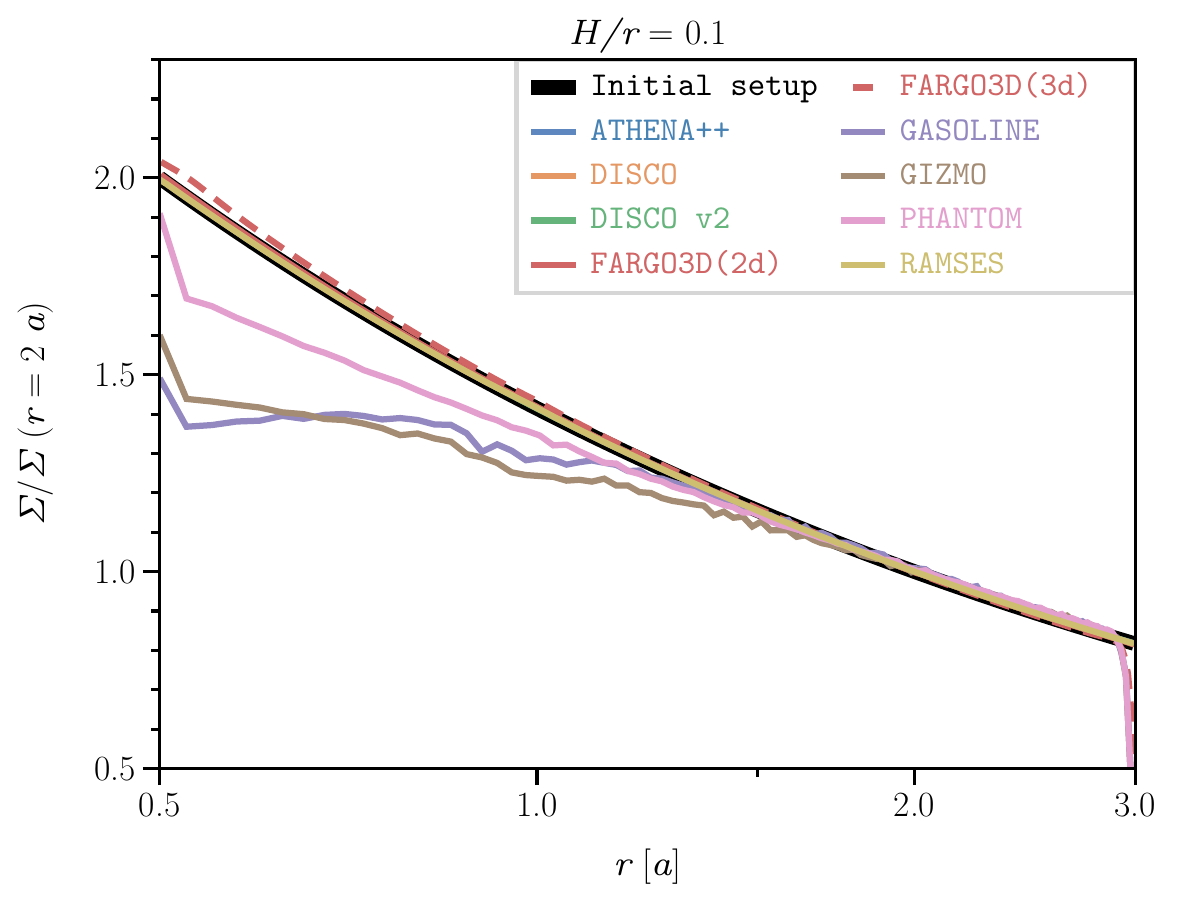}
    \caption{Initial and azimuthally-averaged density profiles after $100$ orbits in the alignment run ($H/r=0.1$) for the codes that reached that point (data for \texttt{GASOLINE} is plotted after 33 orbits). Particle-based code data have been averaged over one orbit.}
    \label{fig:r_density}
\end{figure}

Figure~\ref{fig:r_density} shows the radial profile of the azimuthally-averaged surface density after $100$ orbits ($33$ orbits for \texttt{GASOLINE}) together with the initial profile. Because the accretion discs surface density in the particle-based codes evolved significantly over time, we normalize these profiles by the surface density value at $r=2a$ to more easily compare the disc profiles. In this setup ($H/r=0.1$), the profile remains relatively close to the initial one, well within $\pm 10\%$ for codes that used Dirichlet boundary conditions. The particle-based codes (\texttt{GIZMO}, \texttt{PHANTOM}, and \texttt{GASOLINE}), on the other hand, simulated finite and viscously spreading discs, which depleted over time. However, this depletion does not appear to affect the torque convergence in each code. 

\subsubsection{Radial torque profiles}\label{sec:r_torque}

\begin{figure}
    \includegraphics[width=.49\textwidth]{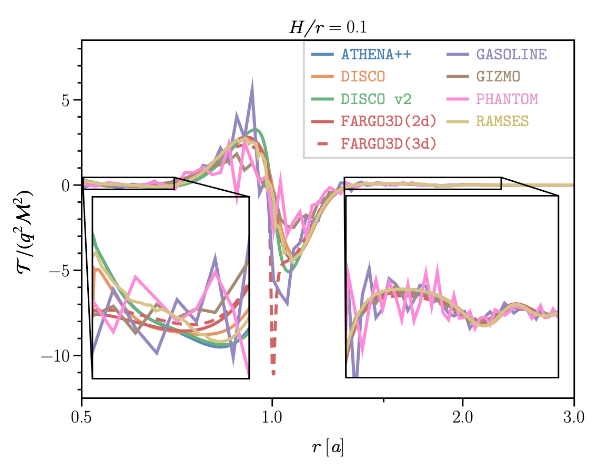}
    \caption{Azimuthally-averaged torque profiles after $100$ orbits in the alignment run ($H/r=0.1$), excepting the results of the \texttt{GASOLINE} simulation, which are plotted after 33 orbits. Particle-based code data have been averaged over one orbit.}
    \label{fig:r_torque}
\end{figure}

While in Section~\ref{sec:torque} we reported the total torque on the secondary over time, we focus here on the radial torque profile to understand which regions of the disc contribute to the total torque, and how these are resolved differently by various codes. To this end, we calculated the 2D (vertically integrated) torque density $\Sigma\partial_\phi\Phi$ in individual snapshots and then calculated azimuthal averages of this quantity on a radial grid (with radial bin size $\Delta r$) according to

\begin{equation}\label{eq:torque_r} 
\mathcal{T} = \frac{\int_{r}^{r+\Delta r} {\rm d}A\,\Sigma\partial_\phi\Phi}{\int_{r}^{r+\Delta r}\,{\rm d}A}.
\end{equation}

While this differs from the usual linear torque density by geometric factors, it serves to cleanly highlight how the two-dimensional torque density changes as a function of radius. 

Figure~\ref{fig:r_torque} shows the radial torque profile computed using Equation~\eqref{eq:torque_r}. The magnitude of the torque profile is greatest near the secondary, within a few pressure scale heights, and peters off at large distances. 
Zooming in, as do the insets in Figure~\ref{fig:r_torque}, reveals subtler oscillations in the torque profile caused by constructive interference of multiple azimuthal modes excited in the disc along the binary separation axis \citep{2024MNRAS.529..425C}. In the inner regions, however, the disc response is strongly affected by the inner boundary (when present), since there were only a few scale heights of the disc between the inner boundary and the secondary.

The same general structure of the torque profile is present in each simulation, though each code has its own individual characteristics. The outer Lindblad region, for example, is captured similarly by each code, although the torque profiles from the particle-based codes are somewhat noisier. The inner regions show much more variation among the codes because the torque profile is truncated by the inner boundary: the different boundary conditions used between codes lead to visible discrepancies. The inner profiles from \texttt{ATHENA++} and the different versions of \texttt{DISCO} agree well with each other and with \texttt{RAMSES}, while the inner regions of the \texttt{GIZMO}, \texttt{PHANTOM} and \texttt{GASOLINE} simulations are quite similar to each other. As expected based on Figure~\ref{fig:surfacedensitymap_vr_hr01}, the torque profile of \texttt{FARGO3D} (2D) disagrees with the other codes in the inner regions of the disc. The magnitude of the corotation torque is notably smaller in the particle-based codes simulations than in the various grid-based codes.

In order to determine whether these discrepancies might follow from comparing 2D and 3D simulations, we conducted a 3D grid-based simulation using \texttt{FARGO3D}. 
For this 3D simulation, the potential was softened as described by Eq.~\eqref{eq:4}, but contrary to 2D simulations in which the softening distance is a fraction of the pressure length scale so as to somehow capture the effects of the disc vertical extent, the softening distance is here comparable to the size of the cells and is used exclusively to avoid a divergence of the potential in the vicinity of the secondary (see more detail in Appendix~\ref{app:codes}). 
Interestingly, this simulation yielded an intense, negative torque density in the immediate vicinity of the secondary, but was otherwise more similar to the other grid-based codes (despite their two-dimensional nature) rather than the other 3D (particle-based codes), except for the corotation region discussed above, suggesting that other differences like the boundary conditions and discretization of the equations of hydrodynamics also play a role, at least in this specific case. When the gas within $0.4 R_{\rm H}$ is excluded from the torque calculation in \texttt{FARGO3D}, the measured torque agrees well with the value from the \texttt{GIZMO} run, suggesting similarities between the two (see Section~\ref{sec:convergence}).

We can conclude on the alignment run that all codes converge on a similar gas morphology, with larger discrepancies coming from finite versus infinite discs. All codes agree that the torque on the secondary should be negative, and that its GW-driven inspiral should be accelerated. The qualitative features of the excited torque profile are consistent between codes, and different boundary conditions typically lead to more significant differences than different \textit{numerical} choices between codes.

\subsection{Thin discs}\label{sec:thinDisc}

In the following, we discuss the set of simulations of discs with $H/r=0.03$. 
In this regime, the Hill sphere of the secondary exceeds the scale height of the disc, limiting the applicability of linear theory such that there are no appropriate analytical predictions for what the torque should be. Because these runs used a smaller softening than the previous runs, in this case $\epsilon=0.6h=0.018a$, the torque was much more sensitive to small fluctuations in the flow near the secondary. Oscillations in torques often arise in this case due to the nonlinear dynamical gas flow close to the secondary. 

\begin{figure*}
    \centering
    \includegraphics[width=.95\textwidth]{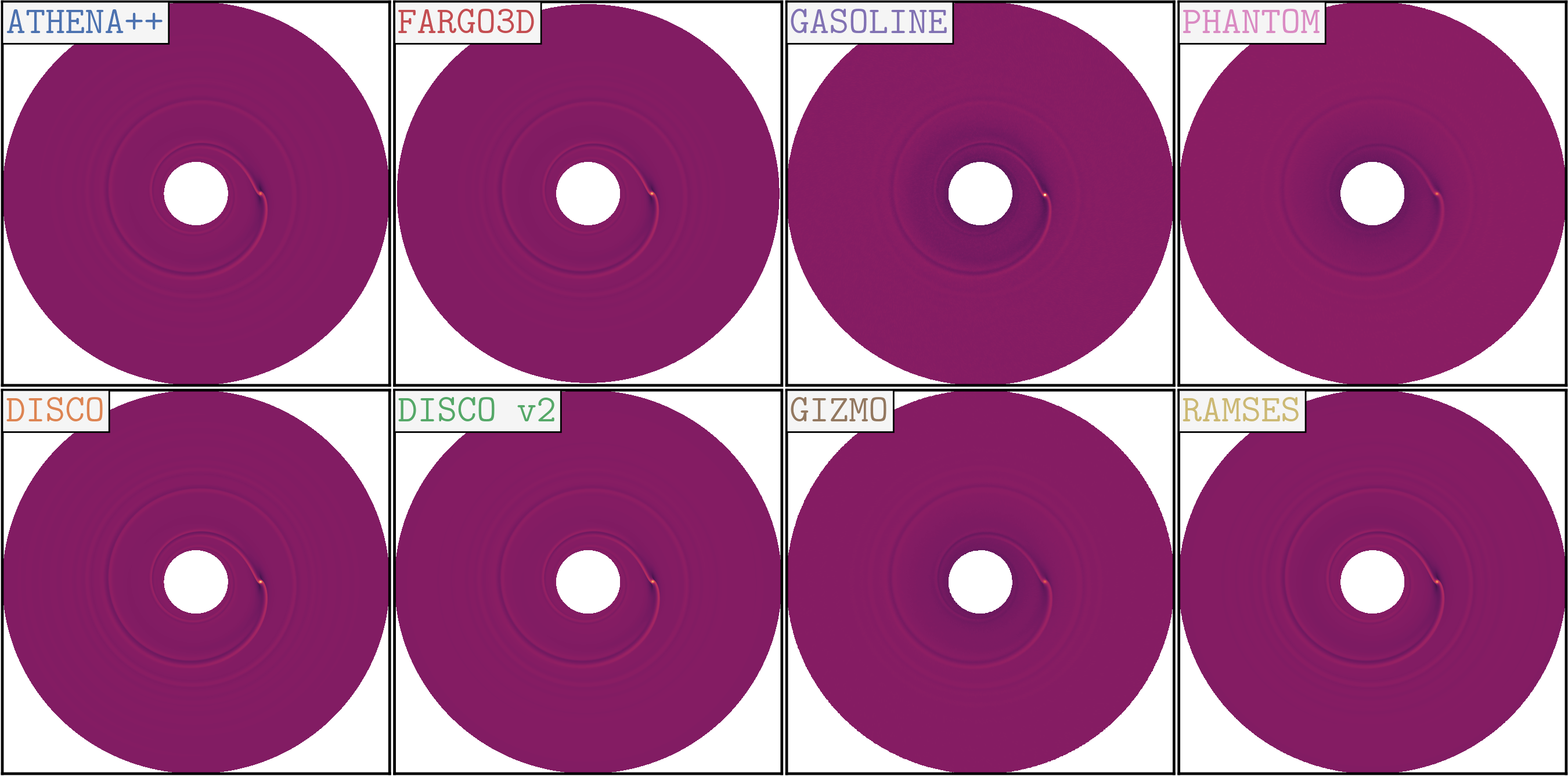}
    \includegraphics[width=.95\textwidth]{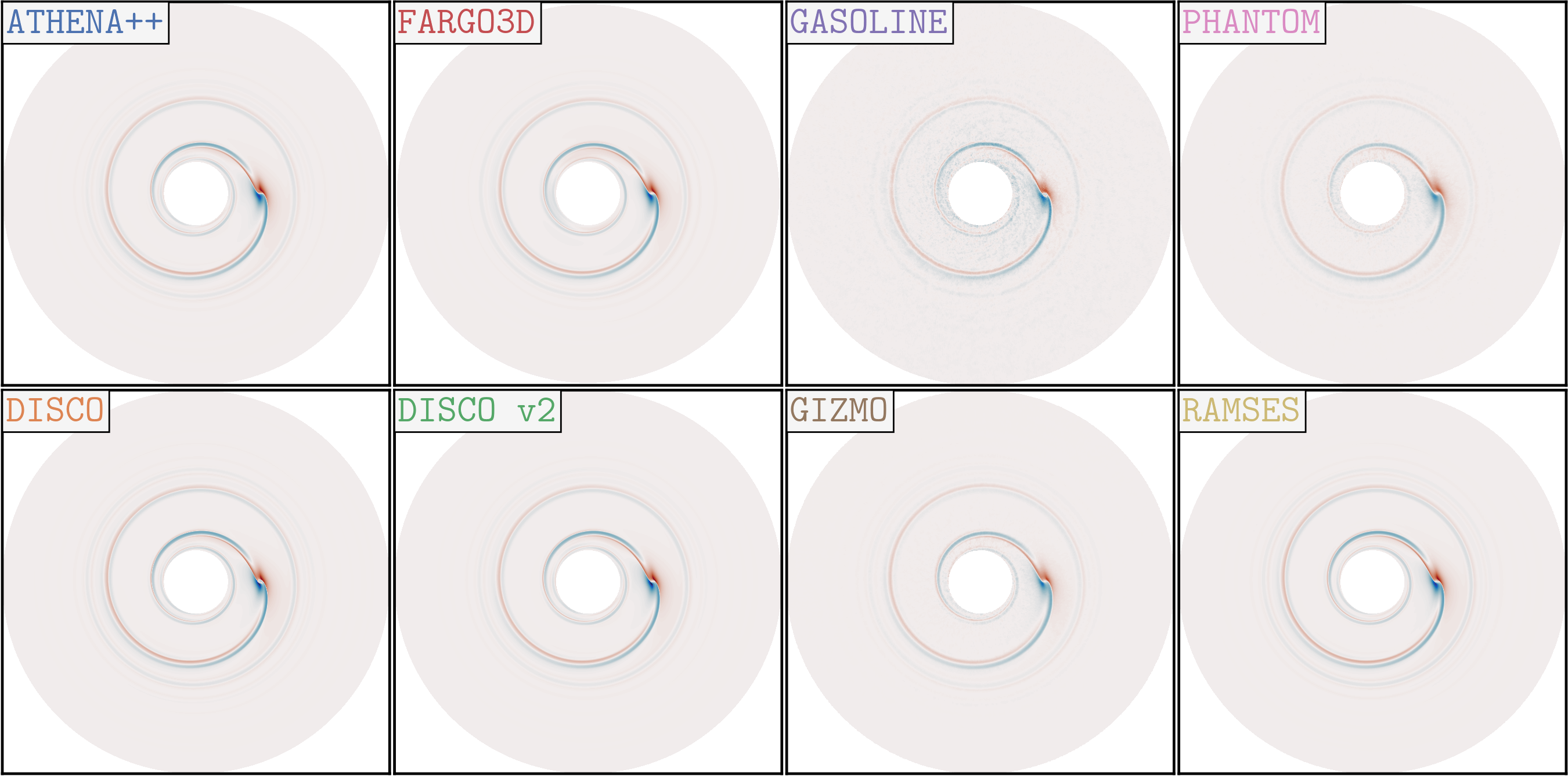}
    \caption{Surface density map (top two rows) and radial velocity map (bottom two rows) for the thin-disc run ($H/r=0.03$) after 100 orbits (50 for \texttt{GASOLINE}). Specifically, the top panels plot $\log_{10}{(\Sigma/\Sigma_0)}$ over a range [-0.5,1.5]; the bottom panels plot $v_r$ on a scale of $[-0.05,0.05]$, with red indicating negative velocities. Typical deviations of the surface density due to the forcing from the secondary are much larger than in the thick-disc case, from a depression by about $\sim$50\% in the coorbital region to an excess of $\sim$150\% within the Hill sphere of the secondary. 
    A slightly deeper gap opens within the coorbital region of the \texttt{GASOLINE} simulation.
    The \texttt{DISCO} and \texttt{GASOLINE} simulations also find much more dense circumsecondary discs. Both maps illustrate the trailing spiral arm launched by the perturber, which winds more tightly than in the thick-disc case thanks to the lower sound speed. A complimentary view in $(r,\phi)$ is provided in Appendix~\ref{app:extra} in Fig.\ref{fig:surfacedensitymap_hr01_unrolled}.}
    \label{fig:surfacedensitymap_vr_hr003}
\end{figure*}

\begin{figure*}
    \centering
    \includegraphics[width=.95\textwidth]{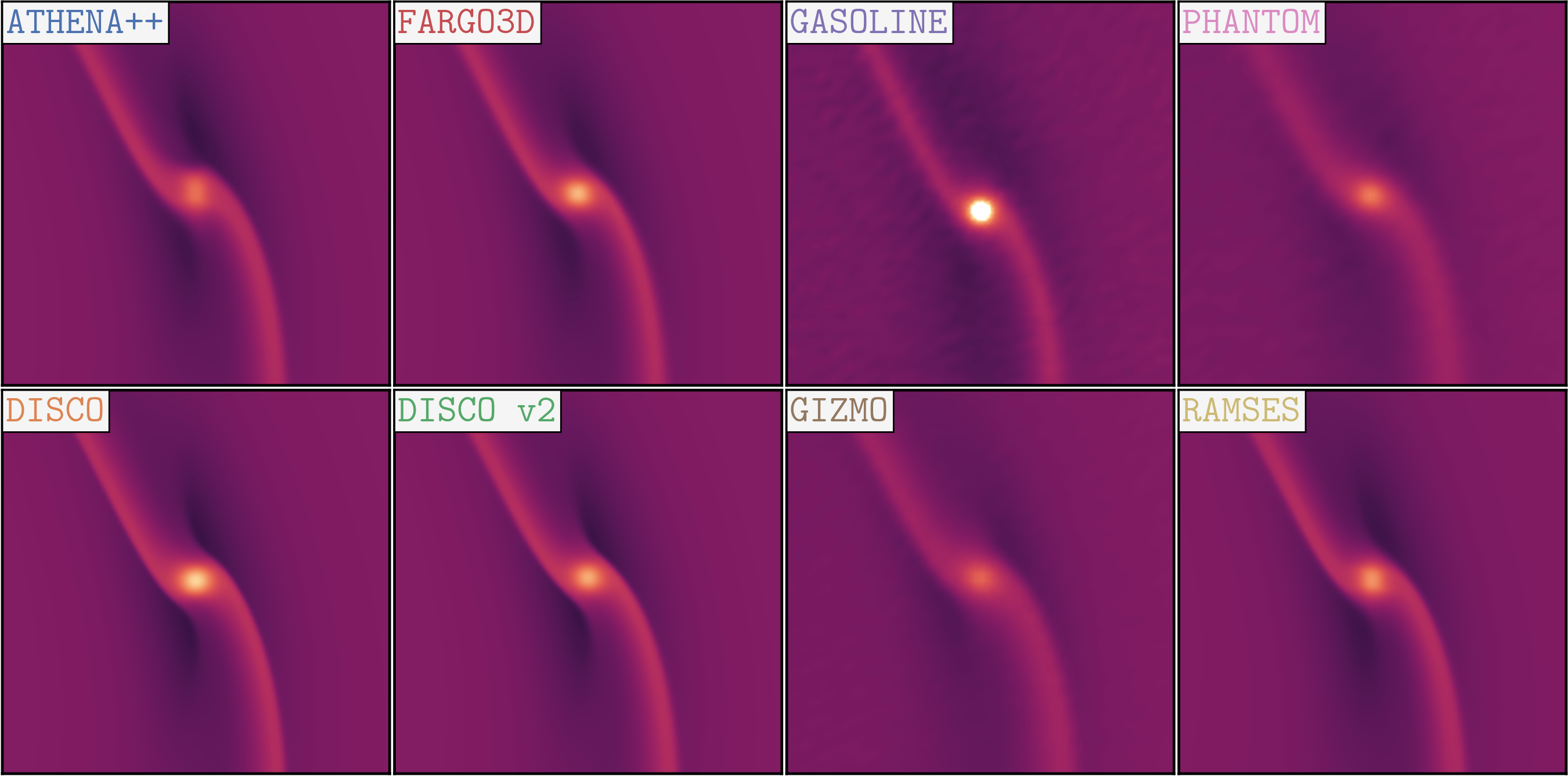}
    \includegraphics[width=.95\textwidth]{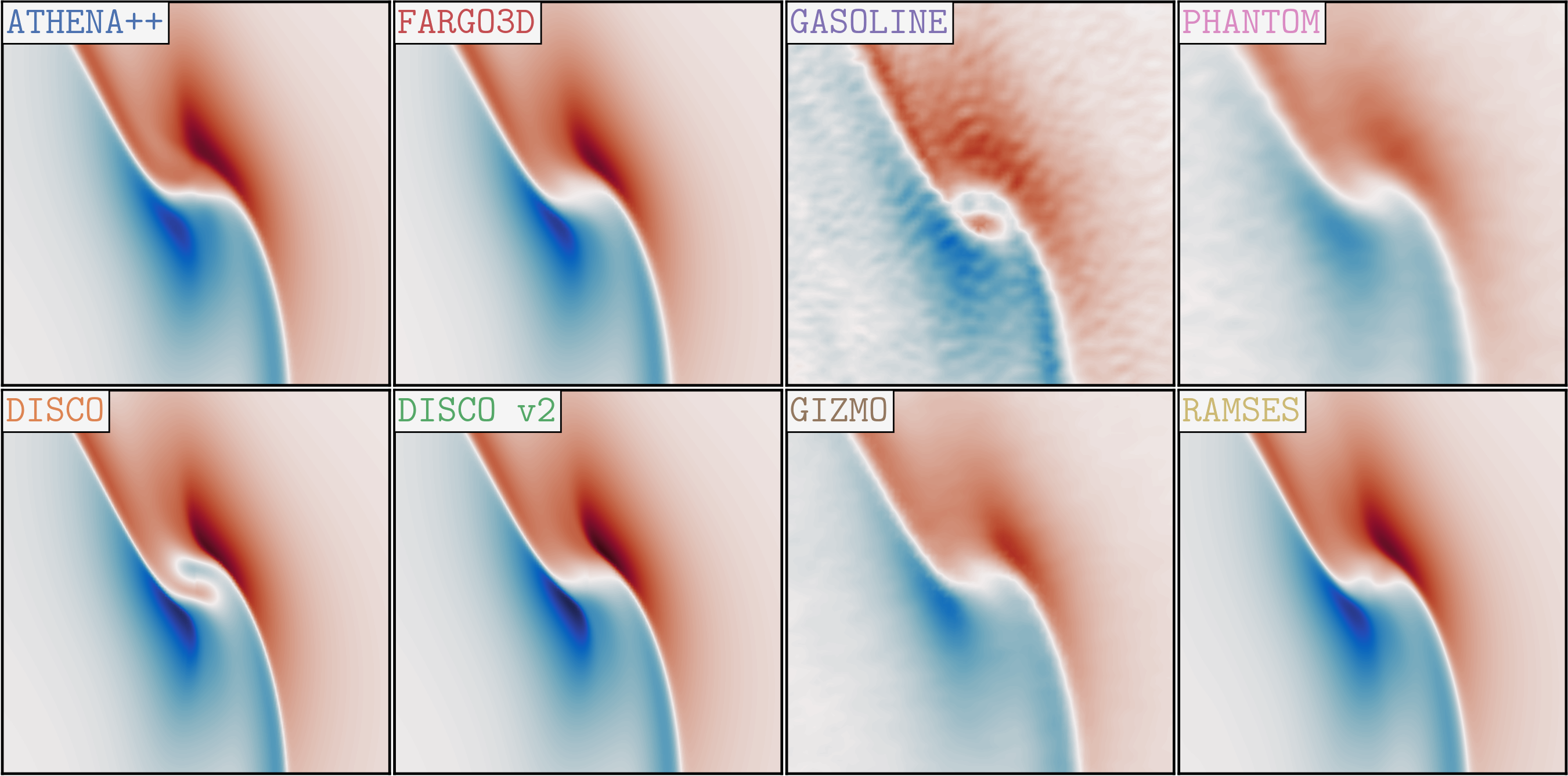}
    \caption{Surface density map zoomed around the secondary for the thin-disc run ($H/r=0.03$). The top panels plot $\log_{10}{(\Sigma/\Sigma_0)}$ over a range $[-0.5,1.5]$; the bottom panels plot $v_r$ on a scale of $[-0.05,0.05]$ with red indicating negative velocities. The overdensity within the Hill sphere of the secondary is more azimuthally extended in the \texttt{RAMSES} and \texttt{ATHENA++} simulations, while \texttt{DISCO} and \texttt{GASOLINE} produce much greater overdensities. Generally, the spiral arms are captured much more sharply by the grid codes than the particle-based codes, but the shapes are consistent across the board. The velocity field within the Hill sphere of the secondary is also quite different in the \texttt{DISCO} and \texttt{GASOLINE} simulations compared to the other codes, suggesting some sort of circulation associated with their overdensities.}
    \label{fig:surfacedensitymap_hr003_hill}
\end{figure*}

\subsubsection{Gas morphology}

The lower sound speed of the thinner disc leads to a more tightly wound spiral, visualized using both the surface density perturbations ($\log_{10}{\Sigma/\Sigma_0}$) and radial velocity in Figure~\ref{fig:surfacedensitymap_vr_hr003}. The wave damps much more significantly before reaching either boundary than in the thin-disc case. The maps show the relevant quantities after $100$ orbits for all codes, except for \texttt{GASOLINE} shown after $55$ orbits.
On a global scale, relatively few features are visible in the surface density maps: \texttt{GASOLINE} 
and \texttt{DISCO} produce more prominent overdensities near the secondary and the \texttt{GASOLINE} simulation features a mild gap in the co-orbital region. Similarly to the alignment run shown above, the spiral arms feature prominently in the radial velocity as well, largely following the surface density. 

Subtle differences between the codes become more apparent when examining the flow within the Hill sphere of the secondary, as displayed in Figure~\ref{fig:surfacedensitymap_hr003_hill}. In the surface density plots shown there, we see the degree of the overdensity in the \texttt{GASOLINE} and \texttt{DISCO} simulation, and also the azimuthal extent of the gas around the secondary in the \texttt{RAMSES} and \texttt{ATHENA++} simulations. The morphology of the gas in the other four simulations is quite consistent, although the spiral arms are less sharp in the particle-based simulations. The radial velocity maps near the secondary reveal associated kinematic features: recalling that blue denotes outflow and red denotes inflow, \texttt{GASOLINE} and \texttt{DISCO} produce very different circulation patterns within their Hill spheres, in comparison to the other codes. As we illustrate in Appendix~\ref{app:extra}, the differences in the flow within the Hill sphere between \texttt{DISCO} and \texttt{DISCO v2} are due to the viscosity approximations made in \citet{Duffell2016}; Appendix~\ref{app:extra} also shows that the misshapen gas flow around the secondary in the \texttt{ATHENA++} simulation is related to the low resolution (and lack of orbital advection) used in these particular simulations (see Appendix \ref{app:codes}). The Hill-sphere overdensity in the \texttt{GASOLINE} simulation seems most likely to be a result of the disc self-gravity considered in that simulation which is more relevant in this case thanks to the higher disc density.

\subsubsection{Torques}

\begin{figure}
    \centering
    \includegraphics[width=\linewidth]{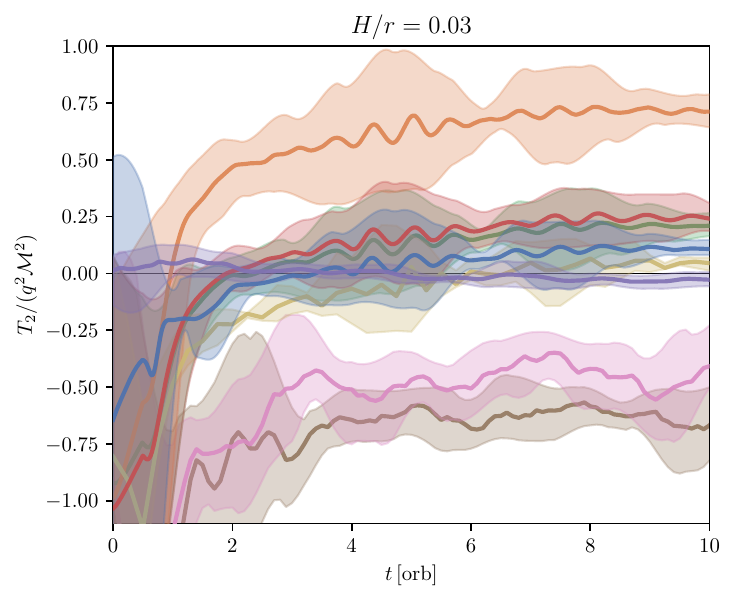}\\
    \includegraphics[width=\linewidth]{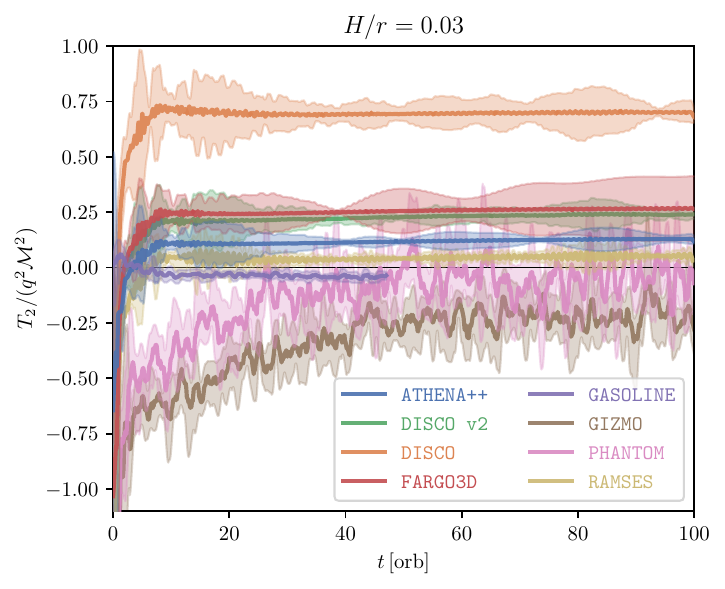}
    \caption{
    Torque on the secondary 
    for the thin disc case with $H/r=0.03$.  For this run, there are no appropriate analytical predictions, as discussed in Section~\ref{sec:disk-second-inter}. The torques are smoothed over a 1-orbit window, and the envelopes outline the variability seen in the unsmoothed torque. For this choice of disc parameters, the total torque on the secondary is close to zero, and results from near cancellation between the torques from various parts of the disc, making the results acutely sensitive to differences between codes.}
    \label{fig:torque_h03_M2}
\end{figure}

The torques on the secondary 
from each thin-disc simulation are plotted in Figure~\ref{fig:torque_h03_M2}, highlighting both the first ten and first hundred orbits (except for the \texttt{GASOLINE} run, showing the first fifty orbits). While most codes settle into quasi-steady values of the torque within about ten orbital periods, the torques in \texttt{PHANTOM} and \texttt{GIZMO} evolve over much longer timescales, reaching a quasi-steady values after about fifty orbits. Across all codes, the torque exhibits greater variability than in the alignment run.

For this more challenging problem, differences between codes become more apparent. This weakly nonlinear regime, for discs with surface density profiles like the one studied here, is known to result in near-cancellation of the negative Lindblad torques and positive corotation torques, sometimes resulting in outward migration \citep[e.g.,][]{2006ApJ...652..730M,2015ApJ...806..182D}. \texttt{GIZMO} and \texttt{PHANTOM} exhibit negative torques on the secondary, while \texttt{GASOLINE} shows a nearly-zero torque. The grid-based codes find positive torques on the secondary. \texttt{RAMSES} finds a positive torque on the secondary, but as we will show in Section~\ref{sec:convergence}, the inclusion of the torque on the primary results in a nearly-zero torque on the binary, while all other 2D grid codes find a positive torque on the binary.

Beyond the sign of the torque, many codes also disagree on its magnitude. Comparing the grid-based codes, Figure~\ref{fig:surfacedensitymap_hr003_hill} reveals that \texttt{FARGO3D} and \texttt{DISCO v2} resolve the flow within the Hill sphere quite similarly, while the flow in \texttt{ATHENA++} and \texttt{RAMSES} is more diffuse, and the surface density in \texttt{DISCO} is more highly peaked; these features map to the difference in torques measured between codes, \texttt{DISCO} finding the greatest positive torque, \texttt{ATHENA++} and \texttt{RAMSES} finding lower values, and \texttt{FARGO3D} and \texttt{DISCO v2} finding good mutual agreement. 
Sharp flow features appear to be captured with higher fidelity by \texttt{GIZMO} than by \texttt{PHANTOM}, potentially leading to the differences in torque observed in Figure~\ref{fig:torque_h03_M2}. We investigate this further in the following sections.

\subsubsection{Radial Profiles}

\begin{figure}
    \centering
    \includegraphics[width=.49\textwidth]{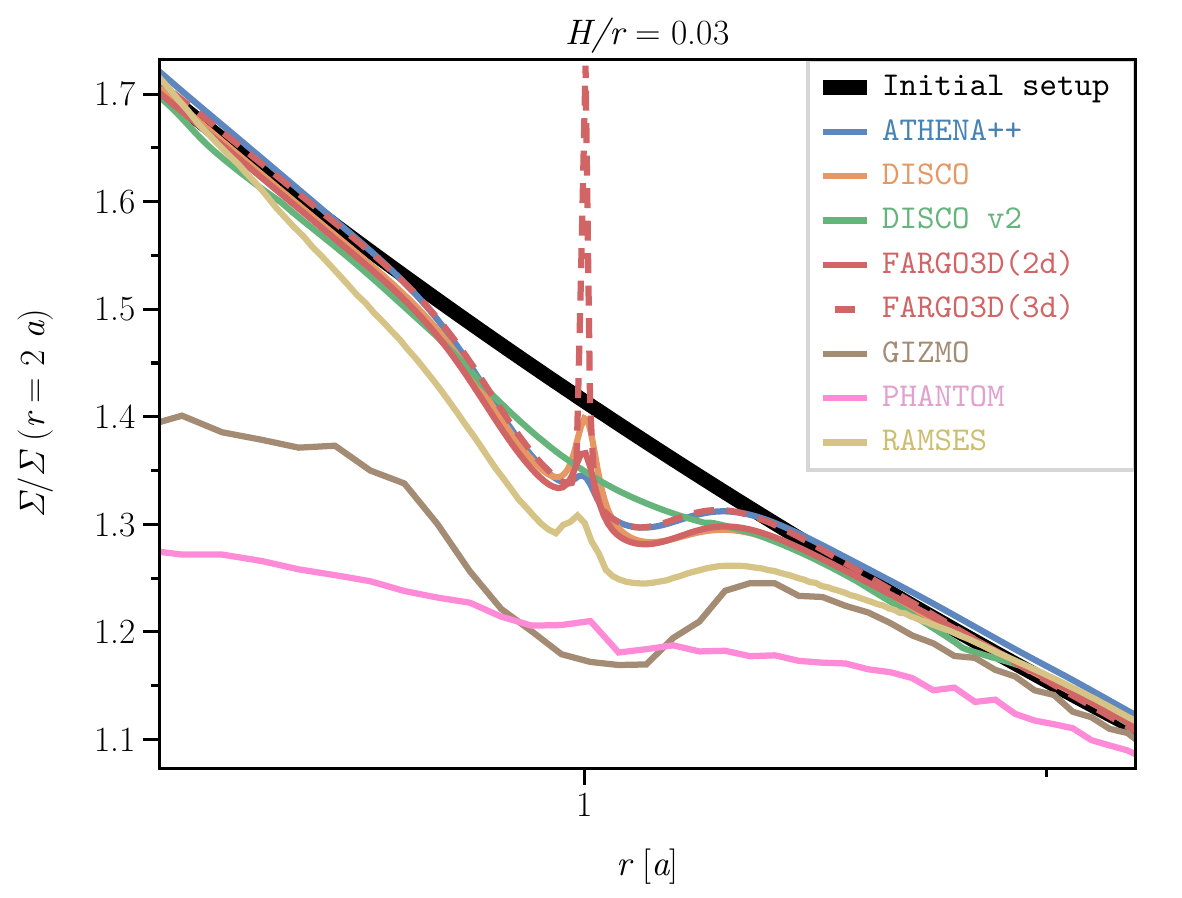}\\
    \includegraphics[width=.49\textwidth]{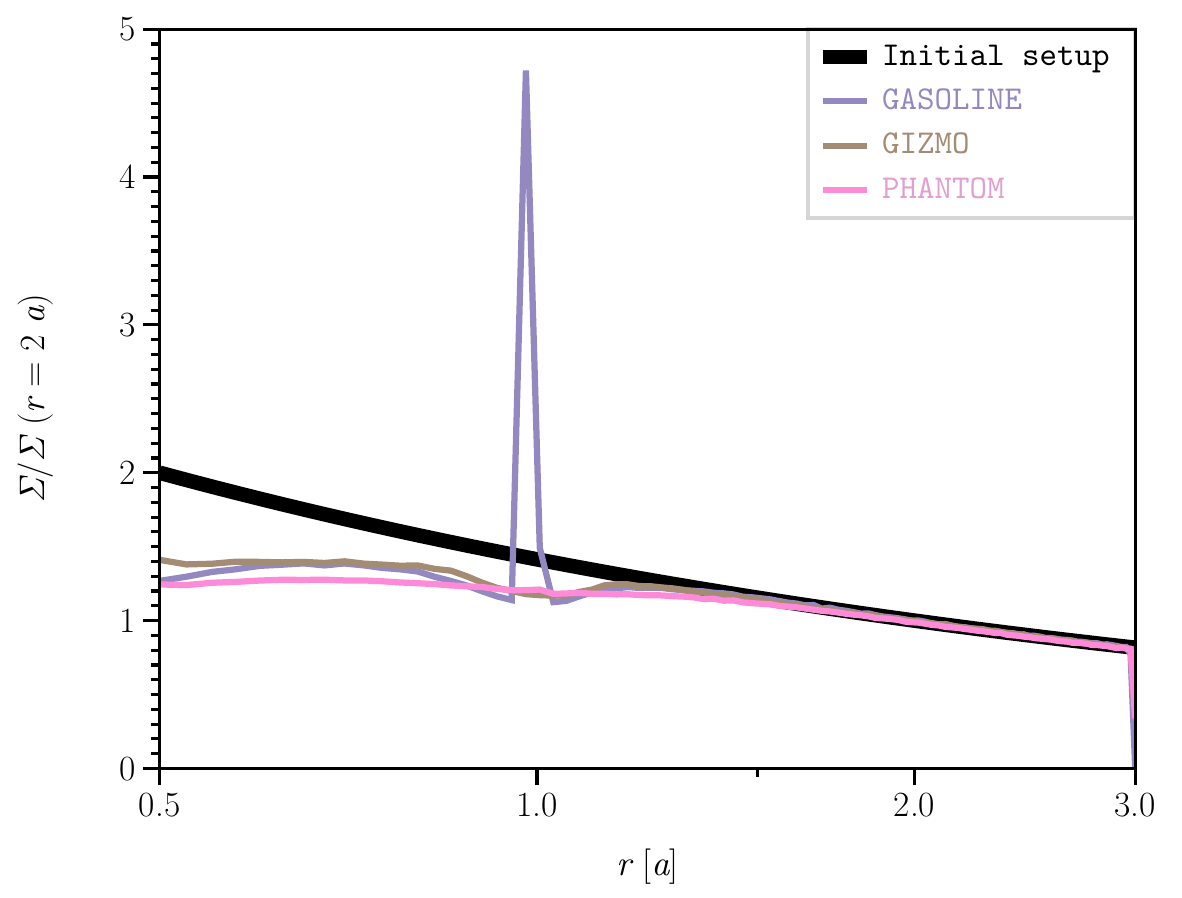}
    \caption{Initial and azimuthally-averaged density profiles after $100$ orbits in the thin-disc ($H/r=0.03$) run, excepting the \texttt{GASOLINE} data which are plotted after $50$ orbits. The top panel shows data from all codes excluding \texttt{GASOLINE}, and the bottom panel compares the \texttt{GASOLINE} data to the other particle-based codes. 
    Particle-based code data have been averaged over one orbit.    }
    \label{fig:r_density_thin}
\end{figure}

Figure~\ref{fig:r_density_thin} shows the azimuthally-averaged surface density profile after $100$ orbits (except for the GASOLINE run, showing the first 50 orbits). In contrast with the alignment run, as shown by Figure~\ref{fig:r_density}, the density substantially deviates from its initial profile, especially near the secondary; this follows from the weakly nonlinear regime probed by these simulations ($qh^{-3}\approx3.7$). The finite discs simulated by particle-based codes, which spread viscously over time, naturally deplete, particularly in their inner regions, similarly to the alignment run, albeit to a lesser extent. The grid-based codes generally agree well.

\begin{figure}
    \centering
    \includegraphics[width=.49\textwidth]{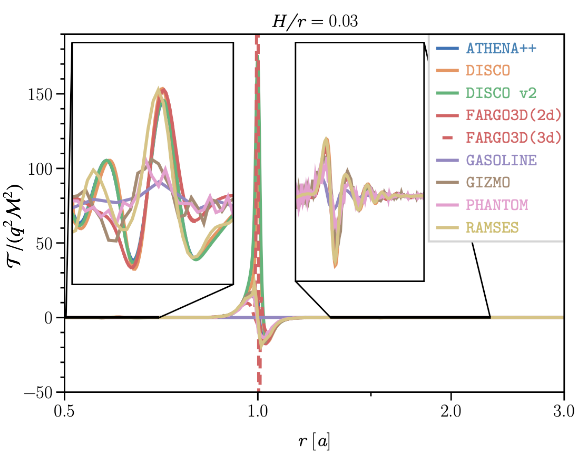}
    \caption{Azimuthally-averaged torque profiles after $100$ orbits in the thin-disc ($H/r \, {=} \, 0.03$) run, or after 50 orbits for \texttt{GASOLINE}. Many of the codes overlap substantially, making it difficult to distinguish them: The curve corresponding to \texttt{ATHENA++} sits behind the two largely-overlapping \texttt{DISCO} curves, as well as the \texttt{FARGO3D} curves throughout most of the domain; the \texttt{FARGO3D} curves become discrepant near the inner boundary, because of its different inner boundary and reference frame; the 3D \texttt{FARGO3D} results overlap completely with the 2D \texttt{FARGO3D} results away from corotation. The particle-based codes exhibit noisier torque profiles but generally match the grid-based codes well in the outer regions of the disc, though the magnitude of the torque in the inner regions differs owing to the differences in the surface density distribution (Figure~\ref{fig:r_density_thin}) following from the different boundary conditions in those simulations. 
    Particle-based code data have been averaged over one orbit.
    }
    \label{fig:r_torque_thin}
\end{figure}

Figure~\ref{fig:r_torque_thin} shows the radial profile of the azimuthally-averaged torque density after $100$ orbits (except for the GASOLINE run, showing the torque after 50 orbits). Differences in the torque between codes largely follow differences in the surface density shown in Figure~\ref{fig:r_density_thin}. Most of the grid-based codes agree with each other, though the corotating frame and boundary conditions used by \texttt{FARGO3D} lead to discrepancies in the inner regions. The particle-based codes  
find lower-magnitude torques. This is related to the natural spreading of their finite discs, which depletes the mass in the region inside the secondary's orbit, leading to a weaker positive torque contribution.
\texttt{GASOLINE} has a much lower torque density than all other codes, potentially as a result of the disc self-gravity included in that simulation (although see Section~\ref{sec:discrepancies}).
\texttt{GIZMO} and \texttt{PHANTOM} find qualitatively similar results to each other, and agree with the grid-based codes in the outer region of the disc; as expected based on Figure~\ref{fig:r_density_thin},  \texttt{GIZMO} and \texttt{PHANTOM} find weaker torques in the inner regions of the disc. These may contribute to the more negative overall torques measured by those codes, as Lindblad resonances in the inner disc tend to add angular momentum to the binary while Lindblad resonances in the outer disc tend to sap angular momentum from the perturber. 
As we show by comparison with a three-dimensional \texttt{FARGO3D} simulation in Figure~\ref{fig:resolution}, another reason for different codes finding either positive or negative torques is the flow of gas very near the secondary. 

\subsubsection{Sensitive corotation torques}

As discussed in Section~\ref{sec:disk-second-inter}, IMRIs evolution in thin discs are difficult to predict analytically both because of nonlinear forcing ($q\gtrsim h^3$) and because the viscous diffusion timescale through the co-orbital region can become comparable to the libration timescale of fluid elements in that region. In particular, the latter effect has been found to cause nonlinear corotation torques to counteract, and sometimes overwhelm entirely, other torques in the system and drive migrating objects outward \citep[e.g.,][]{2006ApJ...652..730M,2015ApJ...806..182D}. This appears to be the case for the IMRI problem studied here. The relatively lower gas densities in the co-orbital region of the particle-based codes reduce the impact of these torques, potentially contributing to the discrepancies between the total torques measured by those codes and the others; the discrepant circulation patterns and surface density within the Hill sphere in the \texttt{DISCO} simulation likely contribute to the much larger torque measured by that code compared to the other grid-based codes. 

It is worth noting that simple torque prescriptions attempting to extend linear results into the nonlinear regime \citep[e.g.,][Equation~\ref{eq:kanagawa}]{Kanagawa2018} fail here. Though they disagree slightly amongst themselves, the codes all find $-0.5\lesssim T/\Gamma_0\lesssim0.7$, while Equations~\eqref{eq:linear2D} and \eqref{eq:kanagawa} suggest $T/\Gamma_0\approx-1.8$. Simulations thus appear essential to properly model the nonlinearlities in this problem. However, the importance of both Lindblad resonances and the flow within the co-orbital region make this problem especially challenging, and differences in the boundary conditions and viscosity implementations between various codes that were unimportant in the thicker disc case studied in Section~\ref{sec:thickDisc} are extremely important for these thinner and less viscous discs. 

\subsection{Dimensionality and Convergence}\label{sec:convergence}

\begin{figure}
    \centering
    \includegraphics[width=.47\textwidth]{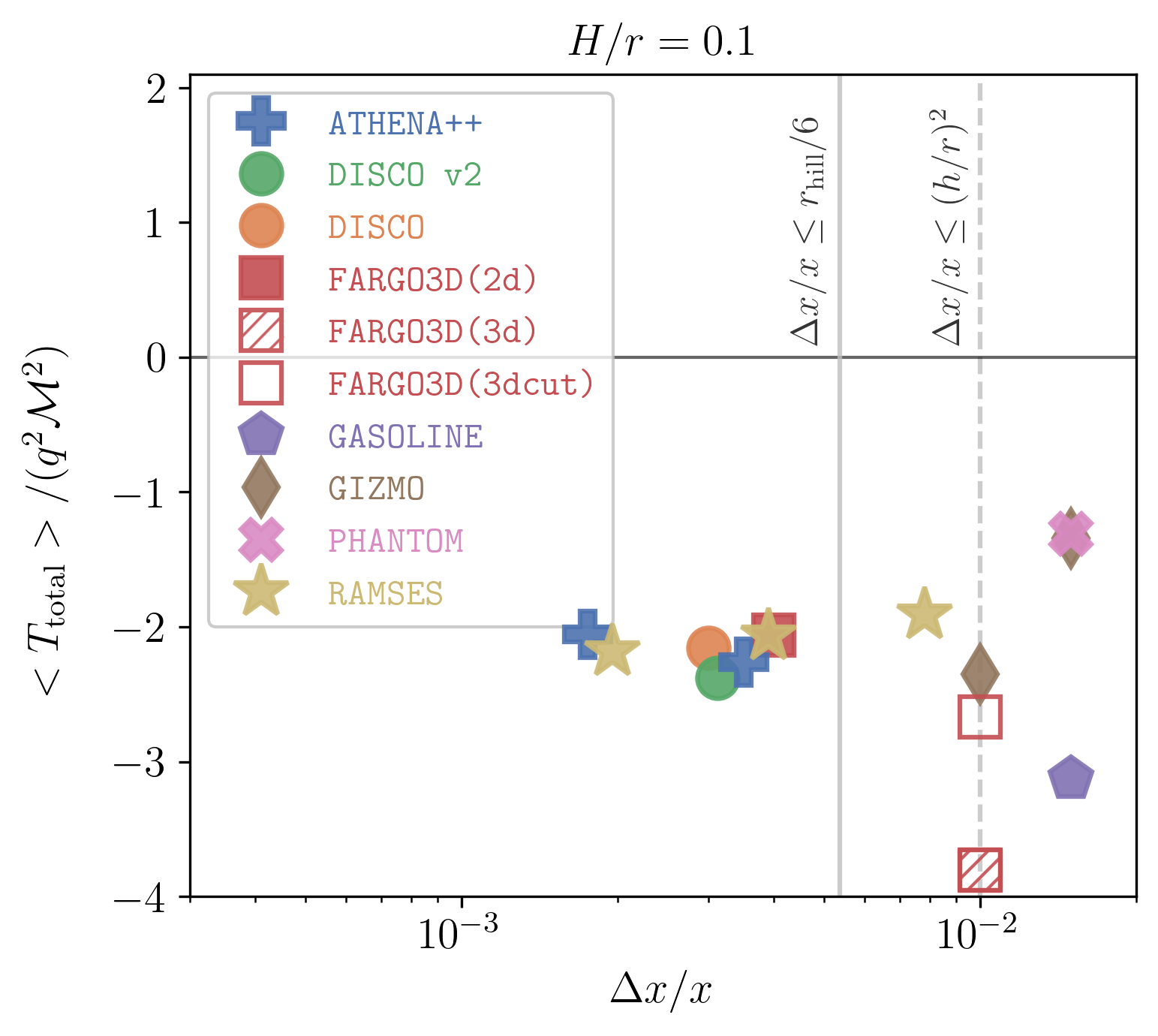}\\
    \hspace*{-1.mm}  
      \includegraphics[width=.485\textwidth]{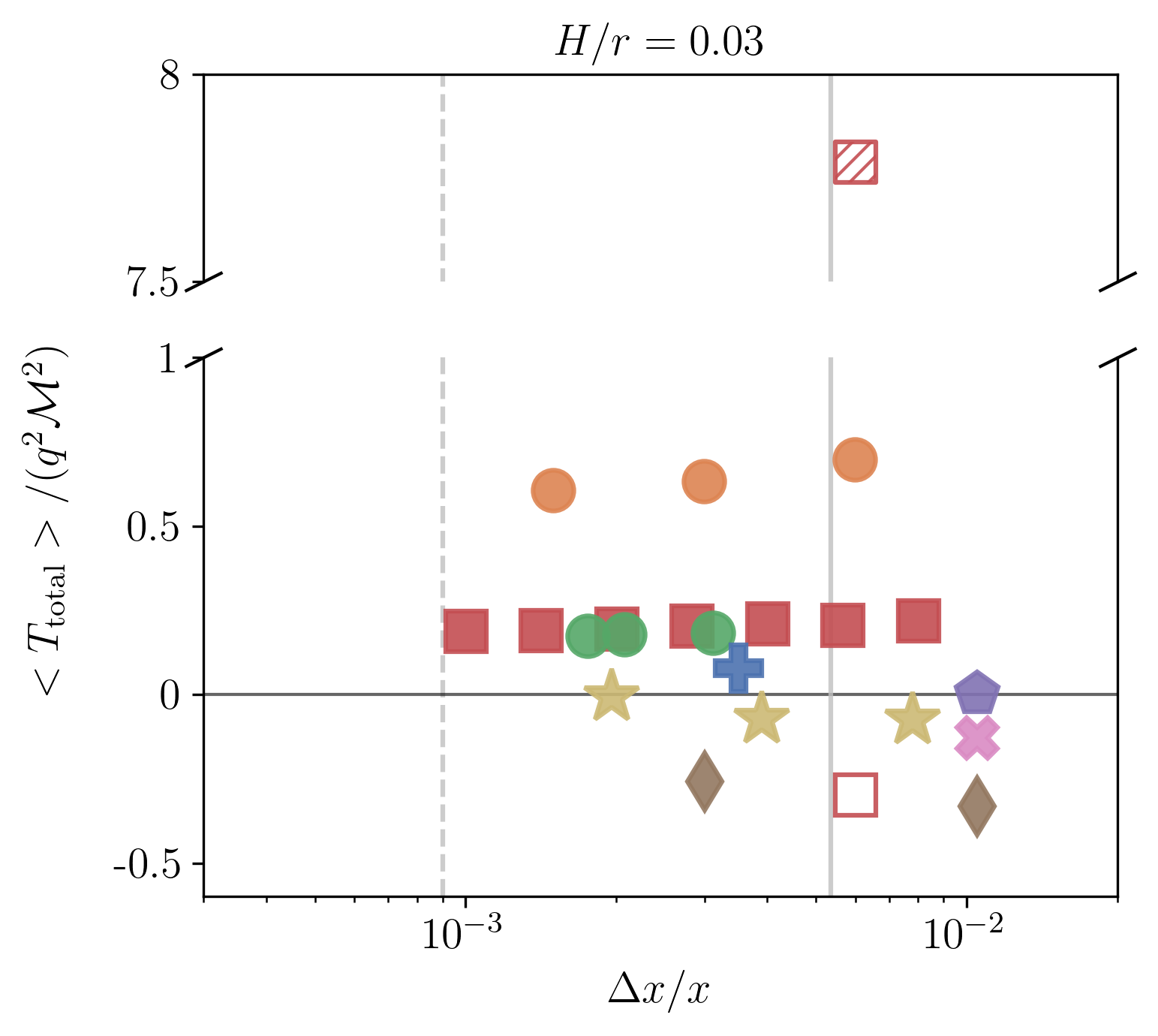}
    \caption{
    Converged values of the total torque on the binary for the runs with disc thickness $H/r=0.1$ (top panel) and $H/r=0.03$ (bottom panel) versus spatial resolution near the secondary. The averaged values exclude the first 10 orbits of the simulation. The vertical lines correspond to resolution limits as labeled. A horizontal line line is added at $T_{\rm total} = 0$ to highlight the transition from negative to positive values. Of note is the difference between \texttt{FARGO3D} runs in 2D (solid squares), 3D (hatched squares), and 3D with the Hill region excised (hollow squares). In the thin disc run in particular, the particle runs agree on a negative torque value. The \texttt{FARGO3D} grid simulation \emph{only} agrees with the particle codes if the gas in the Hill region is excluded from the torque calculation. Otherwise, this run finds a strongly positive torque.}
    \label{fig:resolution}
\end{figure}

A major difference between the particle-based and grid-based simulations in this study is that all of the particle-based codes conducted 3D simulations, while all the grid-based codes performed two-dimensional simulations. Another difference is that the particle codes simulated finite, viscously spreading discs, while the grid-based codes continuously fed material into the disc from the outer boundary. Given the multiple differences between these families of simulations, it is difficult to determine whether the discrepant results between them are caused by the dimensionality of the calculations, their boundary conditions, or different details of the numerical schemes. To disentangle these possibilities, we conducted a pair of 3D simulations using the grid-based code \texttt{FARGO3D}, which use the same boundary conditions as the two-dimensional simulations (the late-time surface density profile is nearly identical to the 2D case, as illustrated in Figure~\ref{fig:r_density_thin}). When measuring torques from these runs, we show measurements using the total gas distribution, labeled \texttt{FARGO3D} (3D), and measurements with the Hill region excised (specifically the gas within $0.4R_{\rm H}$ is excluded), labeled \texttt{FARGO3D} (3Dcut). Additionally, to assess the robustness of our results and the resolution requirements for these and similar simulations, we have performed a number of simulations using different resolution settings. The time-averaged torques from these simulations, as well as our standard suite, are presented in Figure~\ref{fig:resolution}.

In the thin-disc case, Figure \ref{fig:resolution} reveals systematic discrepancies between the grid-based and particle-based codes, but also sheds light on possible reasons for those discrepancies. Because all of the particle-based simulations were conducted in 3D, the 3D \texttt{FARGO3D} simulation provides a direct comparison. Intriguingly, the `3Dcut' \texttt{FARGO3D} measurement, which excluded gas within $0.4R_{\rm H}$ of the secondary from the torque calculations, displays good agreement with the \texttt{GIZMO} results. However, when the entire domain is included in the 3D \texttt{FARGO3D} calculation, the net torque becomes positive, and larger than in the 2D simulations by almost an order of magnitude, suggesting that gas within the Hill spheres of these objects plays a crucial role in determining their migration. Although we expect dimensionality to be important given the challenges in mapping three dimensional gravitational interactions onto two dimensions \citep[e.g.,][]{2016ApJ...817...19M,2024MNRAS.534...39B,2025arXiv250904282C}, none of the simulations in this study can claim convergence, leaving the true answer, and thus which code approaches it most closely, ultimately uncertain. 
We note that \texttt{GIZMO} gives very similar torque magnitudes for two different resolutions, one of which comparable to the resolution achieved by the 2D grid-based codes. Furthermore, there is a slightly decreasing trend on the torque magnitude with resolution also in 2D codes, e.g. , \texttt{FARGO3D} (red squares), \texttt{DISCO} (orange dots) and \texttt{DISCOv2} (green dots), except for \texttt{Ramses} (yellow stars) which shows an increasing trend with resolution. Note that while the \texttt{RAMSES} runs find a positive torque on the secondary, this is offset by the torque on the primary, leading to a negative torque on the binary for the lower resolution runs.

In the thick disc case, the results from the particle-based \texttt{PHANTOM} and \texttt{GIZMO} codes agree very well when using the standard secondary softening, i.e. $\epsilon=0.06a$, as shown in the top panel of Figure~\ref{fig:resolution}, although they, \texttt{GASOLINE}, and \texttt{FARGO3D} are in mutual disagreement. The \texttt{GIZMO} simulation performed with a much smaller softening ($\epsilon=0.018a$) agrees instead fairly well with the grid-based 3D \texttt{FARGO3D} that used ($\epsilon=0.06a$) when the Hill region is excluded. Because of the greater cost of higher-resolution simulations, especially in three dimensions, we have not been able to perform resolution tests with many codes.

\subsection{Torque variability}\label{sec:torquevar}

One potentially surprising result from these simulations is the variability of the torque on the secondary: if the fluid were able to reach a steady state, one could imagine moving to a frame corotating with the binary, in which all time variability would vanish. However, we observe torque variability over a wide range of timescales, even after one hundred orbits, as shown in Figures~\ref{fig:torque_h01_M2} and \ref{fig:torque_h03_M2}. To gain a more quantitative sense of time-averaged torques and their variability, we collect their values in Table~\ref{tab:converged_torques}. Similar variability has been observed in the past: for example, \citet{derdzinski_evolution_2021} found that the variability of gas-induced torques on an IMRI increased with the binary mass ratio and decreased as the accretion disc became thinner. If this variability is realistic, it could lead to appreciable binary dephasing over the course of a LISA observation \citep{2022MNRAS.511.6143Z}.

For particle-based codes, some of the torque variability is caused by the shot noise intrinsically related both to the number of particles used to discretize the disc and also the number of particles within a given kernel \citep{2015ApJ...800....6Z}. The particle-based codes tended to have greater variability in the thick-disc case, possibly related to the lower particle densities (at fixed particle number) and commensurately reduced resolution within the Hill sphere. Grid-based codes, on the other hand, exhibit greater variability in the thin-disc case. To understand the origin of the torque variability, we conducted a small number of 2D thin-disc \texttt{FARGO3D} simulations where instead of introducing the secondary at $t=0$, the mass of the secondary was slowly increased from $q=0$ to $q=10^{-4}$ over the first couple of orbits; in this `tapered' case, the torque variability was significantly reduced, suggesting that most -- if not all -- of the short-timescale variability is related to initial transients ringing through the disc. The smaller scale height of the disc in relation to the Hill radius of the binary also plays a role, since individual fluctuations can undergo temporary but relatively strong interactions with the secondary.

\begin{table}[ht]
\centering
\setlength{\extrarowheight}{5pt}
\begin{tabular}{|c|c|c|}
\hline
\textbf{Code Name} & \textbf{Torque} & \textbf{Torque } \\
 & \textbf{($H/r=0.1$)} & \textbf{($H/r=0.03$)} \\
\hline
\texttt{ATHENA++}       & $-2.26\pm 0.10$ & $0.07\pm0.09$ \\
\hline
\texttt{DISCO}          & $-2.15\pm 0.09$ & $0.62\pm0.14$ \\
\hline
\texttt{DISCO v2}       & $-2.38\pm0.07$ & $0.18\pm0.07$ \\
\hline
\texttt{FARGO3D} (2D)  & -$2.06\pm0.05$ & $0.21\pm0.06$ \\
\hline
\texttt{FARGO3D} (3D)  & $-3.80\pm0.27$ & $7.79\pm1.22$ \\
\hline
\texttt{FARGO3D} (3Dcut)  & $-2.67\pm0.19$ & $-0.30\pm0.10$ \\
\hline
\texttt{GASOLINE}      & $-3.11\pm1.18$ & $0.01\pm0.02$ \\
\hline
\texttt{GIZMO}         & $-1.34\pm0.40$ & $-0.38\pm0.22$ \\
\hline
\texttt{GIZMO \!($\epsilon = 0.018a$)} & $-2.35\pm1.00$ & \\
\hline
\texttt{PHANTOM}       & $-1.31\pm0.26$ & $-0.18\pm0.23$ \\
\hline
\texttt{RAMSES}        & $-2.18\pm0.12$ & $-0.01\pm0.12$\\
\hline
\end{tabular}
\vspace{0.5em}
\caption{Time-averaged total torque values (averages and standard deviations), normalized by $q^2 \mach^2$, for different codes in both the alignment run ($H/r=0.1$) and thin-disc case ($H/r=0.03$) with fiducial resolutions. \texttt{FARGO3D} (3Dcut) corresponds to the \texttt{FARGO3D} (3D) run with the gas within $0.4R_{\rm H}$ excised from the torque calculation. For reference, least-squares fits to modified shearing box linear theory yield a value of $-2.1$ in three dimensional discs \citep{TanakaOkada2024}, and fitting formulae to linear theory with complementary nonlinear simulations in two dimensional discs yields a value of $-1.9$ \citep{2010MNRAS.401.1950P}. In most codes, the absolute level of torque variability (relative to $\Gamma_0$) is quite similar for both disc thicknesses; for the thin disc case however, the standard deviation of the torque can approach or even exceed its average.}
\label{tab:converged_torques}
\end{table}

\subsection{Unresolved discrepancies}
We see clear discrepancies in the behavior of \texttt{GASOLINE} results in both the thick and thin disc runs. For the thick disc case, aside from the torque value differing from the other Lagrangian codes (\texttt{GIZMO} and \texttt{PHANTOM}), there are stronger amplitude fluctuations. 
In the thin disc case, however, the torque amplitude fluctuations are notably weaker, and the magnitude is also not in agreement with \texttt{GIZMO} and \texttt{PHANTOM}.  
Currently we are unable to isolate the underlying cause of these differences, although further tests are in progress. For the time
being,  we provide a few potential reasons behind such discrepancy.

First is the implementation of  viscosity in \texttt{GASOLINE}, which includes a physical and numerical viscosity. The latter includes a correction for gradients via the Cullen and Dehnen formalism \citep{Cullen_Dehnen_2010}. Parameter choices in this formalism can cause particle penetration in the disc midplane, leading to significant noise in the particle distribution \citep{2012MNRAS.427.2022M}. 
Based on tests available at the time of writing, we cannot confirm that the viscosity implementation in \texttt{GASOLINE} is performing as expected. 
We also see in Fig.~\ref{fig:torque_h03_M2} a rapid convergence in the time evolution of the torque (within $\sim 5$ orbits), as opposed to the slower convergence observed in the other particle codes ($\sim 50$ orbits). 
This suggests a potential viscosity discrepancy, although the timescale of convergence to a near-stationary torque value is typically controlled by a sound crossing time. Further motivation for viscosity testing is seen in Fig.~\ref{fig:surfacedensitymap_hr003_hill}, in which the \texttt{GASOLINE} velocity map around the secondary shows similar features to \texttt{DISCO}, the latter of which was found to be due to an oversimplified implementation of viscous fluxes (see Appendix~\ref{app:codes}). 
An additional difference in the \texttt{GASOLINE} runs is the inclusion of self-gravity. Although the disc is gravitationally stable, the ratio of disc mass to secondary mass is of order $10^{-2}$.  
Since the migration torque follows from the sensitive balance of similar-magnitude effects, the non-negligible mass of the disk might
affect the behaviour of the torque through self-gravity.
Moreover, as we have previously noted, we observe a significant accumulation of gas around the secondary, leading to
a more prominent overdensity compared to the other codes. 
This high density peak at the secondary location may reflect the
non-negligible impact of self-gravity, which in turn would affect
the local, co-orbital component of the torque. The effect of
self-gravity is mediated by the gravitational softening of gas
particles, which then would become another sensitive parameter
in this context.

\section{Discussion}\label{sec:discussion}

\subsection{Gas-Embedded IMRIs as Gravitational Wave Sources}
\label{sec:discrepancies}
EMRIs and IMRIs will be crucial targets for LISA and other GW detectors. They will be important for testing general relativity \citep[e.g.,][]{2024arXiv240108085C}, measuring SMBH masses and spins, and understanding the dynamics of nuclear star clusters. However, these efforts will be stymied, and their results potentially quite biased, if the environment of each IMRI and EMRI cannot be modeled with sufficient accuracy.

Due to the long inspiral of IMRIs in the LISA band, we expect thousands of detectable GW cycles. Therefore, even a small perturbation by gas to the binary dynamics can accumulate and produce a detectable GW dephasing \citep[e.g.,][]{Yunes2011,KocsisYunesLoeb:2011,derdzinski_evolution_2021,Garg2022,Garg2025}. However, things may become even more complicated if binaries can sustain appreciable eccentricity near merger, as the effects of eccentricity and gas on the GW waveform can be somewhat degenerate \citep[e.g.,][]{Garg2024b} or falsely violate general relativity \citep[e.g.,][]{Garg2024d}. This is indeed the case suggested by our 3D simulations, in which the gas-induced torque on the binary is negative.

\begin{table*}
    \centering
    \setlength{\extrarowheight}{5.pt}
    \begin{tabular}{|c|c|c|c|c|c|c|}
    \hline
    Code & 2D/3D & Country & Core number/type & Runtime [hrs] & CO$_2\rm e$ [kg] & CO$_2{\rm e}|_{\rm Austria}$ [kg] \\
    \hline
     \texttt{ATHENA++} & 2D & USA (MD) & 128 CPU & 2.16 ($h=0.03$) & 0.78238 & 0.28\\
     \hline
     \texttt{DISCO}  & 2D & USA (TN) & 32 CPU & 29.8 ($h=0.1$), 9.8 ($h=0.03$) &  4.56, 1.50 & 1.86, 0.76  \\
     \hline
     \texttt{DISCO v2} & 2D & USA (NJ) & 96 CPU & 2.39 ($h=0.03$) & 0.47543 & 0.225 \\
     \hline
     \texttt{FARGO3D} & 2D & Mexico & 1 GPU (P100) & 2.00 ($h=0.1$), 0.21 ($h=0.03$) & 0.22, 0.024 & 0.057, 0.0062\\
     \hline
     \texttt{FARGO3D} & 3D & Mexico & 1 GPU (P100) & 430 ($h=0.1$), 57 $(h=0.03)$ & 47.2, 6.26 & 12.16, 1.6\\
     \hline
     \texttt{GIZMO} & 3D & Switzerland & 128 CPU & 328 ($h=0.1$), 141 ($h=0.03$) & 3.46, 1.6 & 33.4, 15.4\\
     \hline
     \texttt{GASOLINE} & 3D & Switzerland & 640 CPU & 10560 ($h=0.1$), 4560 ($h=0.03$) & 518.62, 223.95 & 5010, 2160\\
     \hline
     \texttt{RAMSES} & 2D & France & 80 CPU &  150 ($h=0.1$), 50 ($h=0.03$)  & 8.43, 2.81  & 18.3, 6.1  \\
     \hline
     \texttt{PHANTOM} & 3D & Italy & 48 CPU & 5688 ($h=0.1$), 360 ($h=0.03$)& 11760, 744 & 4037.4, 255.4 \\
    \hline
    \end{tabular}
    \vspace{2mm}
    \caption{Where, on how many GPUs or CPU cores, and for how long each simulation was run, along with an estimate of the associated CO$_2$ emissions. The costs listed here are for simulations of 100 (actual or extrapolated, for \texttt{GASOLINE}) orbital periods in duration. In the rightmost column, we show the estimated CO$_2$ footprint if all simulations were ran in the same country. The total computing impact of the project is much larger than the sum of all the simulations in the table: many codes were run multiple times, and oftentimes human error or omission necessitated running the same simulation more than once. We also note that, in contrast with all other codes, the \texttt{GASOLINE} simulations compute self-gravity for all particles, thereby increasing the computational cost significantly.
    \vspace{2mm}  }
    \label{tab:CO2}
\end{table*}

Whether or not gas will significantly affect a given IMRI depends sensitively on the binary mass ratio and disc properties, especially the surface density of gas within the inner regions of the disc: while thin and dense discs might appreciably alter a given inspiral, a more tenuous environment could leave no trace at all \citep[e.g.,][]{derdzinski_evolution_2021}. Rigorous modeling of IMRIs and EMRIs will thus place constraints on the environments around SMBHs. If environments are stochastic, this can complicate the constraint of the disc properties \citep{2025PhRvD.111j4079C}. However, periodicities in the torque evolution can provide additional information \citep{2024PhRvD.110j3005Z}.  For IMRIs, when the disc-BH interaction can be weakly nonlinear, simulations such as those carried out in this work will be crucial. For thicker discs (and systems safely in the linear regime, with $q\lesssim h^3$), most codes agree with each other and with linear estimates in terms of the magnitude of the torque, given they are run with sufficient resolution (see Figures~\ref{fig:torque_h01_M2} and \ref{fig:resolution}). The variable features, however, warrant further investigation. 

For thinner discs, where simulations are more essential and linear theory fails outright, differences between codes also become more substantial. While the grid-based codes \texttt{ATHENA++}, \texttt{RAMSES}, \texttt{DISCO v2} and \texttt{FARGO3D} agree with each other, \texttt{DISCO} disagrees with these codes because of the assumptions made in its viscosity implementation.\footnote{A previous implementation of the viscosity in \texttt{RAMSES}, with assumptions in part similar to the ones made in \texttt{DISCO}, also led to non-convergent results that were different from all the other codes.}

At lower resolution, the particle-based codes \texttt{PHANTOM} and \texttt{GIZMO} agree on the sign of the migration direction but disagree on the value of the torque, while the particle-based code \texttt{GASOLINE} finds an almost-zero torque value. For comparison, we ran a small number of low-resolution 3D simulations using \texttt{FARGO3D}: when gas within $<0.4R_{\rm H}$ is excluded from the torque calculation (hollow red squares), the results agree fairly well with \texttt{GIZMO}; however, when the torque from the entire domain is included (as in every other data point) the torque was found to be positive and more than an order of magnitude larger than in the 2D case. While many of the 2D grid-based simulations appear to tend towards convergence (despite at different values of the torque), we have been unable to demonstrate convergence in any of the 3D simulations. 
Still, we hope the results of this study will help guide researchers in future studies of IMRI-disc interactions.

\subsection{Potential Electromagnetic Counterparts}

If detected, electromagnetic (EM) counterparts to IMRIs could provide intriguing multi-messenger targets and allow astronomers to directly probe the inner structure of SMBH accretion discs. Here we list two examples of EM features that could be observed from IMRIs embedded in AGN discs.\\
\indent As a simple model, we can imagine that the optical/ultraviolet/X-ray monochromatic flux of an AGN is produced by concentric rings in the accretion disc of inwardly increasing temperatures, which integrated form the spectral energy distribution (SED). If IMRIs open a gap in the disc, the spectral energy distribution will be altered by the dearth of material at certain temperatures \citep[e.g.,][]{mckernan14}.\\
\indent Another relevant EM feature is the effect of such gaps on the shape of X-ray lines, such as Fe K$\alpha$ \citep[e.g.,][]{gultekin12,mckernan14}. The iron line is typically produced by fluorescence of incident coronal X-rays on material surrounding the black hole, particularly the accretion disc \citep[e.g.,][]{matt93}. Especially in the inner regions of the disc, as the orbit of the secondary becomes relativistic, gap opening would produce a dip in the iron line profile due to the lack of emission at one specific energy. Additionally, both the energy and the intensity of the dip might be modulated on the orbital period of the secondary.

As we found in Section~\ref{sec:thinDisc}, gap opening in simulations can vary significantly between codes: \texttt{GASOLINE}, for example, found noticeably deeper gaps than the other codes. 
Detailed spectral modeling may rely on accurately modeling the velocity profile of the disc as well, making \texttt{DISCO} (but not \texttt{DISCO v2}) unreliable owing  to its approximate viscosity treatment and the corresponding errors in the radial velocity of the gas near the secondary.
Variability of the torque betrays general variability in the accretion flow; based on Section~\ref{sec:thinDisc}, it seems that there may be numerical code-to-code systematic uncertainties in any variability study, with \texttt{GIZMO}, \texttt{GASOLINE}, and \texttt{PHANTOM} especially susceptible to particle noise. 
We have also found that the amount of variability depends sensitively on how each simulation is initialized, which should be kept in mind when attempting to model variability numerically. Still, it is also worth keeping in mind that AGN discs are expected to be highly turbulent \citep[e.g.][]{kara_supermassive_2025}, as opposed to the viscous largely-laminar models used here, so one should also consider which physical ingredients are captured by a given code.

\subsection{Efficiency and Environmental Impact}

As we have shown above, analytical estimates fail completely in the weakly nonlinear regime of disc-BH interactions probed by IMRIs, necessitating simulations. However, simulations can consume substantial time and computing resources, which can lead to deleterious greenhouse gas emissions. More efficient codes (at a given level of accuracy) should be preferred over less efficient codes, given the limited computational resources available to the community and their lesser harm to the environment. We caution that each computer system can be quite unique, and code performance and efficiency can be affected by numerous factors, from hardware details like interconnect and file system input/output speeds, to the compilers available on each cluster and used with each code. These factors preclude a fair and direct comparison between the codes in this study, and one should test the performance and scaling of their code of choice before starting a large suite of simulations on the cluster available to them. Still, in an effort to increase transparency in the environmental impact of simulation studies, we report the estimated costs of the runs developed for this paper.

Table~\ref{tab:CO2} lists the runtime and resources consumed by each simulation. We have estimated the corresponding carbon footprints (or CO$_2$ equivalent emission, CO$_2\rm e$) based on the core-hours, the CPU model, and the geographical location where each simulation was conducted using the calculator provided by \url{https://calculator.green-algorithms.org/} \citep{LannelongueGreen}. Because the carbon intensity of a simulation can vary enormously based on the power source used for a particular cluster, we have also included a column normalizing each estimate to what they might have been in a fiducial reference country (the calculator default, Austria). For reference, a typical one-way trans-Atlantic flight produces $\sim$300~kg~CO$_2\rm e$ per passenger.\footnote{According to the International Civil Aviation Organization estimate for a flight from New York City to London (\url{www.icao.int}).}

When interpreting the results shown in Table~\ref{tab:CO2}, one must keep in mind that three-dimensional simulations are much more expensive than two-dimensional simulations at comparable resolutions. If 2D simulations are sufficient, they should be strongly preferred, although they will lack the valuable information provided by three dimensional runs. If 3D simulations are necessary, they should be approached with much greater prudence. \texttt{FARGO3D} was the most efficient code out of those tested in this work, in both 2D (by a modest margin) and 3D (by a great margin). However, \texttt{FARGO3D} relied on GPUs, while if one is limited to CPUs then \texttt{ATHENA++} or \texttt{DISCO v2} would be similarly applicable.\footnote{Three dimensional effects are clearly quite important in the thin-disc case. However, the only 3D grid-based simulation we have included above was run using GPUs, making it difficult to directly compare with the 3D particle-based codes. Taking the ratio of the costs of the 3D and 2D \texttt{FARGO3D} simulations, we can estimate the approximate relative cost of a 3D vs 2D grid-based simulation to be a factor of $\sim260$. Applying this to the \texttt{ATHENA++} and \texttt{DISCOv2} CO$_2$ emissions, and accounting for the factor of $3$ in linear resolution difference between the fiducial particle-based and grid-based simulations, suggests that applying the grid-based cylindrical-coordinate CPU codes in 3D would result in CO$_2$ emissions of the order of $\sim$1.5--2.5$\,\rm{kg}$, or $\sim$124--204$\,\rm{kg}$ at the typical resolution of the 2D grid-based codes.} It is important to consider that the comparison shown here is based on the application of these codes to rotationally supersonic accretion discs, and the suitability of each code will vary depending on the system being modeled. 

Although not included in this code comparison, the performance-portable code \texttt{IDEFIX} \citep{2023A&A...677A...9L} should also be kept in mind, given its ability to run on GPUs which tend to be much more energy efficient (per floating point operation) \citep[e.g.,][]{2025FrP....1342474S}. Next-generation particle-based codes like \texttt{SHAMROCK} \citep{2025MNRAS.539....1D} and \texttt{SPH-EXA} \citep{2025arXiv250310273C} may also provide efficiency improvements over those included in this study. The good performance of \texttt{RAMSES} in both setups suggests that Cartesian-grid codes can also produce accurate results, and accordingly Cartesian GPU-enabled codes such as \texttt{AthenaK} \citep{2024arXiv240916053S} or \texttt{DYABLO} \citep{2025JPhCS2997a2014D}
 might be useful for future investigations into this problem, particularly those requiring three-dimensional or relativistic simulations. 

\section{Conclusions}

We have conducted a series of simulations of disc-embedded intermediate mass ratio ($q=10^{-4}$) binaries. We tested eight different hydrodynamics codes and measured the torques, surface density profiles, and disc velocity signatures of the binary in discs of two thicknesses, $H/r=0.1$ and $H/r=0.03$. The thicker-disc case allowed us to compare the performance of each code against analytical predictions, while the thinner-disc was more representative of an AGN disc and necessitates numerical simulations, since linear theory breaks down. Although EMRIs and IMRIs will be exciting sources with which to probe the inner regions of galactic nuclei and stress-test general relativity, extracting unbiased information from these sources will require accurate models of how their environments can affect their orbits. The studies presented here, in addition to providing two data points on how accretion discs may affect IMRIs, should aid future studies by contextualizing the cost and efficiency of various codes. 

In the thicker-disc case, we find qualitative agreement between all codes at higher resolutions, but the lowest-resolution 3D simulations deviate by up to $\sim$50\% on the magnitude of the torque but still agree on the (negative) sign. Note that using a smaller softening in 3D particle-based codes leads to a far better match with the (3D) analytical prediction.
The thinner-disc case is more challenging numerically, and most codes were unable to achieve agreement: omissions in \texttt{DISCO}'s viscosity implementation lead to errors in the coorbital gas flow, though the other grid based codes (including \texttt{DISCOv2} and its complete viscosity implementation) are in fairly close agreement.
The particle codes (all in 3D) find a qualitatively different result, that gas drives the secondary to inspiral, while a 3D \texttt{FARGO3D} simulation found that the contribution from within $0.4R_{\rm H}$ might drive the torque to flip sign, becoming positive and an order of magnitude higher, therefore leading to outspiral at an even more rapid rate than in the 2D simulations.
We therefore conclude that 3D simulations are necessary in order to properly model the flow around the secondary as this is essentially the region that governs the outcome in terms of IMRI migration direction.
Further simulations, both 2D and 3D are needed in order to ensure that the final answer does not depend on resolution.

We also investigated the computational and environmental cost of these simulations. To minimize environmental costs, our results suggest that GPU-based codes should be preferred if possible. Otherwise, it appears that grid-based codes, using techniques such as moving meshes or Lagrangian remapping/orbital advection, can provide accurate and still fairly efficient results for this problem. Out of the particle-based codes, \texttt{GIZMO} seems to be the most efficient at addressing this problem, especially when leveraging particle-splitting to better resolve gas around the secondary.

\section*{Description of author contributions} 

Coordination of this project was led by A.~Derdzinski, A.~J.~Dittmann, A.~Franchini, and A.~Lupi. The simulations in this study were performed by N.~Brucy, P.~R.~Capelo, A.~Derdzinski, A.~J.~Dittmann, A.~Franchini, F.~S.~Masset, R.~Mignon-Risse, E.~Santiago~Leandro, M.~Rizzo~Smith, M.~Toscani, D.~Velasco-Romero, and R.~Wissing. Simulation analysis was performed by A.~Derdzinski, A.~J.~Dittmann, M.~Garg, and R.~Mignon-Risse. Significant writing of the manuscript was done by A.~Derdzinski, A.~J.~Dittmann, A.~Franchini, F.~S.~Masset, L.~Mayer, R.~Mignon-Risse, and R.~Serafinelli. Supportive input to the writing was provided by N.~Brucy, P.~R.~Capelo, M.~Garg, J.~Menu, and R.~Wissing. This project was shaped by regular collaborative discussions featuring input from N.~Brucy, A.~Derdzinski, A.~J.~Dittmann, A.~Franchini, M.~Garg, A.~Lupi, F.~S.~Masset, L.~Mayer, R.~Mignon-Risse, and M.~Toscani. Documentation of these meetings was provided by N.~Brucy. The calculation of CO2 impact of this study was overseen by L.~Souvaitzis. 
Comments on the final manuscript were provided by D. J. D'Orazio and J. Menu.

\section*{Data Availability}

We encourage readers to further investigate the data that were produced for this study. 
Snapshot\soutPC{s} files containing the fluid positions, densities, and velocities along with timeseries files containing calculated torque components are publicly available on \url{https://zenodo.org/records/17372154}. 

\begin{acknowledgements}
This project benefited from insightful discussions with several members of the LISA Astrophysics Working Group. AD thanks Kelly Holley-Bockelmann for building a radical environment where even a code comparison study can feel fulfilling. AF acknowledges financial support from the Unione europea-Next Generation EU, Missione 4 Componente 1
CUP G43C24002290001. AF also acknowledges support provided by the “GW-learn” grant agreement CRSII5 213497 and the Tomalla Foundation. AF acknowledges computational support provided by the Swiss National Computing System (CSCS).
NB acknowledges support from the ANR BRIDGES grant (ANR-23-CE31-0005) and from the European Research Council synergy grant ECOGAL (Grant: 855130).
Some of the \texttt{RAMSES} simulations were produced using the TGGC (Très Grand Centre de Calcul), through the GENCI allocation grant A0170411111.
NB also gratefully acknowledge support from the CBPsmn (PSMN, Pôle Scientifique de Modélisation Numérique) of the ENS de Lyon for the computing resources. The platform operates the SIDUS solution \citep{quemenerSIDUSSolutionExtreme2013} developed by Emmanuel Quemener. PRC acknowledges support from the Swiss National Science Foundation under the Sinergia Grant CRSII5\_213497 (GW-Learn). AD and MRS acknowledge financial support from the National Science Foundation (NSF) EMIT Program (NSF-2125764). MRS also acknowledges support from the NSF Graduate Research Fellowship Program under Grant No. NSF-2444112. AJD was supported by NASA through the NASA Hubble Fellowship grant No. HST-HF2-51553.001, awarded by the Space Telescope Science Institute, which is operated by the Association of Universities for Research in Astronomy, Inc., for NASA, under contract NAS5-26555.
RMR has received funding from the European Research Concil (ERC) under the European Union Horizon 2020 research and innovation programme (grant agreement number No. 101002352, PI: M. Linares) and acknowledges the project Ref. PID2024-157196NB-I00 (MICIU/AEI/10.13039/501100011033). RS acknowledges funding from the CAS-ANID grant No. CAS220016. JM thanks the Belgian Federal Science Policy Office (BELSPO) for the provision of financial support in the framework of the PRODEX Programme of the European Space Agency (ESA) under contract number PEA4000144253.
\end{acknowledgements}

\bibliographystyle{aa} 
\bibliography{refs,bibliography} 

@ARTICLE{2022MNRAS.514.3886M,
       author = {{McKernan}, B. and {Ford}, K.~E.~S. and {Callister}, T. and {Farr}, W.~M. and {O'Shaughnessy}, R. and {Smith}, R. and {Thrane}, E. and {Vajpeyi}, A.},
        title = "{LIGO-Virgo correlations between mass ratio and effective inspiral spin: testing the active galactic nuclei channel}",
      journal = {\mnras},
         year = 2022,
        month = aug,
       volume = {514},
       number = {3},
        pages = {3886-3893},
          doi = {10.1093/mnras/stac1570},
archivePrefix = {arXiv},
       eprint = {2107.07551},
 primaryClass = {astro-ph.HE},
       adsurl = {https://ui.adsabs.harvard.edu/abs/2022MNRAS.514.3886M}
}

@ARTICLE{matt93,
       author = {{Matt}, G. and {Fabian}, A.~C. and {Ross}, R.~R.},
        title = "{Iron K-alpha lines from X-ray photoionized accretion discs.}",
      journal = {\mnras},
         year = 1993,
        month = may,
       volume = {262},
        pages = {179-186},
          doi = {10.1093/mnras/262.1.179},
       adsurl = {https://ui.adsabs.harvard.edu/abs/1993MNRAS.262..179M}
}

@ARTICLE{2022MNRAS.517.5827F,
       author = {{Ford}, K.~E. Saavik and {McKernan}, Barry},
        title = "{Binary black hole merger rates in AGN discs versus nuclear star clusters: loud beats quiet}",
      journal = {\mnras},
         year = 2022,
        month = dec,
       volume = {517},
       number = {4},
        pages = {5827-5834},
          doi = {10.1093/mnras/stac2861},
archivePrefix = {arXiv},
       eprint = {2109.03212},
 primaryClass = {astro-ph.HE},
       adsurl = {https://ui.adsabs.harvard.edu/abs/2022MNRAS.517.5827F}
}

@ARTICLE{2025arXiv250923897L,
       author = {{Li}, Yin-Jie and {Wang}, Yuan-Zhu and {Tang}, Shao-Peng and {Fan}, Yi-Zhong},
        title = "{Aligned Hierarchical Black Hole Mergers in AGN disks revealed by GWTC-4}",
      journal = {arXiv e-prints},
         year = 2025,
        month = sep,
          eid = {arXiv:2509.23897},
        pages = {arXiv:2509.23897},
          doi = {10.48550/arXiv.2509.23897},
archivePrefix = {arXiv},
       eprint = {2509.23897},
 primaryClass = {astro-ph.HE},
       adsurl = {https://ui.adsabs.harvard.edu/abs/2025arXiv250923897L}
}

@ARTICLE{2025arXiv250818082T,
       author = {{The LIGO Scientific Collaboration} and {the Virgo Collaboration} and {the KAGRA Collaboration} and {Abac}, A.~G. and {Abouelfettouh}, I. and {Acernese}, F. and {Ackley}, K. and {Adamcewicz}, C. and {Adhicary}, S. and {Adhikari}, D. and {Adhikari}, N. and {Adhikari}, R.~X. and {Adkins}, V.~K. and {Afroz}, S. and {Agapito}, A. and {Agarwal}, D. and {Agathos}, M. and {Aggarwal}, N. and {Aggarwal}, S. and {Aguiar}, O.~D. and {Ahrend}, I. -L. and {Aiello}, L. and {Ain}, A. and {Ajith}, P. and {Akutsu}, T. and {Albanesi}, S. and {Ali}, W. and {Al-Kershi}, S. and {All{\'e}n{\'e}}, C. and {Allocca}, A. and {Al-Shammari}, S. and {Altin}, P.~A. and {Alvarez-Lopez}, S. and {Amar}, W. and {Amarasinghe}, O. and {Amato}, A. and {Amicucci}, F. and {Amra}, C. and {Ananyeva}, A. and {Anderson}, S.~B. and {Anderson}, W.~G. and {Andia}, M. and {Ando}, M. and {Andr{\'e}s-Carcasona}, M. and {Andri{\'c}}, T. and {Anglin}, J. and {Ansoldi}, S. and {Antelis}, J.~M. and {Antier}, S. and {Aoumi}, M. and {Appavuravther}, E.~Z. and {Appert}, S. and {Apple}, S.~K. and {Arai}, K. and {Araya}, A. and {Araya}, M.~C. and {Arca Sedda}, M. and {Areeda}, J.~S. and {Aritomi}, N. and {Armato}, F. and {Armstrong}, S. and {Arnaud}, N. and {Arogeti}, M. and {Aronson}, S.~M. and {Arun}, K.~G. and {Ashton}, G. and {Aso}, Y. and {Asprea}, L. and {Assiduo}, M. and {Assis de Souza Melo}, S. and {Aston}, S.~M. and {Astone}, P. and {Attadio}, F. and {Aubin}, F. and {AultONeal}, K. and {Avallone}, G. and {Avila}, E.~A. and {Babak}, S. and {Badger}, C. and {Bae}, S. and {Bagnasco}, S. and {Baiotti}, L. and {Bajpai}, R. and {Baka}, T. and {Baker}, A.~M. and {Baker}, K.~A. and {Baker}, T. and {Baldi}, G. and {Baldicchi}, N. and {Ball}, M. and {Ballardin}, G. and {Ballmer}, S.~W. and {Banagiri}, S. and {Banerjee}, B. and {Bankar}, D. and {Baptiste}, T.~M. and {Baral}, P. and {Baratti}, M. and {Barayoga}, J.~C. and {Barish}, B.~C. and {Barker}, D. and {Barman}, N. and {Barneo}, P. and {Barone}, F. and {Barr}, B. and {Barsotti}, L. and {Barsuglia}, M. and {Barta}, D. and {Bartoletti}, A.~M. and {Barton}, M.~A. and {Bartos}, I. and {Basalaev}, A. and {Bassiri}, R. and {Basti}, A. and {Bawaj}, M. and {Baxi}, P. and {Bayley}, J.~C. and {Baylor}, A.~C. and {Baynard}, II, P.~A. and {Bazzan}, M. and {Bedakihale}, V.~M. and {Beirnaert}, F. and {Bejger}, M. and {Belardinelli}, D. and {Bell}, A.~S. and {Bellie}, D.~S. and {Bellizzi}, L. and {Benoit}, W. and {Bentara}, I. and {Bentley}, J.~D. and {Ben Yaala}, M. and {Bera}, S. and {Bergamin}, F. and {Berger}, B.~K. and {Bernuzzi}, S. and {Beroiz}, M. and {Berry}, C.~P.~L. and {Bersanetti}, D. and {Bertheas}, T. and {Bertolini}, A. and {Betzwieser}, J. and {Beveridge}, D. and {Bevilacqua}, G. and {Bevins}, N. and {Bhandare}, R. and {Bhatt}, R. and {Bhattacharjee}, D. and {Bhattacharyya}, S. and {Bhaumik}, S. and {Biancalana}, V. and {Bianchi}, A. and {Bilenko}, I.~A. and {Billingsley}, G. and {Binetti}, A. and {Bini}, S. and {Binu}, C. and {Biot}, S. and {Birnholtz}, O. and {Biscoveanu}, S. and {Bisht}, A. and {Bitossi}, M. and {Bizouard}, M. -A. and {Blaber}, S. and {Blackburn}, J.~K. and {Blagg}, L.~A. and {Blair}, C.~D. and {Blair}, D.~G. and {Bode}, N. and {Boettner}, N. and {Boileau}, G. and {Boldrini}, M. and {Bolingbroke}, G.~N. and {Bolliand}, A. and {Bonavena}, L.~D. and {Bondarescu}, R. and {Bondu}, F. and {Bonilla}, E. and {Bonilla}, M.~S. and {Bonino}, A. and {Bonnand}, R. and {Borchers}, A. and {Borhanian}, S. and {Boschi}, V. and {Bose}, S. and {Bossilkov}, V. and {Bothra}, Y. and {Boudon}, A. and {Bourg}, L. and {Boyle}, M. and {Bozzi}, A. and {Bradaschia}, C. and {Brady}, P.~R. and {Branch}, A. and {Branchesi}, M. and {Braun}, I. and {Briant}, T. and {Brillet}, A. and {Brinkmann}, M. and {Brockill}, P. and {Brockmueller}, E.},
        title = "{GWTC-4.0: Updating the Gravitational-Wave Transient Catalog with Observations from the First Part of the Fourth LIGO-Virgo-KAGRA Observing Run}",
      journal = {arXiv e-prints},
         year = 2025,
        month = aug,
          eid = {arXiv:2508.18082},
        pages = {arXiv:2508.18082},
          doi = {10.48550/arXiv.2508.18082},
archivePrefix = {arXiv},
       eprint = {2508.18082},
 primaryClass = {gr-qc},
       adsurl = {https://ui.adsabs.harvard.edu/abs/2025arXiv250818082T}
}

@ARTICLE{2022PhRvL.129x1103C,
       author = {{Cardoso}, Vitor and {Destounis}, Kyriakos and {Duque}, Francisco and {Macedo}, Rodrigo Panosso and {Maselli}, Andrea},
        title = "{Gravitational Waves from Extreme-Mass-Ratio Systems in Astrophysical Environments}",
      journal = {\prl},
         year = 2022,
        month = dec,
       volume = {129},
       number = {24},
          eid = {241103},
        pages = {241103},
          doi = {10.1103/PhysRevLett.129.241103},
archivePrefix = {arXiv},
       eprint = {2210.01133},
 primaryClass = {gr-qc},
       adsurl = {https://ui.adsabs.harvard.edu/abs/2022PhRvL.129x1103C}
}

@ARTICLE{2024PhRvD.110j3005Z,
       author = {{Zwick}, Lorenz and {Tiede}, Christopher and {Trani}, Alessandro A. and {Derdzinski}, Andrea and {Haiman}, Zoltan and {D'Orazio}, Daniel J. and {Samsing}, Johan},
        title = "{Novel category of environmental effects on gravitational waves from binaries perturbed by periodic forces}",
      journal = {\prd},
         year = 2024,
        month = nov,
       volume = {110},
       number = {10},
          eid = {103005},
        pages = {103005},
          doi = {10.1103/PhysRevD.110.103005},
archivePrefix = {arXiv},
       eprint = {2405.05698},
 primaryClass = {gr-qc},
       adsurl = {https://ui.adsabs.harvard.edu/abs/2024PhRvD.110j3005Z}
}

@ARTICLE{2025PhRvD.111h4006D,
       author = {{Duque}, Francisco and {Kejriwal}, Shubham and {Sberna}, Laura and {Speri}, Lorenzo and {Gair}, Jonathan},
        title = "{Constraining accretion physics with gravitational waves from eccentric extreme-mass-ratio inspirals}",
      journal = {\prd},
         year = 2025,
        month = apr,
       volume = {111},
       number = {8},
          eid = {084006},
        pages = {084006},
          doi = {10.1103/PhysRevD.111.084006},
archivePrefix = {arXiv},
       eprint = {2411.03436},
 primaryClass = {gr-qc},
       adsurl = {https://ui.adsabs.harvard.edu/abs/2025PhRvD.111h4006D}
}

@ARTICLE{2025PhRvD.111j4079C,
       author = {{Copparoni}, Lorenzo and {Barausse}, Enrico and {Speri}, Lorenzo and {Sberna}, Laura and {Derdzinski}, Andrea},
        title = "{Implications of stochastic gas torques for asymmetric binaries in the LISA band}",
      journal = {\prd},
         year = 2025,
        month = may,
       volume = {111},
       number = {10},
          eid = {104079},
        pages = {104079},
          doi = {10.1103/PhysRevD.111.104079},
archivePrefix = {arXiv},
       eprint = {2502.10087},
 primaryClass = {gr-qc},
       adsurl = {https://ui.adsabs.harvard.edu/abs/2025PhRvD.111j4079C}
}

@ARTICLE{2025PhRvD.112f3005Z,
       author = {{Zwick}, Lorenz and {Hendriks}, Kai and {O'Neill}, David and {Tak{\'a}tsy}, J{\'a}nos and {Kirkeberg}, Philip and {Tiede}, Christopher and {Stegmann}, Jakob and {Samsing}, Johan and {D'Orazio}, Daniel J.},
        title = "{Dissecting environmental effects with eccentric gravitational wave sources}",
      journal = {\prd},
         year = 2025,
        month = sep,
       volume = {112},
       number = {6},
          eid = {063005},
        pages = {063005},
          doi = {10.1103/lz7k-bvjf},
archivePrefix = {arXiv},
       eprint = {2506.09140},
 primaryClass = {astro-ph.HE},
       adsurl = {https://ui.adsabs.harvard.edu/abs/2025PhRvD.112f3005Z}
}

@ARTICLE{2025arXiv251002433D,
       author = {{Duque}, Francisco and {Sberna}, Laura and {Spiers}, Andrew and {Vicente}, Rodrigo},
        title = "{Extreme-mass-ratio inspirals in relativistic accretion discs}",
      journal = {arXiv e-prints},
         year = 2025,
        month = oct,
          eid = {arXiv:2510.02433},
        pages = {arXiv:2510.02433},
          doi = {10.48550/arXiv.2510.02433},
archivePrefix = {arXiv},
       eprint = {2510.02433},
 primaryClass = {gr-qc},
       adsurl = {https://ui.adsabs.harvard.edu/abs/2025arXiv251002433D}
}

@ARTICLE{2012ARA&A..50..211K,
       author = {{Kley}, W. and {Nelson}, R.~P.},
        title = "{Planet-Disk Interaction and Orbital Evolution}",
      journal = {\araa},
         year = 2012,
        month = sep,
       volume = {50},
        pages = {211-249},
          doi = {10.1146/annurev-astro-081811-125523},
archivePrefix = {arXiv},
       eprint = {1203.1184},
 primaryClass = {astro-ph.EP},
       adsurl = {https://ui.adsabs.harvard.edu/abs/2012ARA&A..50..211K}
}

@ARTICLE{2025arXiv250510488S,
       author = {{Suzuguchi}, Tomoya and {Omiya}, Hidetoshi and {Takeda}, Hiroki},
        title = "{Possibility of Multi-Messenger Observations of Quasi-Periodic Eruptions with X-rays and Gravitational Waves}",
      journal = {arXiv e-prints},
         year = 2025,
        month = may,
          eid = {arXiv:2505.10488},
        pages = {arXiv:2505.10488},
          doi = {10.48550/arXiv.2505.10488},
archivePrefix = {arXiv},
       eprint = {2505.10488},
 primaryClass = {astro-ph.HE},
       adsurl = {https://ui.adsabs.harvard.edu/abs/2025arXiv250510488S}
}

@ARTICLE{2024arXiv240108085C,
       author = {{C{\'a}rdenas-Avenda{\~n}o}, Alejandro and {Sopuerta}, Carlos F.},
        title = "{Testing gravity with Extreme-Mass-Ratio Inspirals}",
      journal = {arXiv e-prints},
         year = 2024,
        month = jan,
          eid = {arXiv:2401.08085},
        pages = {arXiv:2401.08085},
          doi = {10.48550/arXiv.2401.08085},
archivePrefix = {arXiv},
       eprint = {2401.08085},
 primaryClass = {gr-qc},
       adsurl = {https://ui.adsabs.harvard.edu/abs/2024arXiv240108085C}
}

@ARTICLE{2011PhRvD..84b4032K,
       author = {{Kocsis}, Bence and {Yunes}, Nicol{\'a}s and {Loeb}, Abraham},
        title = "{Observable signatures of extreme mass-ratio inspiral black hole binaries embedded in thin accretion disks}",
      journal = {\prd},
         year = 2011,
        month = jul,
       volume = {84},
       number = {2},
          eid = {024032},
        pages = {024032},
          doi = {10.1103/PhysRevD.84.024032},
archivePrefix = {arXiv},
       eprint = {1104.2322},
 primaryClass = {astro-ph.GA},
       adsurl = {https://ui.adsabs.harvard.edu/abs/2011PhRvD..84b4032K}
}

@ARTICLE{2025arXiv250310273C,
       author = {{Cabez{\'o}n}, Rub{\'e}n M. and {Garc{\'\i}a-Senz}, Domingo and {Seckin Simsek}, Osman and {Keller}, Sebastian and {Sanz}, Axel and {Zhu}, Yiqing and {Mayer}, Lucio and {Klessen}, Ralf and {Ciorba}, Florina M.},
        title = "{Modelling subsonic turbulence with SPH-EXA}",
      journal = {arXiv e-prints},
         year = 2025,
        month = mar,
          eid = {arXiv:2503.10273},
        pages = {arXiv:2503.10273},
          doi = {10.48550/arXiv.2503.10273},
archivePrefix = {arXiv},
       eprint = {2503.10273},
 primaryClass = {astro-ph.IM},
       adsurl = {https://ui.adsabs.harvard.edu/abs/2025arXiv250310273C}
}

@ARTICLE{Fletcher2019,
       author = {{Fletcher}, M. and {Nayakshin}, S. and {Stamatellos}, D. and {Dehnen}, W. and {Meru}, F. and {Mayer}, L. and {Deng}, H. and {Rice}, K.},
        title = "{Giant planets and brown dwarfs on wide orbits: a code comparison project}",
      journal = {\mnras},
         year = 2019,
        month = jul,
       volume = {486},
       number = {3},
        pages = {4398-4413},
          doi = {10.1093/mnras/stz1123},
archivePrefix = {arXiv},
       eprint = {1901.08089},
 primaryClass = {astro-ph.EP},
       adsurl = {https://ui.adsabs.harvard.edu/abs/2019MNRAS.486.4398F}
}

@ARTICLE{LaiDongMunoz2023,
       author = {{Lai}, Dong and {Mu{\~n}oz}, Diego J.},
        title = "{Circumbinary Accretion: From Binary Stars to Massive Binary Black Holes}",
      journal = {\araa},
         year = 2023,
        month = aug,
       volume = {61},
        pages = {517-560},
          doi = {10.1146/annurev-astro-052622-022933},
archivePrefix = {arXiv},
       eprint = {2211.00028},
 primaryClass = {astro-ph.HE},
       adsurl = {https://ui.adsabs.harvard.edu/abs/2023ARA&A..61..517L}
}

@ARTICLE{Duffell2016,
       author = {{Duffell}, Paul C.},
        title = "{DISCO: A 3D Moving-mesh Magnetohydrodynamics Code Designed for the Study of Astrophysical Disks}",
      journal = {\apjs},
         year = 2016,
        month = sep,
       volume = {226},
       number = {1},
          eid = {2},
        pages = {2},
          doi = {10.3847/0067-0049/226/1/2},
archivePrefix = {arXiv},
       eprint = {1605.03577},
 primaryClass = {physics.comp-ph},
       adsurl = {https://ui.adsabs.harvard.edu/abs/2016ApJS..226....2D}
}

@ARTICLE{1992ApJS...80..753S,
       author = {{Stone}, James M. and {Norman}, Michael L.},
        title = "{ZEUS-2D: A Radiation Magnetohydrodynamics Code for Astrophysical Flows in Two Space Dimensions. I. The Hydrodynamic Algorithms and Tests}",
      journal = {\apjs},
         year = 1992,
        month = jun,
       volume = {80},
        pages = {753},
          doi = {10.1086/191680},
       adsurl = {https://ui.adsabs.harvard.edu/abs/1992ApJS...80..753S}
}

@BOOK{1984frh..book.....M,
       author = {{Mihalas}, D. and {Mihalas}, B.~W.},
        title = "{Foundations of radiation hydrodynamics}",
         year = 1984,
        publisher="{Oxford}",
       adsurl = {https://ui.adsabs.harvard.edu/abs/1984frh..book.....M}
}

@ARTICLE{2001NewA....6...79S,
       author = {{Springel}, Volker and {Yoshida}, Naoki and {White}, Simon D.~M.},
        title = "{GADGET: a code for collisionless and gasdynamical cosmological simulations}",
      journal = {\na},
         year = 2001,
        month = apr,
       volume = {6},
       number = {2},
        pages = {79-117},
          doi = {10.1016/S1384-1076(01)00042-2},
archivePrefix = {arXiv},
       eprint = {astro-ph/0003162},
 primaryClass = {astro-ph},
       adsurl = {https://ui.adsabs.harvard.edu/abs/2001NewA....6...79S}
}

@ARTICLE{LISAastrophysics2023,
       author = {{Amaro-Seoane}, Pau and {Andrews}, Jeff and {Arca Sedda}, Manuel and {Askar}, Abbas and {Baghi}, Quentin and {Balasov}, Razvan and {Bartos}, Imre and {Bavera}, Simone S. and {Bellovary}, Jillian and {Berry}, Christopher P.~L. and {Berti}, Emanuele and {Bianchi}, Stefano and {Blecha}, Laura and {Blondin}, St{\'e}phane and {Bogdanovi{\'c}}, Tamara and {Boissier}, Samuel and {Bonetti}, Matteo and {Bonoli}, Silvia and {Bortolas}, Elisa and {Breivik}, Katelyn and {Capelo}, Pedro R. and {Caramete}, Laurentiu and {Cattorini}, Federico and {Charisi}, Maria and {Chaty}, Sylvain and {Chen}, Xian and {Chru{\'s}li{\'n}ska}, Martyna and {Chua}, Alvin J.~K. and {Church}, Ross and {Colpi}, Monica and {D'Orazio}, Daniel and {Danielski}, Camilla and {Davies}, Melvyn B. and {Dayal}, Pratika and {De Rosa}, Alessandra and {Derdzinski}, Andrea and {Destounis}, Kyriakos and {Dotti}, Massimo and {Du{\c{t}}an}, Ioana and {Dvorkin}, Irina and {Fabj}, Gaia and {Foglizzo}, Thierry and {Ford}, Saavik and {Fouvry}, Jean-Baptiste and {Franchini}, Alessia and {Fragos}, Tassos and {Fryer}, Chris and {Gaspari}, Massimo and {Gerosa}, Davide and {Graziani}, Luca and {Groot}, Paul and {Habouzit}, Melanie and {Haggard}, Daryl and {Haiman}, Zoltan and {Han}, Wen-Biao and {Istrate}, Alina and {Johansson}, Peter H. and {Khan}, Fazeel Mahmood and {Kimpson}, Tomas and {Kokkotas}, Kostas and {Kong}, Albert and {Korol}, Valeriya and {Kremer}, Kyle and {Kupfer}, Thomas and {Lamberts}, Astrid and {Larson}, Shane and {Lau}, Mike and {Liu}, Dongliang and {Lloyd-Ronning}, Nicole and {Lodato}, Giuseppe and {Lupi}, Alessandro and {Ma}, Chung-Pei and {Maccarone}, Tomas and {Mandel}, Ilya and {Mangiagli}, Alberto and {Mapelli}, Michela and {Mathis}, St{\'e}phane and {Mayer}, Lucio and {McGee}, Sean and {McKernan}, Berry and {Miller}, M. Coleman and {Mota}, David F. and {Mumpower}, Matthew and {Nasim}, Syeda S. and {Nelemans}, Gijs and {Noble}, Scott and {Pacucci}, Fabio and {Panessa}, Francesca and {Paschalidis}, Vasileios and {Pfister}, Hugo and {Porquet}, Delphine and {Quenby}, John and {Ricarte}, Angelo and {R{\"o}pke}, Friedrich K. and {Regan}, John and {Rosswog}, Stephan and {Ruiter}, Ashley and {Ruiz}, Milton and {Runnoe}, Jessie and {Schneider}, Raffaella and {Schnittman}, Jeremy and {Secunda}, Amy and {Sesana}, Alberto and {Seto}, Naoki and {Shao}, Lijing and {Shapiro}, Stuart and {Sopuerta}, Carlos and {Stone}, Nicholas C. and {Suvorov}, Arthur and {Tamanini}, Nicola and {Tamfal}, Tomas and {Tauris}, Thomas and {Temmink}, Karel and {Tomsick}, John and {Toonen}, Silvia and {Torres-Orjuela}, Alejandro and {Toscani}, Martina and {Tsokaros}, Antonios and {Unal}, Caner and {V{\'a}zquez-Aceves}, Ver{\'o}nica and {Valiante}, Rosa and {van Putten}, Maurice and {van Roestel}, Jan and {Vignali}, Christian and {Volonteri}, Marta and {Wu}, Kinwah and {Younsi}, Ziri and {Yu}, Shenghua and {Zane}, Silvia and {Zwick}, Lorenz and {Antonini}, Fabio and {Baibhav}, Vishal and {Barausse}, Enrico and {Bonilla Rivera}, Alexander and {Branchesi}, Marica and {Branduardi-Raymont}, Graziella and {Burdge}, Kevin and {Chakraborty}, Srija and {Cuadra}, Jorge and {Dage}, Kristen and {Davis}, Benjamin and {de Mink}, Selma E. and {Decarli}, Roberto and {Doneva}, Daniela and {Escoffier}, Stephanie and {Gandhi}, Poshak and {Haardt}, Francesco and {Lousto}, Carlos O. and {Nissanke}, Samaya and {Nordhaus}, Jason and {O'Shaughnessy}, Richard and {Portegies Zwart}, Simon and {Pound}, Adam and {Schussler}, Fabian and {Sergijenko}, Olga and {Spallicci}, Alessandro and {Vernieri}, Daniele and {Vigna-G{\'o}mez}, Alejandro},
        title = "{Astrophysics with the Laser Interferometer Space Antenna}",
      journal = {Living Reviews in Relativity},
         year = 2023,
        month = dec,
       volume = {26},
       number = {1},
          eid = {2},
        pages = {2},
          doi = {10.1007/s41114-022-00041-y},
archivePrefix = {arXiv},
       eprint = {2203.06016},
 primaryClass = {gr-qc},
       adsurl = {https://ui.adsabs.harvard.edu/abs/2023LRR....26....2A}
}

@ARTICLE{Ziampras2023,
       author = {{Ziampras}, Alexandros and {Paardekooper}, Sijme-Jan and {Nelson}, Richard P.},
        title = "{Buoyancy response of a disc to an embedded planet: a cross-code comparison at high resolution}",
      journal = {\mnras},
         year = 2023,
        month = nov,
       volume = {525},
       number = {4},
        pages = {5893-5904},
          doi = {10.1093/mnras/stad2692},
archivePrefix = {arXiv},
       eprint = {2308.14896},
 primaryClass = {astro-ph.EP},
       adsurl = {https://ui.adsabs.harvard.edu/abs/2023MNRAS.525.5893Z}
}

@ARTICLE{2012MNRAS.427.2022M,
       author = {{Meru}, Farzana and {Bate}, Matthew R.},
        title = "{On the convergence of the critical cooling time-scale for the fragmentation of self-gravitating discs}",
      journal = {\mnras},
         year = 2012,
        month = dec,
       volume = {427},
       number = {3},
        pages = {2022-2046},
          doi = {10.1111/j.1365-2966.2012.22035.x},
archivePrefix = {arXiv},
       eprint = {1209.1107},
 primaryClass = {astro-ph.EP},
       adsurl = {https://ui.adsabs.harvard.edu/abs/2012MNRAS.427.2022M}
}

@ARTICLE{TanakaOkada2024,
       author = {{Tanaka}, Hidekazu and {Okada}, Kohei},
        title = "{Three-dimensional Interaction between a Planet and an Isothermal Gaseous Disk. III. Locally Isothermal Cases}",
      journal = {\apj},
         year = 2024,
        month = jun,
       volume = {968},
       number = {1},
          eid = {28},
        pages = {28},
          doi = {10.3847/1538-4357/ad410d},
archivePrefix = {arXiv},
       eprint = {2404.12521},
 primaryClass = {astro-ph.EP},
       adsurl = {https://ui.adsabs.harvard.edu/abs/2024ApJ...968...28T}
}

@ARTICLE{1979MNRAS.186..799L,
       author = {{Lin}, D.~N.~C. and {Papaloizou}, J.},
        title = "{Tidal torques on accretion discs in binary systems with extreme mass ratios.}",
      journal = {\mnras},
         year = 1979,
        month = mar,
       volume = {186},
        pages = {799-812},
          doi = {10.1093/mnras/186.4.799},
       adsurl = {https://ui.adsabs.harvard.edu/abs/1979MNRAS.186..799L}
}

@ARTICLE{2025arXiv250904282C,
       author = {{Cordwell}, Amelia J. and {Ziampras}, Alexandros and {Brown}, Joshua J. and {Rafikov}, Roman R.},
        title = "{How two-dimensional are planet--disc interactions? I. Locally isothermal discs}",
      journal = {arXiv e-prints},
         year = 2025,
        month = sep,
          eid = {arXiv:2509.04282},
        pages = {arXiv:2509.04282},
          doi = {10.48550/arXiv.2509.04282},
archivePrefix = {arXiv},
       eprint = {2509.04282},
 primaryClass = {astro-ph.EP},
       adsurl = {https://ui.adsabs.harvard.edu/abs/2025arXiv250904282C}
}

@ARTICLE{Tanaka2002,
       author = {{Tanaka}, Hidekazu and {Takeuchi}, Taku and {Ward}, William R.},
        title = "{Three-Dimensional Interaction between a Planet and an Isothermal Gaseous Disk. I. Corotation and Lindblad Torques and Planet Migration}",
      journal = {\apj},
         year = 2002,
        month = feb,
       volume = {565},
       number = {2},
        pages = {1257-1274},
          doi = {10.1086/324713},
       adsurl = {https://ui.adsabs.harvard.edu/abs/2002ApJ...565.1257T}
}

@ARTICLE{2024MNRAS.534...39B,
       author = {{Brown}, Joshua J. and {Ogilvie}, Gordon I.},
        title = "{Horseshoes and spiral waves: capturing the 3D flow induced by a low-mass planet analytically}",
      journal = {\mnras},
         year = 2024,
        month = oct,
       volume = {534},
       number = {1},
        pages = {39-55},
          doi = {10.1093/mnras/stae2060},
archivePrefix = {arXiv},
       eprint = {2409.02687},
 primaryClass = {astro-ph.EP},
       adsurl = {https://ui.adsabs.harvard.edu/abs/2024MNRAS.534...39B}
}

@ARTICLE{2016ApJ...817...19M,
       author = {{Masset}, F.~S. and {Ben{\'\i}tez-Llambay}, P.},
        title = "{Horseshoe Drag in Three-dimensional Globally Isothermal Disks}",
      journal = {\apj},
         year = 2016,
        month = jan,
       volume = {817},
       number = {1},
          eid = {19},
        pages = {19},
          doi = {10.3847/0004-637X/817/1/19},
archivePrefix = {arXiv},
       eprint = {1511.07946},
 primaryClass = {astro-ph.EP},
       adsurl = {https://ui.adsabs.harvard.edu/abs/2016ApJ...817...19M}
}

@ARTICLE{Korycansky1993,
       author = {{Korycansky}, D.~G. and {Pollack}, J.~B.},
        title = "{Numerical Calculations of the Linear Response of a Gaseous Disk to a Protoplanet}",
      journal = {\icarus},
         year = 1993,
        month = mar,
       volume = {102},
       number = {1},
        pages = {150-165},
          doi = {10.1006/icar.1993.1039},
       adsurl = {https://ui.adsabs.harvard.edu/abs/1993Icar..102..150K}
}

@ARTICLE{Goldreich1980,
       author = {{Goldreich}, P. and {Tremaine}, S.},
        title = "{Disk-satellite interactions.}",
      journal = {\apj},
         year = 1980,
        month = oct,
       volume = {241},
        pages = {425-441},
          doi = {10.1086/158356},
       adsurl = {https://ui.adsabs.harvard.edu/abs/1980ApJ...241..425G}
}

@ARTICLE{Kanagawa2018,
       author = {{Kanagawa}, Kazuhiro D. and {Tanaka}, Hidekazu and {Szuszkiewicz}, Ewa},
        title = "{Radial Migration of Gap-opening Planets in Protoplanetary Disks. I. The Case of a Single Planet}",
      journal = {\apj},
         year = 2018,
        month = jul,
       volume = {861},
       number = {2},
          eid = {140},
        pages = {140},
          doi = {10.3847/1538-4357/aac8d9},
archivePrefix = {arXiv},
       eprint = {1805.11101},
 primaryClass = {astro-ph.EP},
       adsurl = {https://ui.adsabs.harvard.edu/abs/2018ApJ...861..140K}
}

@INPROCEEDINGS{Paardekooper2023,
       author = {{Paardekooper}, S. and {Dong}, R. and {Duffell}, P. and {Fung}, J. and {Masset}, F.~S. and {Ogilvie}, G. and {Tanaka}, H.},
        title = "{Planet-Disk Interactions and Orbital Evolution}",
    booktitle = {Protostars and Planets VII},
         year = 2023,
       editor = {{Inutsuka}, S. and {Aikawa}, Y. and {Muto}, T. and {Tomida}, K. and {Tamura}, M.},
       series = {Astronomical Society of the Pacific Conference Series},
       volume = {534},
        month = jul,
        pages = {685},
          doi = {10.48550/arXiv.2203.09595},
archivePrefix = {arXiv},
       eprint = {2203.09595},
 primaryClass = {astro-ph.EP},
       adsurl = {https://ui.adsabs.harvard.edu/abs/2023ASPC..534..685P}
}

@ARTICLE{2024ApJ...967...12D,
       author = {{Dittmann}, Alexander J. and {Ryan}, Geoffrey},
        title = "{The Evolution of Accreting Binaries: From Brown Dwarfs to Supermassive Black Holes}",
      journal = {\apj},
         year = 2024,
        month = may,
       volume = {967},
       number = {1},
          eid = {12},
        pages = {12},
          doi = {10.3847/1538-4357/ad2f1e},
archivePrefix = {arXiv},
       eprint = {2310.07758},
 primaryClass = {astro-ph.GA},
       adsurl = {https://ui.adsabs.harvard.edu/abs/2024ApJ...967...12D}
}

@ARTICLE{2008ApJS..178..137S,
       author = {{Stone}, James M. and {Gardiner}, Thomas A. and {Teuben}, Peter and {Hawley}, John F. and {Simon}, Jacob B.},
        title = "{Athena: A New Code for Astrophysical MHD}",
      journal = {\apjs},
         year = 2008,
        month = sep,
       volume = {178},
       number = {1},
        pages = {137-177},
          doi = {10.1086/588755},
archivePrefix = {arXiv},
       eprint = {0804.0402},
 primaryClass = {astro-ph},
       adsurl = {https://ui.adsabs.harvard.edu/abs/2008ApJS..178..137S}
}

@ARTICLE{1974JCoPh..14..361V,
       author = {{van Leer}, Bram},
        title = "{Towards the Ultimate Conservation Difference Scheme. II. Monotonicity and Conservation Combined in a Second-Order Scheme}",
      journal = {Journal of Computational Physics},
         year = 1974,
        month = mar,
       volume = {14},
       number = {4},
        pages = {361-370},
          doi = {10.1016/0021-9991(74)90019-9},
       adsurl = {https://ui.adsabs.harvard.edu/abs/1974JCoPh..14..361V}
}

@ARTICLE{2024MNRAS.534.1394C,
       author = {{Cordwell}, Amelia J. and {Rafikov}, Roman R.},
        title = "{Early stages of gap opening by planets in protoplanetary discs}",
      journal = {\mnras},
         year = 2024,
        month = oct,
       volume = {534},
       number = {2},
        pages = {1394-1413},
          doi = {10.1093/mnras/stae2089},
archivePrefix = {arXiv},
       eprint = {2407.01728},
 primaryClass = {astro-ph.EP},
       adsurl = {https://ui.adsabs.harvard.edu/abs/2024MNRAS.534.1394C}
}

@ARTICLE{2020ApJS..249....4S,
       author = {{Stone}, James M. and {Tomida}, Kengo and {White}, Christopher J. and {Felker}, Kyle G.},
        title = "{The Athena++ Adaptive Mesh Refinement Framework: Design and Magnetohydrodynamic Solvers}",
      journal = {\apjs},
         year = 2020,
        month = jul,
       volume = {249},
       number = {1},
          eid = {4},
        pages = {4},
          doi = {10.3847/1538-4365/ab929b},
archivePrefix = {arXiv},
       eprint = {2005.06651},
 primaryClass = {astro-ph.IM},
       adsurl = {https://ui.adsabs.harvard.edu/abs/2020ApJS..249....4S}
}

@article{Artymowicz_1993,
	title = {On the {Wave} {Excitation} and a {Generalized} {Torque} {Formula} for {Lindblad} {Resonances} {Excited} by {External} {Potential}},
	volume = {419},
	issn = {0004-637X},
	url = {https://ui.adsabs.harvard.edu/abs/1993ApJ...419..155A},
	doi = {10.1086/173469},
	urldate = {2022-12-20},
	journal = {\apj},
	author = {Artymowicz, Pawel},
	month = dec,
	year = {1993},
	note = {ADS Bibcode: 1993ApJ...419..155A},
	pages = {155},
}

@BOOK{2002apa..book.....F,
       author = {{Frank}, Juhan and {King}, Andrew and {Raine}, Derek J.},
        title = "{Accretion Power in Astrophysics: Third Edition}",
         year = 2002,
       adsurl = {https://ui.adsabs.harvard.edu/abs/2002apa..book.....F},
        publisher = "Cambridge University Press"
}

@ARTICLE{deValBorro2006,
       author = {{de Val-Borro}, M. and {Edgar}, R.~G. and {Artymowicz}, P. and {Ciecielag}, P. and {Cresswell}, P. and {D'Angelo}, G. and {Delgado-Donate}, E.~J. and {Dirksen}, G. and {Fromang}, S. and {Gawryszczak}, A. and {Klahr}, H. and {Kley}, W. and {Lyra}, W. and {Masset}, F. and {Mellema}, G. and {Nelson}, R.~P. and {Paardekooper}, S. -J. and {Peplinski}, A. and {Pierens}, A. and {Plewa}, T. and {Rice}, K. and {Sch{\"a}fer}, C. and {Speith}, R.},
        title = "{A comparative study of disc-planet interaction}",
      journal = {\mnras},
         year = 2006,
        month = aug,
       volume = {370},
       number = {2},
        pages = {529-558},
          doi = {10.1111/j.1365-2966.2006.10488.x},
archivePrefix = {arXiv},
       eprint = {astro-ph/0605237},
 primaryClass = {astro-ph},
       adsurl = {https://ui.adsabs.harvard.edu/abs/2006MNRAS.370..529D}
}

@ARTICLE{2024arXiv240916053S,
       author = {{Stone}, James M. and {Mullen}, Patrick D. and {Fielding}, Drummond and {Grete}, Philipp and {Guo}, Minghao and {Kempski}, Philipp and {Most}, Elias R. and {White}, Christopher J. and {Wong}, George N.},
        title = "{AthenaK: A Performance-Portable Version of the Athena++ AMR Framework}",
      journal = {arXiv e-prints},
         year = 2024,
        month = sep,
          eid = {arXiv:2409.16053},
        pages = {arXiv:2409.16053},
          doi = {10.48550/arXiv.2409.16053},
archivePrefix = {arXiv},
       eprint = {2409.16053},
 primaryClass = {astro-ph.IM},
       adsurl = {https://ui.adsabs.harvard.edu/abs/2024arXiv240916053S}
}

@article{FairbairnRafikov_2025,
  title = {Eccentric Planet--Disc Interactions: Orbital Migration and Eccentricity Evolution},
  shorttitle = {Eccentric Planet--Disc Interactions},
  author = {Fairbairn, Callum W and Rafikov, Roman R},
  year = {2025},
  month = feb,
  journal = {\mnras},
  volume = {537},
  number = {2},
  pages = {1779--1806},
  issn = {0035-8711},
  doi = {10.1093/mnras/staf117},
  urldate = {2025-04-04},
}

@ARTICLE{2025MNRAS.543..565F,
       author = {{Fairbairn}, Callum W. and {Dittmann}, Alexander J.},
        title = "{Pushing the limits of eccentricity in planet{\textendash}disc interactions}",
      journal = {\mnras},
         year = 2025,
        month = oct,
       volume = {543},
       number = {1},
        pages = {565-586},
          doi = {10.1093/mnras/staf1399},
archivePrefix = {arXiv},
       eprint = {2506.19917},
 primaryClass = {astro-ph.EP},
       adsurl = {https://ui.adsabs.harvard.edu/abs/2025MNRAS.543..565F}
}

@ARTICLE{Crida2009,
       author = {{Crida}, A. and {Baruteau}, C. and {Kley}, W. and {Masset}, F.},
        title = "{The dynamical role of the circumplanetary disc in planetary migration}",
      journal = {\aap},
         year = 2009,
        month = aug,
       volume = {502},
       number = {2},
        pages = {679-693},
          doi = {10.1051/0004-6361/200811608},
archivePrefix = {arXiv},
       eprint = {0906.0888},
 primaryClass = {astro-ph.EP},
       adsurl = {https://ui.adsabs.harvard.edu/abs/2009A&A...502..679C}
}

@ARTICLE{Price2018,
       author = {{Price}, Daniel J. and {Wurster}, James and {Tricco}, Terrence S. and {Nixon}, Chris and {Toupin}, St{\'e}ven and {Pettitt}, Alex and {Chan}, Conrad and {Mentiplay}, Daniel and {Laibe}, Guillaume and {Glover}, Simon and {Dobbs}, Clare and {Nealon}, Rebecca and {Liptai}, David and {Worpel}, Hauke and {Bonnerot}, Cl{\'e}ment and {Dipierro}, Giovanni and {Ballabio}, Giulia and {Ragusa}, Enrico and {Federrath}, Christoph and {Iaconi}, Roberto and {Reichardt}, Thomas and {Forgan}, Duncan and {Hutchison}, Mark and {Constantino}, Thomas and {Ayliffe}, Ben and {Hirsh}, Kieran and {Lodato}, Giuseppe},
        title = "{Phantom: A Smoothed Particle Hydrodynamics and Magnetohydrodynamics Code for Astrophysics}",
      journal = {\pasa},
         year = 2018,
        month = sep,
       volume = {35},
          eid = {e031},
        pages = {e031},
          doi = {10.1017/pasa.2018.25},
archivePrefix = {arXiv},
       eprint = {1702.03930},
 primaryClass = {astro-ph.IM},
       adsurl = {https://ui.adsabs.harvard.edu/abs/2018PASA...35...31P}
}

@ARTICLE{Duffell2013,
       author = {{Duffell}, Paul C. and {MacFadyen}, Andrew I.},
        title = "{Gap Opening by Extremely Low-mass Planets in a Viscous Disk}",
      journal = {\apj},
         year = 2013,
        month = may,
       volume = {769},
       number = {1},
          eid = {41},
        pages = {41},
          doi = {10.1088/0004-637X/769/1/41},
archivePrefix = {arXiv},
       eprint = {1302.1934},
 primaryClass = {astro-ph.EP},
       adsurl = {https://ui.adsabs.harvard.edu/abs/2013ApJ...769...41D}
}

@ARTICLE{Franchini2022,
       author = {{Franchini}, Alessia and {Lupi}, Alessandro and {Sesana}, Alberto},
        title = "{Resolving Massive Black Hole Binary Evolution via Adaptive Particle Splitting}",
      journal = {\apjl},
         year = 2022,
        month = apr,
       volume = {929},
       number = {1},
          eid = {L13},
        pages = {L13},
          doi = {10.3847/2041-8213/ac63a2},
archivePrefix = {arXiv},
       eprint = {2201.05619},
 primaryClass = {astro-ph.HE},
       adsurl = {https://ui.adsabs.harvard.edu/abs/2022ApJ...929L..13F}
}

@ARTICLE{Hopkins2015,
       author = {{Hopkins}, Philip F.},
        title = "{A new class of accurate, mesh-free hydrodynamic simulation methods}",
      journal = {\mnras},
         year = 2015,
        month = jun,
       volume = {450},
       number = {1},
        pages = {53-110},
          doi = {10.1093/mnras/stv195},
archivePrefix = {arXiv},
       eprint = {1409.7395},
 primaryClass = {astro-ph.CO},
       adsurl = {https://ui.adsabs.harvard.edu/abs/2015MNRAS.450...53H}
}

@ARTICLE{Bate1995,
       author = {{Bate}, Matthew R. and {Bonnell}, Ian A. and {Price}, Nigel M.},
        title = "{Modelling accretion in protobinary systems}",
      journal = {\mnras},
         year = 1995,
        month = nov,
       volume = {277},
       number = {2},
        pages = {362-376},
          doi = {10.1093/mnras/277.2.362},
archivePrefix = {arXiv},
       eprint = {astro-ph/9510149},
 primaryClass = {astro-ph},
       adsurl = {https://ui.adsabs.harvard.edu/abs/1995MNRAS.277..362B}
}

@ARTICLE{Derdzinski2023,
       author = {{Derdzinski}, Andrea and {Mayer}, Lucio},
        title = "{In situ extreme mass ratio inspirals via subparsec formation and migration of stars in thin, gravitationally unstable AGN discs}",
      journal = {\mnras},
         year = 2023,
        month = may,
       volume = {521},
       number = {3},
        pages = {4522-4543},
          doi = {10.1093/mnras/stad749},
archivePrefix = {arXiv},
       eprint = {2205.10382},
 primaryClass = {astro-ph.GA},
       adsurl = {https://ui.adsabs.harvard.edu/abs/2023MNRAS.521.4522D}
}

@ARTICLE{Yunes2011,
       author = {{Yunes}, Nicol{\'a}s and {Kocsis}, Bence and {Loeb}, Abraham and {Haiman}, Zolt{\'a}n},
        title = "{Imprint of Accretion Disk-Induced Migration on Gravitational Waves from Extreme Mass Ratio Inspirals}",
      journal = {\prl},
         year = 2011,
        month = oct,
       volume = {107},
       number = {17},
          eid = {171103},
        pages = {171103},
          doi = {10.1103/PhysRevLett.107.171103},
archivePrefix = {arXiv},
       eprint = {1103.4609},
 primaryClass = {astro-ph.CO},
       adsurl = {https://ui.adsabs.harvard.edu/abs/2011PhRvL.107q1103Y}
}

@ARTICLE{Garg2024b,
       author = {{Garg}, Mudit and {Derdzinski}, Andrea and {Tiwari}, Shubhanshu and {Gair}, Jonathan and {Mayer}, Lucio},
        title = "{Measuring eccentricity and gas-induced perturbation from gravitational waves of LISA massive black hole binaries}",
      journal = {\mnras},
         year = 2024,
        month = aug,
       volume = {532},
       number = {4},
        pages = {4060-4074},
          doi = {10.1093/mnras/stae1764},
archivePrefix = {arXiv},
       eprint = {2402.14058},
 primaryClass = {astro-ph.GA},
       adsurl = {https://ui.adsabs.harvard.edu/abs/2024MNRAS.532.4060G}
}

@ARTICLE{Garg2025,
       author = {{Garg}, Mudit and {Franchini}, Alessia and {Lupi}, Alessandro and {Bonetti}, Matteo and {Mayer}, Lucio},
        title = "{Gas-induced Perturbations on the Gravitational Wave Inspiral of Live Post-Newtonian LISA Massive Black Hole Binaries}",
      journal = {\apj},
         year = 2025,
        month = nov,
       volume = {993},
       number = {1},
          eid = {145},
        pages = {145},
          doi = {10.3847/1538-4357/ae10b4},
archivePrefix = {arXiv},
       eprint = {2410.17305},
 primaryClass = {astro-ph.HE},
       adsurl = {https://ui.adsabs.harvard.edu/abs/2025ApJ...993..145G}
}

@ARTICLE{Garg2022,
       author = {{Garg}, Mudit and {Derdzinski}, Andrea and {Zwick}, Lorenz and {Capelo}, Pedro R. and {Mayer}, Lucio},
        title = "{The imprint of gas on gravitational waves from LISA intermediate-mass black hole binaries}",
      journal = {\mnras},
         year = 2022,
        month = nov,
       volume = {517},
       number = {1},
        pages = {1339-1354},
          doi = {10.1093/mnras/stac2711},
archivePrefix = {arXiv},
       eprint = {2206.05292},
 primaryClass = {astro-ph.GA},
       adsurl = {https://ui.adsabs.harvard.edu/abs/2022MNRAS.517.1339G}
}

@ARTICLE{Garg2024d,
       author = {{Garg}, Mudit and {Sberna}, Laura and {Speri}, Lorenzo and {Duque}, Francisco and {Gair}, Jonathan},
        title = "{Systematics in tests of general relativity using LISA massive black hole binaries}",
      journal = {\mnras},
         year = 2024,
        month = dec,
       volume = {535},
       number = {4},
        pages = {3283-3292},
          doi = {10.1093/mnras/stae2605},
archivePrefix = {arXiv},
       eprint = {2410.02910},
 primaryClass = {astro-ph.GA},
       adsurl = {https://ui.adsabs.harvard.edu/abs/2024MNRAS.535.3283G}
}

@ARTICLE{Pan2021a,
       author = {{Pan}, Zhen and {Yang}, Huan},
        title = "{Formation rate of extreme mass ratio inspirals in active galactic nuclei}",
      journal = {\prd},
         year = 2021,
        month = may,
       volume = {103},
       number = {10},
          eid = {103018},
        pages = {103018},
          doi = {10.1103/PhysRevD.103.103018},
archivePrefix = {arXiv},
       eprint = {2101.09146},
 primaryClass = {astro-ph.HE},
       adsurl = {https://ui.adsabs.harvard.edu/abs/2021PhRvD.103j3018P}
}

@ARTICLE{1994ShWav...4...25T,
       author = {{Toro}, E.~F. and {Spruce}, M. and {Speares}, W.},
        title = "{Restoration of the contact surface in the HLL-Riemann solver}",
      journal = {Shock Waves},
         year = 1994,
        month = jul,
       volume = {4},
       number = {1},
        pages = {25-34},
          doi = {10.1007/BF01414629},
       adsurl = {https://ui.adsabs.harvard.edu/abs/1994ShWav...4...25T}
}

@ARTICLE{1998MaCom..67...73G,
       author = {{Gottlieb}, S. and {Shu}, C.~W.},
        title = "{Total variation diminishing Runge-Kutta schemes}",
      journal = {Mathematics of Computation},
         year = 1998,
        month = jan,
       volume = {67},
       number = {221},
        pages = {73-85},
       adsurl = {https://ui.adsabs.harvard.edu/abs/1998MaCom..67...73G}
}

@article{Teyssier2002,
  title = {Cosmological Hydrodynamics with Adaptive Mesh Refinement. {{A}} New High Resolution Code Called {{RAMSES}}},
  author = {Teyssier, R.},
  year = {2002},
  month = apr,
  journal = {A\&A},
  volume = {385},
  eprint = {arXiv:astro-ph/0111367},
  pages = {337--364},
  doi = {10/fcr8x4},
  adsurl = {http://adsabs.harvard.edu/abs/2002A\%26A...385..337T},
}

@ARTICLE{Wadsley_et_al_2004,
       author = {{Wadsley}, J.~W. and {Stadel}, J. and {Quinn}, T.},
        title = "{Gasoline: a flexible, parallel implementation of TreeSPH}",
      journal = {\na},
         year = 2004,
        month = feb,
       volume = {9},
       number = {2},
        pages = {137-158},
          doi = {10.1016/j.newast.2003.08.004},
archivePrefix = {arXiv},
       eprint = {astro-ph/0303521},
 primaryClass = {astro-ph},
       adsurl = {https://ui.adsabs.harvard.edu/abs/2004NewA....9..137W}
}

@ARTICLE{Wadsley_et_al_2017,
       author = {{Wadsley}, James W. and {Keller}, Benjamin W. and {Quinn}, Thomas R.},
        title = "{Gasoline2: a modern smoothed particle hydrodynamics code}",
      journal = {\mnras},
         year = 2017,
        month = oct,
       volume = {471},
       number = {2},
        pages = {2357-2369},
          doi = {10.1093/mnras/stx1643},
archivePrefix = {arXiv},
       eprint = {1707.03824},
 primaryClass = {astro-ph.IM},
       adsurl = {https://ui.adsabs.harvard.edu/abs/2017MNRAS.471.2357W}
}

@ARTICLE{Wissing_Shen_2020,
       author = {{Wissing}, Robert and {Shen}, Sijing},
        title = "{Smoothed particle magnetohydrodynamics with the geometric density average force expression}",
      journal = {\aap},
         year = 2020,
        month = jun,
       volume = {638},
          eid = {A140},
        pages = {A140},
          doi = {10.1051/0004-6361/201936739},
archivePrefix = {arXiv},
       eprint = {1909.09650},
 primaryClass = {astro-ph.IM},
       adsurl = {https://ui.adsabs.harvard.edu/abs/2020A&A...638A.140W}
}

@PHDTHESIS{Stadel_2001,
       author = {{Stadel}, Joachim Gerhard},
        title = "{Cosmological N-body simulations and their analysis}",
       school = {University of Washington, Seattle},
         year = 2001,
        month = jan,
       adsurl = {https://ui.adsabs.harvard.edu/abs/2001PhDT........21S}
}

@ARTICLE{Monaghan_1992,
       author = {{Monaghan}, J.~J.},
        title = "{Smoothed particle hydrodynamics.}",
      journal = {\araa},
         year = 1992,
        month = jan,
       volume = {30},
        pages = {543-574},
          doi = {10.1146/annurev.aa.30.090192.002551},
       adsurl = {https://ui.adsabs.harvard.edu/abs/1992ARA&A..30..543M}
}

@ARTICLE{Ritchie_Thomas_2001,
       author = {{Ritchie}, Benedict W. and {Thomas}, Peter A.},
        title = "{Multiphase smoothed-particle hydrodynamics}",
      journal = {\mnras},
         year = 2001,
        month = may,
       volume = {323},
       number = {3},
        pages = {743-756},
          doi = {10.1046/j.1365-8711.2001.04268.x},
archivePrefix = {arXiv},
       eprint = {astro-ph/0005357},
 primaryClass = {astro-ph},
       adsurl = {https://ui.adsabs.harvard.edu/abs/2001MNRAS.323..743R}
}

@ARTICLE{Keller_et_al_2014,
       author = {{Keller}, B.~W. and {Wadsley}, J. and {Benincasa}, S.~M. and {Couchman}, H.~M.~P.},
        title = "{A superbubble feedback model for galaxy simulations}",
      journal = {\mnras},
         year = 2014,
        month = aug,
       volume = {442},
       number = {4},
        pages = {3013-3025},
          doi = {10.1093/mnras/stu1058},
archivePrefix = {arXiv},
       eprint = {1405.2625},
 primaryClass = {astro-ph.GA},
       adsurl = {https://ui.adsabs.harvard.edu/abs/2014MNRAS.442.3013K}
}

@ARTICLE{Wendland_1995,
   author = {{Wendland}, H.},
    title = "{Piecewise polynomial, positive definite and compactly supported radial functions of minimal degree}",
  journal = {Adv. in Comput. Math.},
     year = 1995,
    month = dec,
   volume = 4,
    pages = {389-396},
      doi = {10.1007/BF02123482}
}

@ARTICLE{Dehnen_Aly_2012,
       author = {{Dehnen}, Walter and {Aly}, Hossam},
        title = "{Improving convergence in smoothed particle hydrodynamics simulations without pairing instability}",
      journal = {\mnras},
         year = 2012,
        month = sep,
       volume = {425},
       number = {2},
        pages = {1068-1082},
          doi = {10.1111/j.1365-2966.2012.21439.x},
archivePrefix = {arXiv},
       eprint = {1204.2471},
 primaryClass = {astro-ph.IM},
       adsurl = {https://ui.adsabs.harvard.edu/abs/2012MNRAS.425.1068D}
}

@ARTICLE{Morris_Monaghan_1997,
       author = {{Morris}, J.~P. and {Monaghan}, J.~J.},
        title = "{A Switch to Reduce SPH Viscosity}",
      journal = {Journal of Computational Physics},
         year = 1997,
        month = sep,
       volume = {136},
       number = {1},
        pages = {41-50},
          doi = {10.1006/jcph.1997.5690},
       adsurl = {https://ui.adsabs.harvard.edu/abs/1997JCoPh.136...41M}
}

@ARTICLE{Cullen_Dehnen_2010,
       author = {{Cullen}, Lee and {Dehnen}, Walter},
        title = "{Inviscid smoothed particle hydrodynamics}",
      journal = {\mnras},
         year = 2010,
        month = oct,
       volume = {408},
       number = {2},
        pages = {669-683},
          doi = {10.1111/j.1365-2966.2010.17158.x},
archivePrefix = {arXiv},
       eprint = {1006.1524},
 primaryClass = {astro-ph.IM},
       adsurl = {https://ui.adsabs.harvard.edu/abs/2010MNRAS.408..669C}
}

@ARTICLE{2000A&AS..141..165M,
       author = {{Masset}, F.},
        title = "{FARGO: A fast eulerian transport algorithm for differentially rotating disks}",
      journal = {\aaps},
         year = 2000,
        month = jan,
       volume = {141},
        pages = {165-173},
          doi = {10.1051/aas:2000116},
archivePrefix = {arXiv},
       eprint = {astro-ph/9910390},
 primaryClass = {astro-ph},
       adsurl = {https://ui.adsabs.harvard.edu/abs/2000A&AS..141..165M}
}

@ARTICLE{2016ApJS..223...11B,
       author = {{Ben{\'\i}tez-Llambay}, Pablo and {Masset}, Fr{\'e}d{\'e}ric S.},
        title = "{FARGO3D: A New GPU-oriented MHD Code}",
      journal = {\apjs},
         year = 2016,
        month = mar,
       volume = {223},
       number = {1},
          eid = {11},
        pages = {11},
          doi = {10.3847/0067-0049/223/1/11},
archivePrefix = {arXiv},
       eprint = {1602.02359},
 primaryClass = {astro-ph.IM},
       adsurl = {https://ui.adsabs.harvard.edu/abs/2016ApJS..223...11B}
}

@ARTICLE{2015ApJ...800....6Z,
       author = {{Zhu}, Qirong and {Hernquist}, Lars and {Li}, Yuexing},
        title = "{Numerical Convergence In Smoothed Particle Hydrodynamics}",
      journal = {\apj},
         year = 2015,
        month = feb,
       volume = {800},
       number = {1},
          eid = {6},
        pages = {6},
          doi = {10.1088/0004-637X/800/1/6},
archivePrefix = {arXiv},
       eprint = {1410.4222},
 primaryClass = {astro-ph.CO},
       adsurl = {https://ui.adsabs.harvard.edu/abs/2015ApJ...800....6Z}
}

@ARTICLE{2022MNRAS.511.6143Z,
       author = {{Zwick}, Lorenz and {Derdzinski}, Andrea and {Garg}, Mudit and {Capelo}, Pedro R. and {Mayer}, Lucio},
        title = "{Dirty waveforms: multiband harmonic content of gas-embedded gravitational wave sources}",
      journal = {\mnras},
         year = 2022,
        month = apr,
       volume = {511},
       number = {4},
        pages = {6143-6159},
          doi = {10.1093/mnras/stac299},
archivePrefix = {arXiv},
       eprint = {2110.09097},
 primaryClass = {astro-ph.HE},
       adsurl = {https://ui.adsabs.harvard.edu/abs/2022MNRAS.511.6143Z}
}

@ARTICLE{derdzinski_evolution_2019,
       author = {{Derdzinski}, A.~M. and {D'Orazio}, D. and {Duffell}, P. and {Haiman}, Z. and {MacFadyen}, A.},
        title = "{Probing gas disc physics with LISA: simulations of an intermediate mass ratio inspiral in an accretion disc}",
      journal = {\mnras},
         year = 2019,
        month = jun,
       volume = {486},
       number = {2},
        pages = {2754-2765},
          doi = {10.1093/mnras/stz1026},
archivePrefix = {arXiv},
       eprint = {1810.03623},
 primaryClass = {astro-ph.HE},
       adsurl = {https://ui.adsabs.harvard.edu/abs/2019MNRAS.486.2754D}
}

@article{derdzinski_evolution_2021,
	title = {Evolution of gas disc–embedded intermediate mass ratio inspirals in the \textit{{LISA}} band},
	volume = {501},
	issn = {0035-8711, 1365-2966},
	url = {https://academic.oup.com/mnras/article/501/3/3540/6054833},
	doi = {10.1093/mnras/staa3976},
	language = {en},
	number = {3},
	urldate = {2022-07-28},
	journal = {Monthly Notices of the Royal Astronomical Society},
	author = {Derdzinski, A and D’Orazio, D and Duffell, P and Haiman, Z and MacFadyen, A},
	month = jan,
	year = {2021},
	pages = {3540--3557},
}

@ARTICLE{2024MNRAS.529..425C,
       author = {{Cimerman}, Nicolas P. and {Rafikov}, Roman R. and {Miranda}, Ryan},
        title = "{Torque wiggles - a robust feature of the global disc-planet interaction}",
      journal = {\mnras},
         year = 2024,
        month = mar,
       volume = {529},
       number = {1},
        pages = {425-443},
          doi = {10.1093/mnras/stae467},
archivePrefix = {arXiv},
       eprint = {2306.07341},
 primaryClass = {astro-ph.EP},
       adsurl = {https://ui.adsabs.harvard.edu/abs/2024MNRAS.529..425C}
}

@ARTICLE{2006ApJ...652..730M,
       author = {{Masset}, F.~S. and {D'Angelo}, G. and {Kley}, W.},
        title = "{On the Migration of Protogiant Solid Cores}",
      journal = {\apj},
         year = 2006,
        month = nov,
       volume = {652},
       number = {1},
        pages = {730-745},
          doi = {10.1086/507515},
archivePrefix = {arXiv},
       eprint = {astro-ph/0607155},
 primaryClass = {astro-ph},
       adsurl = {https://ui.adsabs.harvard.edu/abs/2006ApJ...652..730M}
}

@article{MirandaRafikov_2020,
	title = {Planet–{Disk} {Interaction} in {Disks} with {Cooling}: {Basic} {Theory}},
	volume = {892},
	issn = {1538-4357},
	shorttitle = {Planet–{Disk} {Interaction} in {Disks} with {Cooling}},
	url = {https://iopscience.iop.org/article/10.3847/1538-4357/ab791a},
	doi = {10.3847/1538-4357/ab791a},
	language = {en},
	number = {1},
	urldate = {2022-05-27},
	journal = {\apj},
	author = {Miranda, Ryan and Rafikov, Roman R.},
	month = mar,
	year = {2020},
	pages = {65},
}

@ARTICLE{2025MNRAS.539....1D,
       author = {{David-Cl{\'e}ris}, T. and {Laibe}, G. and {Lapeyre}, Y.},
        title = "{The SHAMROCK code: I - smoothed particle hydrodynamics on GPUs}",
      journal = {\mnras},
         year = 2025,
        month = may,
       volume = {539},
       number = {1},
        pages = {1-33},
          doi = {10.1093/mnras/staf444},
archivePrefix = {arXiv},
       eprint = {2503.09713},
 primaryClass = {astro-ph.IM},
       adsurl = {https://ui.adsabs.harvard.edu/abs/2025MNRAS.539....1D}
}

@ARTICLE{Goldreich2003,
       author = {{Goldreich}, Peter and {Sari}, Re'em},
        title = "{Eccentricity Evolution for Planets in Gaseous Disks}",
      journal = {\apj},
         year = 2003,
        month = mar,
       volume = {585},
       number = {2},
        pages = {1024-1037},
          doi = {10.1086/346202},
archivePrefix = {arXiv},
       eprint = {astro-ph/0202462},
 primaryClass = {astro-ph},
       adsurl = {https://ui.adsabs.harvard.edu/abs/2003ApJ...585.1024G}
}

@ARTICLE{2002MNRAS.330..950O,
       author = {{Ogilvie}, G.~I. and {Lubow}, S.~H.},
        title = "{On the wake generated by a planet in a disc}",
      journal = {\mnras},
         year = 2002,
        month = mar,
       volume = {330},
       number = {4},
        pages = {950-954},
          doi = {10.1046/j.1365-8711.2002.05148.x},
archivePrefix = {arXiv},
       eprint = {astro-ph/0111265},
 primaryClass = {astro-ph},
       adsurl = {https://ui.adsabs.harvard.edu/abs/2002MNRAS.330..950O}
}

@ARTICLE{2011MNRAS.410..293P,
       author = {{Paardekooper}, S. -J. and {Baruteau}, C. and {Kley}, W.},
        title = "{A torque formula for non-isothermal Type I planetary migration - II. Effects of diffusion}",
      journal = {\mnras},
         year = 2011,
        month = jan,
       volume = {410},
       number = {1},
        pages = {293-303},
          doi = {10.1111/j.1365-2966.2010.17442.x},
archivePrefix = {arXiv},
       eprint = {1007.4964},
 primaryClass = {astro-ph.EP},
       adsurl = {https://ui.adsabs.harvard.edu/abs/2011MNRAS.410..293P}
}

@ARTICLE{2009MNRAS.394.2283P,
       author = {{Paardekooper}, S. -J. and {Papaloizou}, J.~C.~B.},
        title = "{On corotation torques, horseshoe drag and the possibility of sustained stalled or outward protoplanetary migration}",
      journal = {\mnras},
         year = 2009,
        month = apr,
       volume = {394},
       number = {4},
        pages = {2283-2296},
          doi = {10.1111/j.1365-2966.2009.14511.x},
archivePrefix = {arXiv},
       eprint = {0901.2265},
 primaryClass = {astro-ph.EP},
       adsurl = {https://ui.adsabs.harvard.edu/abs/2009MNRAS.394.2283P}
}

@ARTICLE{2010ApJ...723.1393M,
       author = {{Masset}, F.~S. and {Casoli}, J.},
        title = "{Saturated Torque Formula for Planetary Migration in Viscous Disks with Thermal Diffusion: Recipe for Protoplanet Population Synthesis}",
      journal = {\apj},
         year = 2010,
        month = nov,
       volume = {723},
       number = {2},
        pages = {1393-1417},
          doi = {10.1088/0004-637X/723/2/1393},
archivePrefix = {arXiv},
       eprint = {1009.1913},
 primaryClass = {astro-ph.EP},
       adsurl = {https://ui.adsabs.harvard.edu/abs/2010ApJ...723.1393M}
}

@ARTICLE{2010MNRAS.401.1950P,
       author = {{Paardekooper}, S. -J. and {Baruteau}, C. and {Crida}, A. and {Kley}, W.},
        title = "{A torque formula for non-isothermal type I planetary migration - I. Unsaturated horseshoe drag}",
      journal = {\mnras},
         year = 2010,
        month = jan,
       volume = {401},
       number = {3},
        pages = {1950-1964},
          doi = {10.1111/j.1365-2966.2009.15782.x},
archivePrefix = {arXiv},
       eprint = {0909.4552},
 primaryClass = {astro-ph.EP},
       adsurl = {https://ui.adsabs.harvard.edu/abs/2010MNRAS.401.1950P}
}

@ARTICLE{Hopkins2016,
       author = {{Hopkins}, Philip F.},
        title = "{A constrained-gradient method to control divergence errors in numerical MHD}",
      journal = {\mnras},
         year = 2016,
        month = oct,
       volume = {462},
       number = {1},
        pages = {576-587},
          doi = {10.1093/mnras/stw1578},
archivePrefix = {arXiv},
       eprint = {1509.07877},
 primaryClass = {astro-ph.IM},
       adsurl = {https://ui.adsabs.harvard.edu/abs/2016MNRAS.462..576H}
}

@ARTICLE{2000JCoPh.160..241K,
       author = {{Kurganov}, Alexander and {Tadmor}, Eitan},
        title = "{New High-Resolution Central Schemes for Nonlinear Conservation Laws and Convection-Diffusion Equations}",
      journal = {Journal of Computational Physics},
         year = 2000,
        month = may,
       volume = {160},
       number = {1},
        pages = {241-282},
          doi = {10.1006/jcph.2000.6459},
       adsurl = {https://ui.adsabs.harvard.edu/abs/2000JCoPh.160..241K}
}

@ARTICLE{mckernan14,
       author = {{McKernan}, B. and {Ford}, K.~E.~S. and {Kocsis}, B. and {Lyra}, W. and {Winter}, L.~M.},
        title = "{Intermediate-mass black holes in AGN discs - II. Model predictions and observational constraints}",
      journal = {\mnras},
         year = 2014,
        month = jun,
       volume = {441},
       number = {1},
        pages = {900-909},
          doi = {10.1093/mnras/stu553},
archivePrefix = {arXiv},
       eprint = {1403.6433},
 primaryClass = {astro-ph.GA},
       adsurl = {https://ui.adsabs.harvard.edu/abs/2014MNRAS.441..900M}
}

@ARTICLE{shakura73,
       author = {{Shakura}, N.~I. and {Sunyaev}, R.~A.},
        title = "{Black holes in binary systems. Observational appearance.}",
      journal = {\aap},
         year = 1973,
        month = jan,
       volume = {24},
        pages = {337-355},
       adsurl = {https://ui.adsabs.harvard.edu/abs/1973A&A....24..337S}
}

@ARTICLE{gultekin12,
       author = {{G{\"u}ltekin}, Kayhan and {Miller}, Jon M.},
        title = "{Observable Consequences of Merger-driven Gaps and Holes in Black Hole Accretion Disks}",
      journal = {\apj},
         year = 2012,
        month = dec,
       volume = {761},
       number = {2},
          eid = {90},
        pages = {90},
          doi = {10.1088/0004-637X/761/2/90},
archivePrefix = {arXiv},
       eprint = {1207.0296},
 primaryClass = {astro-ph.HE},
       adsurl = {https://ui.adsabs.harvard.edu/abs/2012ApJ...761...90G}
}

@article{hill1878,
  title={Researches in the lunar theory},
  author={Hill, George William},
  journal={American journal of Mathematics},
  volume={1},
  number={1},
  pages={5--26},
  year={1878},
  publisher={JSTOR}
}

@article{LannelongueGreen,
author = {Lannelongue, Loïc and Grealey, Jason and Inouye, Michael},
title = {Green Algorithms: Quantifying the Carbon Footprint of Computation},
journal = {Advanced Science},
volume = {8},
number = {12},
pages = {2100707},
doi = {https://doi.org/10.1002/advs.202100707},
url = {https://advanced.onlinelibrary.wiley.com/doi/abs/10.1002/advs.202100707},
eprint = {https://advanced.onlinelibrary.wiley.com/doi/pdf/10.1002/advs.202100707},
year = {2021}
}

@ARTICLE{2023A&A...677A...9L,
       author = {{Lesur}, G.~R.~J. and {Baghdadi}, S. and {Wafflard-Fernandez}, G. and {Mauxion}, J. and {Robert}, C.~M.~T. and {Van den Bossche}, M.},
        title = "{IDEFIX: A versatile performance-portable Godunov code for astrophysical flows}",
      journal = {\aap},
         year = 2023,
        month = sep,
       volume = {677},
          eid = {A9},
        pages = {A9},
          doi = {10.1051/0004-6361/202346005},
archivePrefix = {arXiv},
       eprint = {2304.13746},
 primaryClass = {astro-ph.IM},
       adsurl = {https://ui.adsabs.harvard.edu/abs/2023A&A...677A...9L}
}

@ARTICLE{2025FrP....1342474S,
       author = {{Suarez}, Estela and {Amaya}, Jorge and {Frank}, Martin and {Freyermuth}, Oliver and {Girone}, Maria and {Kostrzewa}, Bartosz and {Pfalzner}, Susanne},
        title = "{Energy efficiency trends in HPC: what high-energy and astrophysicists need to know}",
      journal = {Frontiers in Physics},
         year = 2025,
        month = apr,
       volume = {13},
          eid = {1542474},
        pages = {1542474},
          doi = {10.3389/fphy.2025.1542474},
archivePrefix = {arXiv},
       eprint = {2503.17283},
 primaryClass = {cs.DC},
       adsurl = {https://ui.adsabs.harvard.edu/abs/2025FrP....1342474S}
}

@ARTICLE{2015ApJ...806..182D,
       author = {{Duffell}, Paul C.},
        title = "{Halting Migration: Numerical Calculations of Corotation Torques in the Weakly Nonlinear Regime}",
      journal = {\apj},
         year = 2015,
        month = jun,
       volume = {806},
       number = {2},
          eid = {182},
        pages = {182},
          doi = {10.1088/0004-637X/806/2/182},
archivePrefix = {arXiv},
       eprint = {1412.8092},
 primaryClass = {astro-ph.EP},
       adsurl = {https://ui.adsabs.harvard.edu/abs/2015ApJ...806..182D}
}

@ARTICLE{Duffell2024,
       author = {{Duffell}, Paul C. and {Dittmann}, Alexander J. and {D'Orazio}, Daniel J. and {Franchini}, Alessia and {Kratter}, Kaitlin M. and {Penzlin}, Anna B.~T. and {Ragusa}, Enrico and {Siwek}, Magdalena and {Tiede}, Christopher and {Wang}, Haiyang and {Zrake}, Jonathan and {Dempsey}, Adam M. and {Haiman}, Zoltan and {Lupi}, Alessandro and {Pirog}, Michal and {Ryan}, Geoffrey},
        title = "{The Santa Barbara Binary‑disk Code Comparison}",
      journal = {\apj},
         year = 2024,
        month = aug,
       volume = {970},
       number = {2},
          eid = {156},
        pages = {156},
          doi = {10.3847/1538-4357/ad5a7e},
archivePrefix = {arXiv},
       eprint = {2402.13039},
 primaryClass = {astro-ph.SR},
       adsurl = {https://ui.adsabs.harvard.edu/abs/2024ApJ...970..156D}
}

@ARTICLE{2021ApJ...921...71D,
       author = {{Dittmann}, Alexander J. and {Ryan}, Geoffrey},
        title = "{Preventing Anomalous Torques in Circumbinary Accretion Simulations}",
      journal = {\apj},
         year = 2021,
        month = nov,
       volume = {921},
       number = {1},
          eid = {71},
        pages = {71},
          doi = {10.3847/1538-4357/ac1bbd},
archivePrefix = {arXiv},
       eprint = {2102.05684},
 primaryClass = {astro-ph.HE},
       adsurl = {https://ui.adsabs.harvard.edu/abs/2021ApJ...921...71D}
}

@ARTICLE{ahmad_2025,
       author = {{Ahmad}, A. and {Gonz{\'a}lez}, M. and {Hennebelle}, P. and {Lebreuilly}, U. and {Commer{\c{c}}on}, B.},
        title = "{Birth of magnetized low-mass protostars and circumstellar disks}",
      journal = {\aap},
         year = 2025,
        month = apr,
       volume = {696},
          eid = {A238},
        pages = {A238},
          doi = {10.1051/0004-6361/202553663},
archivePrefix = {arXiv},
       eprint = {2503.08637},
 primaryClass = {astro-ph.SR},
       adsurl = {https://ui.adsabs.harvard.edu/abs/2025A&A...696A.238A}
}

@ARTICLE{dubois_2021,
       author = {{Dubois}, Yohan and {Beckmann}, Ricarda and {Bournaud}, Fr{\'e}d{\'e}ric and {Choi}, Hoseung and {Devriendt}, Julien and {Jackson}, Ryan and {Kaviraj}, Sugata and {Kimm}, Taysun and {Kraljic}, Katarina and {Laigle}, Clotilde and {Martin}, Garreth and {Park}, Min-Jung and {Peirani}, S{\'e}bastien and {Pichon}, Christophe and {Volonteri}, Marta and {Yi}, Sukyoung K.},
        title = "{Introducing the NEWHORIZON simulation: Galaxy properties with resolved internal dynamics across cosmic time}",
      journal = {\aap},
         year = 2021,
        month = jul,
       volume = {651},
          eid = {A109},
        pages = {A109},
          doi = {10.1051/0004-6361/202039429},
archivePrefix = {arXiv},
       eprint = {2009.10578},
 primaryClass = {astro-ph.GA},
       adsurl = {https://ui.adsabs.harvard.edu/abs/2021A&A...651A.109D}
}

@ARTICLE{lebreuilly_2021,
       author = {{Lebreuilly}, Ugo and {Hennebelle}, Patrick and {Colman}, Tine and {Commer{\c{c}}on}, Beno{\^\i}t and {Klessen}, Ralf and {Maury}, Ana{\"e}lle and {Molinari}, Sergio and {Testi}, Leonardo},
        title = "{Protoplanetary Disk Birth in Massive Star-forming Clumps: The Essential Role of the Magnetic Field}",
      journal = {\apjl},
         year = 2021,
        month = aug,
       volume = {917},
       number = {1},
          eid = {L10},
        pages = {L10},
          doi = {10.3847/2041-8213/ac158c},
archivePrefix = {arXiv},
       eprint = {2107.08491},
 primaryClass = {astro-ph.SR},
       adsurl = {https://ui.adsabs.harvard.edu/abs/2021ApJ...917L..10L}
}

@ARTICLE{Price_2012,
       author = {{Price}, Daniel J.},
        title = "{Smoothed particle hydrodynamics and magnetohydrodynamics}",
      journal = {Journal of Computational Physics},
         year = 2012,
        month = feb,
       volume = {231},
       number = {3},
        pages = {759-794},
          doi = {10.1016/j.jcp.2010.12.011},
archivePrefix = {arXiv},
       eprint = {1012.1885},
 primaryClass = {astro-ph.IM},
       adsurl = {https://ui.adsabs.harvard.edu/abs/2012JCoPh.231..759P}
}

@ARTICLE{Paardekooper:2008,
       author = {{Paardekooper}, S. -J. and {Papaloizou}, J.~C.~B.},
        title = "{On disc protoplanet interactions in a non-barotropic disc with thermal diffusion}",
      journal = {\aap},
         year = 2008,
        month = jul,
       volume = {485},
       number = {3},
        pages = {877-895},
          doi = {10.1051/0004-6361:20078702},
archivePrefix = {arXiv},
       eprint = {0804.4547},
 primaryClass = {astro-ph},
       adsurl = {https://ui.adsabs.harvard.edu/abs/2008A&A...485..877P}
}

@INPROCEEDINGS{Baruteau:2014,
       author = {{Baruteau}, C. and {Crida}, A. and {Paardekooper}, S. -J. and {Masset}, F. and {Guilet}, J. and {Bitsch}, B. and {Nelson}, R. and {Kley}, W. and {Papaloizou}, J.},
        title = "{Planet-Disk Interactions and Early Evolution of Planetary Systems}",
    booktitle = {Protostars and Planets VI},
         year = 2014,
       editor = {{Beuther}, Henrik and {Klessen}, Ralf S. and {Dullemond}, Cornelis P. and {Henning}, Thomas},
        month = jan,
        pages = {667-689},
          doi = {10.2458/azu_uapress_9780816531240-ch029},
archivePrefix = {arXiv},
       eprint = {1312.4293},
 primaryClass = {astro-ph.EP},
       adsurl = {https://ui.adsabs.harvard.edu/abs/2014prpl.conf..667B}
}

@article{babak_science_2017,
	title = {Science with the space-based interferometer {LISA}. {V}. {Extreme} mass-ratio inspirals},
	volume = {95},
	copyright = {http://link.aps.org/licenses/aps-default-license},
	issn = {2470-0010, 2470-0029},
	url = {http://link.aps.org/doi/10.1103/PhysRevD.95.103012},
	doi = {10.1103/PhysRevD.95.103012},
	language = {en},
	number = {10},
	urldate = {2025-10-06},
	journal = {Phys. Rev. D},
	author = {Babak, Stanislav and Gair, Jonathan and Sesana, Alberto and Barausse, Enrico and Sopuerta, Carlos F. and Berry, Christopher P. L. and Berti, Emanuele and Amaro-Seoane, Pau and Petiteau, Antoine and Klein, Antoine},
	month = may,
	year = {2017},
	pages = {103012},
}

@article{arca_sedda_merging_2021,
	title = {Merging stellar and intermediate-mass black holes in dense clusters: implications for {LIGO}, {LISA}, and the next generation of gravitational wave detectors},
	volume = {652},
	issn = {0004-6361, 1432-0746},
	shorttitle = {Merging stellar and intermediate-mass black holes in dense clusters},
	url = {https://www.aanda.org/10.1051/0004-6361/202037785},
	doi = {10.1051/0004-6361/202037785},
	urldate = {2024-02-07},
	journal = {A\&A},
	author = {Arca Sedda, Manuel and Amaro Seoane, Pau and Chen, Xian},
	month = aug,
	year = {2021},
	pages = {A54}
}

@ARTICLE{Miniutti2019,
       author = {{Miniutti}, G. and {Saxton}, R.~D. and {Giustini}, M. and {Alexander}, K.~D. and {Fender}, R.~P. and {Heywood}, I. and {Monageng}, I. and {Coriat}, M. and {Tzioumis}, A.~K. and {Read}, A.~M. and {Knigge}, C. and {Gandhi}, P. and {Pretorius}, M.~L. and {Ag{\'\i}s-Gonz{\'a}lez}, B.},
        title = "{Nine-hour X-ray quasi-periodic eruptions from a low-mass black hole galactic nucleus}",
      journal = {\nat},
         year = 2019,
        month = sep,
       volume = {573},
       number = {7774},
        pages = {381-384},
          doi = {10.1038/s41586-019-1556-x},
archivePrefix = {arXiv},
       eprint = {1909.04693},
 primaryClass = {astro-ph.HE},
       adsurl = {https://ui.adsabs.harvard.edu/abs/2019Natur.573..381M}
}

@ARTICLE{Giustini2020,
       author = {{Giustini}, Margherita and {Miniutti}, Giovanni and {Saxton}, Richard D.},
        title = "{X-ray quasi-periodic eruptions from the galactic nucleus of RX J1301.9+2747}",
      journal = {\aap},
         year = 2020,
        month = apr,
       volume = {636},
          eid = {L2},
        pages = {L2},
          doi = {10.1051/0004-6361/202037610},
archivePrefix = {arXiv},
       eprint = {2002.08967},
 primaryClass = {astro-ph.HE},
       adsurl = {https://ui.adsabs.harvard.edu/abs/2020A&A...636L...2G}
}

@ARTICLE{Arcodia2021,
       author = {{Arcodia}, R. and {Merloni}, A. and {Nandra}, K. and {Buchner}, J. and {Salvato}, M. and {Pasham}, D. and {Remillard}, R. and {Comparat}, J. and {Lamer}, G. and {Ponti}, G. and {Malyali}, A. and {Wolf}, J. and {Arzoumanian}, Z. and {Bogensberger}, D. and {Buckley}, D.~A.~H. and {Gendreau}, K. and {Gromadzki}, M. and {Kara}, E. and {Krumpe}, M. and {Markwardt}, C. and {Ramos-Ceja}, M.~E. and {Rau}, A. and {Schramm}, M. and {Schwope}, A.},
        title = "{X-ray quasi-periodic eruptions from two previously quiescent galaxies}",
      journal = {\nat},
         year = 2021,
        month = apr,
       volume = {592},
       number = {7856},
        pages = {704-707},
          doi = {10.1038/s41586-021-03394-6},
archivePrefix = {arXiv},
       eprint = {2104.13388},
 primaryClass = {astro-ph.HE},
       adsurl = {https://ui.adsabs.harvard.edu/abs/2021Natur.592..704A}
}

@ARTICLE{Chakraborty2021,
       author = {{Chakraborty}, Joheen and {Kara}, Erin and {Masterson}, Megan and {Giustini}, Margherita and {Miniutti}, Giovanni and {Saxton}, Richard},
        title = "{Possible X-Ray Quasi-periodic Eruptions in a Tidal Disruption Event Candidate}",
      journal = {\apjl},
         year = 2021,
        month = nov,
       volume = {921},
       number = {2},
          eid = {L40},
        pages = {L40},
          doi = {10.3847/2041-8213/ac313b},
archivePrefix = {arXiv},
       eprint = {2110.10786},
 primaryClass = {astro-ph.HE},
       adsurl = {https://ui.adsabs.harvard.edu/abs/2021ApJ...921L..40C}
}

@ARTICLE{Quintin2023,
       author = {{Quintin}, E. and {Webb}, N.~A. and {Guillot}, S. and {Miniutti}, G. and {Kammoun}, E.~S. and {Giustini}, M. and {Arcodia}, R. and {Soucail}, G. and {Clerc}, N. and {Amato}, R. and {Markwardt}, C.~B.},
        title = "{Tormund's return: Hints of quasi-periodic eruption features from a recent optical tidal disruption event}",
      journal = {\aap},
         year = 2023,
        month = jul,
       volume = {675},
          eid = {A152},
        pages = {A152},
          doi = {10.1051/0004-6361/202346440},
archivePrefix = {arXiv},
       eprint = {2306.00438},
 primaryClass = {astro-ph.HE},
       adsurl = {https://ui.adsabs.harvard.edu/abs/2023A&A...675A.152Q}
}

@ARTICLE{2001ApJ...552..793G,
       author = {{Goodman}, J. and {Rafikov}, R.~R.},
        title = "{Planetary Torques as the Viscosity of Protoplanetary Disks}",
      journal = {\apj},
         year = 2001,
        month = may,
       volume = {552},
       number = {2},
        pages = {793-802},
          doi = {10.1086/320572},
archivePrefix = {arXiv},
       eprint = {astro-ph/0010576},
 primaryClass = {astro-ph},
       adsurl = {https://ui.adsabs.harvard.edu/abs/2001ApJ...552..793G}
}

@ARTICLE{Nicholl2024,
       author = {{Nicholl}, M. and {Pasham}, D.~R. and {Mummery}, A. and {Guolo}, M. and {Gendreau}, K. and {Dewangan}, G.~C. and {Ferrara}, E.~C. and {Remillard}, R. and {Bonnerot}, C. and {Chakraborty}, J. and {Hajela}, A. and {Dhillon}, V.~S. and {Gillan}, A.~F. and {Greenwood}, J. and {Huber}, M.~E. and {Janiuk}, A. and {Salvesen}, G. and {van Velzen}, S. and {Aamer}, A. and {Alexander}, K.~D. and {Angus}, C.~R. and {Arzoumanian}, Z. and {Auchettl}, K. and {Berger}, E. and {de Boer}, T. and {Cendes}, Y. and {Chambers}, K.~C. and {Chen}, T. -W. and {Chornock}, R. and {Fulton}, M.~D. and {Gao}, H. and {Gillanders}, J.~H. and {Gomez}, S. and {Gompertz}, B.~P. and {Fabian}, A.~C. and {Herman}, J. and {Ingram}, A. and {Kara}, E. and {Laskar}, T. and {Lawrence}, A. and {Lin}, C. -C. and {Lowe}, T.~B. and {Magnier}, E.~A. and {Margutti}, R. and {McGee}, S.~L. and {Minguez}, P. and {Moore}, T. and {Nathan}, E. and {Oates}, S.~R. and {Patra}, K.~C. and {Ramsden}, P. and {Ravi}, V. and {Ridley}, E.~J. and {Sheng}, X. and {Smartt}, S.~J. and {Smith}, K.~W. and {Srivastav}, S. and {Stein}, R. and {Stevance}, H.~F. and {Turner}, S.~G.~D. and {Wainscoat}, R.~J. and {Weston}, J. and {Wevers}, T. and {Young}, D.~R.},
        title = "{Quasi-periodic X-ray eruptions years after a nearby tidal disruption event}",
      journal = {\nat},
         year = 2024,
        month = oct,
       volume = {634},
       number = {8035},
        pages = {804-808},
          doi = {10.1038/s41586-024-08023-6},
archivePrefix = {arXiv},
       eprint = {2409.02181},
 primaryClass = {astro-ph.HE},
       adsurl = {https://ui.adsabs.harvard.edu/abs/2024Natur.634..804N}
}

@ARTICLE{HernandezGarcia2025,
       author = {{Hern{\'a}ndez-Garc{\'\i}a}, Lorena and {Chakraborty}, Joheen and {S{\'a}nchez-S{\'a}ez}, Paula and {Ricci}, Claudio and {Cuadra}, Jorge and {McKernan}, Barry and {Ford}, K.~E. Saavik and {Ar{\'e}valo}, Patricia and {Rau}, Arne and {Arcodia}, Riccardo and {Kara}, Erin and {Liu}, Zhu and {Merloni}, Andrea and {Bruni}, Gabriele and {Goodwin}, Adelle and {Arzoumanian}, Zaven and {Assef}, Roberto J. and {Baldini}, Pietro and {Bayo}, Amelia and {Bauer}, Franz E. and {Bernal}, Santiago and {Brightman}, Murray and {Calistro Rivera}, Gabriela and {Gendreau}, Keith and {Homan}, David and {Krumpe}, Mirko and {Lira}, Paulina and {Mart{\'\i}nez-Aldama}, Mary Loli and {Salvato}, Mara and {Sotomayor}, Bel{\'e}n},
        title = "{Discovery of extreme quasi-periodic eruptions in a newly accreting massive black hole}",
      journal = {Nature Astronomy},
         year = 2025,
        month = apr,
          doi = {10.1038/s41550-025-02523-9},
archivePrefix = {arXiv},
       eprint = {2504.07169},
 primaryClass = {astro-ph.HE},
       adsurl = {https://ui.adsabs.harvard.edu/abs/2025NatAs.tmp...90H}
}

@ARTICLE{Xian2021,
       author = {{Xian}, Jingtao and {Zhang}, Fupeng and {Dou}, Liming and {He}, Jiasheng and {Shu}, Xinwen},
        title = "{X-Ray Quasi-periodic Eruptions Driven by Star-Disk Collisions: Application to GSN069 and Probing the Spin of Massive Black Holes}",
      journal = {\apjl},
         year = 2021,
        month = nov,
       volume = {921},
       number = {2},
          eid = {L32},
        pages = {L32},
          doi = {10.3847/2041-8213/ac31aa},
archivePrefix = {arXiv},
       eprint = {2110.10855},
 primaryClass = {astro-ph.HE},
       adsurl = {https://ui.adsabs.harvard.edu/abs/2021ApJ...921L..32X}
}

@ARTICLE{Franchini2023,
       author = {{Franchini}, Alessia and {Bonetti}, Matteo and {Lupi}, Alessandro and {Miniutti}, Giovanni and {Bortolas}, Elisa and {Giustini}, Margherita and {Dotti}, Massimo and {Sesana}, Alberto and {Arcodia}, Riccardo and {Ryu}, Taeho},
        title = "{Quasi-periodic eruptions from impacts between the secondary and a rigidly precessing accretion disc in an extreme mass-ratio inspiral system}",
      journal = {\aap},
         year = 2023,
        month = jul,
       volume = {675},
          eid = {A100},
        pages = {A100},
          doi = {10.1051/0004-6361/202346565},
archivePrefix = {arXiv},
       eprint = {2304.00775},
 primaryClass = {astro-ph.HE},
       adsurl = {https://ui.adsabs.harvard.edu/abs/2023A&A...675A.100F}
}

@ARTICLE{Linial2023b,
       author = {{Linial}, Itai and {Metzger}, Brian D.},
        title = "{EMRI + TDE = QPE: Periodic X-Ray Flares from Star-Disk Collisions in Galactic Nuclei}",
      journal = {\apj},
         year = 2023,
        month = nov,
       volume = {957},
       number = {1},
          eid = {34},
        pages = {34},
          doi = {10.3847/1538-4357/acf65b},
archivePrefix = {arXiv},
       eprint = {2303.16231},
 primaryClass = {astro-ph.HE},
       adsurl = {https://ui.adsabs.harvard.edu/abs/2023ApJ...957...34L}
}

@ARTICLE{Tagawa2023,
       author = {{Tagawa}, Hiromichi and {Haiman}, Zolt{\'a}n},
        title = "{Flares from stars crossing active galactic nucleus discs on low-inclination orbits}",
      journal = {\mnras},
         year = 2023,
        month = nov,
       volume = {526},
       number = {1},
        pages = {69-79},
          doi = {10.1093/mnras/stad2616},
archivePrefix = {arXiv},
       eprint = {2304.03670},
 primaryClass = {astro-ph.HE},
       adsurl = {https://ui.adsabs.harvard.edu/abs/2023MNRAS.526...69T}
}

@ARTICLE{HosseinNouri2024,
       author = {{Hossein Nouri}, Fatemeh and {Janiuk}, Agnieszka},
        title = "{Viscous torque in turbulent magnetized active galactic nucleus accretion disks and its effects on the gravitational waves of extreme mass ratio inspirals}",
      journal = {\aap},
         year = 2024,
        month = jul,
       volume = {687},
          eid = {A184},
        pages = {A184},
          doi = {10.1051/0004-6361/202348796},
archivePrefix = {arXiv},
       eprint = {2309.06028},
 primaryClass = {astro-ph.HE},
       adsurl = {https://ui.adsabs.harvard.edu/abs/2024A&A...687A.184H}
}

@INPROCEEDINGS{2025JPhCS2997a2014D,
       author = {{Delorme}, Maxime and {Durocher}, Arnaud and {Sacha Brun}, Allan and {Finley}, Adam J. and {Marchal}, Olivier and {Aubert}, Dominique},
        title = "{Dyablo : A simulation code for astrophysics fluids with adaptive mesh refinement in the exascale era}",
    booktitle = {Journal of Physics Conference Series},
         year = 2025,
       series = {Journal of Physics Conference Series},
       volume = {2997},
        month = apr,
    publisher = {IOP},
          eid = {012014},
        pages = {012014},
          doi = {10.1088/1742-6596/2997/1/012014},
       adsurl = {https://ui.adsabs.harvard.edu/abs/2025JPhCS2997a2014D}
}

@book{frank_accretion_2002,
	address = {Cambridge, UK},
	title = {Accretion {Power} in {Astrophysics}: {Third} {Edition}},
	url = {https://ui.adsabs.harvard.edu/abs/2002apa..book.....F},
	publisher = {Cambridge University Press},
	author = {Frank, Juhan and King, Andrew and Raine, Dereck},
	year = {2002},
}

@ARTICLE{Edwin_paper,
    author = {Santiago-Leandro, Edwin and Mignon-Risse, Raphaël and Batta, Aldo and Motta, David and Brucy, Noé},
    year = {in prep.},
}

@article{tiede_gas-driven_2020,
	title = {Gas-driven {Inspiral} of {Binaries} in {Thin} {Accretion} {Disks}},
	volume = {900},
	issn = {1538-4357},
	url = {https://iopscience.iop.org/article/10.3847/1538-4357/aba432},
	doi = {10.3847/1538-4357/aba432},
	number = {1},
	urldate = {2021-09-10},
	journal = {ApJ},
	author = {Tiede, Christopher and Zrake, Jonathan and MacFadyen, Andrew and Haiman, Zoltan},
	month = aug,
	year = {2020},
	pages = {43},
}

@article{quemenerSIDUSSolutionExtreme2013,
  title = {{{SIDUS}}---the Solution for Extreme Deduplication of an Operating System},
  author = {Quemener, Emmanuel and Corvellec, Marianne},
  year = {2013},
  month = nov,
  journal = {Linux J.},
  volume = {2013},
  number = {235},
  pages = {3:3},
  issn = {1075-3583}
}

@article{lescaudron_dynamical_2023,
	title = {Dynamical friction of a massive black hole in a turbulent gaseous medium},
	volume = {674},
	copyright = {https://creativecommons.org/licenses/by/4.0},
	issn = {0004-6361, 1432-0746},
	url = {https://www.aanda.org/10.1051/0004-6361/202243392},
	doi = {10.1051/0004-6361/202243392},
	urldate = {2025-11-10},
	journal = {A\&A},
	author = {Lescaudron, Sandrine and Dubois, Yohan and Beckmann, Ricarda S. and Volonteri, Marta},
	month = jun,
	year = {2023},
	pages = {A217},
}

@ARTICLE{brucy_2021,
       author = {{Brucy}, No{\'e} and {Hennebelle}, Patrick},
        title = "{A two-step gravitational cascade for the fragmentation of self-gravitating discs}",
      journal = {\mnras},
         year = 2021,
        month = may,
       volume = {503},
       number = {3},
        pages = {4192-4207},
          doi = {10.1093/mnras/stab738},
archivePrefix = {arXiv},
       eprint = {2103.05508},
 primaryClass = {astro-ph.GA},
       adsurl = {https://ui.adsabs.harvard.edu/abs/2021MNRAS.503.4192B}
}

@ARTICLE{2023MNRAS.521.4645Z,
       author = {{Zwick}, Lorenz and {Capelo}, Pedro R. and {Mayer}, Lucio},
        title = "{Priorities in gravitational waveforms for future space-borne detectors: vacuum accuracy or environment?}",
      journal = {\mnras},
         year = 2023,
        month = may,
       volume = {521},
       number = {3},
        pages = {4645-4651},
          doi = {10.1093/mnras/stad707},
archivePrefix = {arXiv},
       eprint = {2209.04060},
 primaryClass = {gr-qc},
       adsurl = {https://ui.adsabs.harvard.edu/abs/2023MNRAS.521.4645Z}
}

@misc{kara_supermassive_2025,
	title = {Supermassive {Black} {Holes} in {X}-rays: {From} {Standard} {Accretion} to {Extreme} {Transients}},
	shorttitle = {Supermassive {Black} {Holes} in {X}-rays},
	url = {http://arxiv.org/abs/2503.22791},
	doi = {10.48550/arXiv.2503.22791},
	urldate = {2025-05-02},
	publisher = {arXiv},
	author = {Kara, Erin and García, Javier},
	month = mar,
	year = {2025},
	note = {arXiv:2503.22791 [astro-ph]},
}

@ARTICLE{Pringle:1981,
   author = {{Pringle}, J.~E.},
    title = "{Accretion discs in astrophysics}",
  journal = {\araa},
     year = 1981,
   volume = 19,
    pages = {137-162},
      doi = {10.1146/annurev.aa.19.090181.001033},
   adsurl = {http://adsabs.harvard.edu/abs/1981ARA%26A..19..137P}
}

@ARTICLE{SS73,
   author = {{Shakura}, N.~I. and {Sunyaev}, R.~A.},
    title = "{Black holes in binary systems. Observational appearance.}",
  journal = {\aap},
     year = 1973,
   volume = 24,
    pages = {337-355},
   adsurl = {http://adsabs.harvard.edu/abs/1973A%26A....24..337S}
}

@ARTICLE{KocsisYunesLoeb:2011,
   author = {{Kocsis}, B. and {Yunes}, N. and {Loeb}, A.},
    title = "{Observable signatures of extreme mass-ratio inspiral black hole binaries embedded in thin accretion disks}",
  journal = {\prd},
archivePrefix = "arXiv",
   eprint = {1104.2322},
 primaryClass = "astro-ph.GA",
     year = 2011,
    month = jul,
   volume = 84,
   number = 2,
      eid = {024032},
    pages = {024032},
      doi = {10.1103/PhysRevD.84.024032},
   adsurl = {http://adsabs.harvard.edu/abs/2011PhRvD..84b4032K}
}

@ARTICLE{Ward:1997,
   author = {{Ward}, W.~R.},
    title = "{Protoplanet Migration by Nebula Tides}",
  journal = {\icarus},
     year = 1997,
    month = apr,
   volume = 126,
    pages = {261-281},
      doi = {10.1006/icar.1996.5647},
   adsurl = {http://adsabs.harvard.edu/abs/1997Icar..126..261W}
}

@ARTICLE{GoodmanRafikov:2001,
   author = {{Goodman}, J. and {Rafikov}, R.~R.},
    title = "{Planetary Torques as the Viscosity of Protoplanetary Disks}",
  journal = {\apj},
   eprint = {astro-ph/0010576},
     year = 2001,
    month = may,
   volume = 552,
    pages = {793-802},
      doi = {10.1086/320572},
   adsurl = {http://adsabs.harvard.edu/abs/2001ApJ...552..793G}
}

@article{1993ApJ...419..166A,
author = {Artymowicz, Pawel},
title = {{Disk-Satellite Interaction via Density Waves and the Eccentricity Evolution of Bodies Embedded in Disks}},
journal = {Astrophysical Journal v.419},
year = {1993},
volume = {419},
pages = {166--},
month = dec
}

@ARTICLE{Ward:1986,
   author = {{Ward}, W.~R.},
    title = "{Density waves in the solar nebula - Differential Lindblad torque}",
  journal = {\icarus},
     year = 1986,
    month = jul,
   volume = 67,
    pages = {164-180},
      doi = {10.1016/0019-1035(86)90182-X},
   adsurl = {http://adsabs.harvard.edu/abs/1986Icar...67..164W}
}

@ARTICLE{TanakaI:2002,
   author = {{Tanaka}, H. and {Takeuchi}, T. and {Ward}, W.~R.},
    title = "{Three-Dimensional Interaction between a Planet and an Isothermal Gaseous Disk. I. Corotation and Lindblad Torques and Planet Migration}",
  journal = {\apj},
     year = 2002,
    month = feb,
   volume = 565,
    pages = {1257-1274},
      doi = {10.1086/324713},
   adsurl = {http://adsabs.harvard.edu/abs/2002ApJ...565.1257T}
}

\appendix
\begin{figure*}
    \centering
    \includegraphics[width=.95\textwidth]{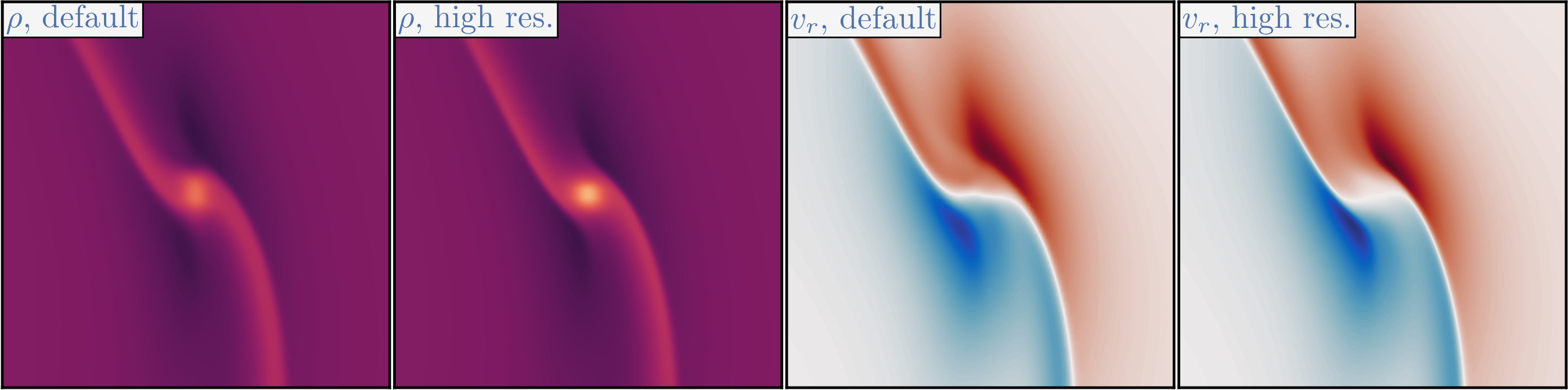}
    \caption{Surface density and radial velocity maps for thin-disc \texttt{ATHENA++} simulations with different resolutions. ``default" resolution is the same shown elsewhere throughout the paper, and ``high res." indicates that a single level of mesh refinement was used, doubling the resolution for $r\in\{0.8, 1.25\}$. The color scale is the same as in Figure \ref{fig:surfacedensitymap_vr_hr003}.}
    \label{fig:athena_resolution}
\end{figure*}

\section{Additional Code Details}\label{app:codes}

Here we describe in detail the setup and parameter choices of each code, so that others can reproduce the data used in this study. 
\subsection{\texttt{ATHENA++}}
\texttt{\texttt{ATHENA++}} implements a wide range of reconstruction algorithms, approximate Riemann solvers, and time-stepping routines. The simulations in this work used piecewise linear reconstruction \citep{1974JCoPh..14..361V} of the primitive variables, the Harten-Lax-van Leer-Contact approximate Riemann solver \citep{1994ShWav...4...25T}, and second-order strong-stability-preserving Runge-Kutta time-stepping \citep{1998MaCom..67...73G} with a Courant–Friedrichs–Lewy safety factor of 0.4. To maintain a locally isothermal profile, these simulations used an ideal gas equation of state with $\gamma=1.0001$ and reset the temperature of each cell to the value prescribed in Section \ref{sec:ICs} every time step. The simulations in this work did not include orbital advection \citep[e.g.][]{2000A&AS..141..165M}, which made them both less efficient and more diffusive. To illustrate the effects of numerical dissipation under these model choices, Figure \ref{fig:athena_resolution} compares the surface density and radial velocity from our default-resolution simulation with that of an additional level of mesh refinement (doubling the effective resolution) near the secondary.

\texttt{\texttt{ATHENA++}} solves the full Navier-Stokes equations as described in \citet{1984frh..book.....M,1992ApJS...80..753S}, assuming zero bulk viscosity. 
Figure~\ref{fig:athenapp_visco} shows the azimuthally-averaged density profile in an axisymmetric test case for comparison with the new viscosity implementation in \texttt{RAMSES} (see below).

\subsection{\texttt{DISCO}}
\texttt{DISCO} is a grid-based code that utilizes a dynamic cylindrical mesh. 
%The domain extends from $0.5-3.0$ radial units with the origin centered on the binary center of mass. The radial resolution is set to $N_{\rm r} = 600$ cells that are spaced logarithmically. This leads to a spatial resolution $\Delta r/r = 10^{(\log{r_{\rm min}} - \log{r_{\rm max}})/N_{\rm r}} - 1$ across the grid. The azimuthal resolution is set such that the aspect ratio of cells is $1$ across the entire domain. 
%Cell motion is set to \texttt{Average}, which in this case essentially follows the Keplerian orbital motion of the gas. 
%An HLLC Riemann solver is used with a slope-limiter of $1.5$. --- AJD: most of this was covered in 3.2.2
We use a Courant–Friedrichs–Lewy condition set to $0.5$ and a generalized minmod slope-limiting parameter of $1.5$ \citep{2000JCoPh.160..241K}.
%Boundary conditions are fixed to the initial conditions. 
The isothermal EOS is acheived with an adiabatic index of $1.00001$ and instantaneous cooling, and no sinks are used.
The resources used for the simulation vary with the parameter choices (the code is slower for thicker discs which have faster sound speeds, which limit the timestep). For the second comparison run, a single simulation ran on 32 MPI processes for $\sim 9$ hours. 

In computing viscous fluxes and source terms, the original version of \texttt{DISCO} makes an assumption of constant dynamic viscosity: when calculating gradients in the viscous stress tensor, gradients in the quantity $\Sigma \nu$ are neglected. This assumption is valid for discs commonly modeled in the literature, in which the dynamic viscosity is nearly constant. Specifically, for a disc in a steady unperturbed state, or only linearly affected by a perturber, constant mass flux implies that the dynamic viscosity is effectively constant. However, this breaks down in the nonlinear regime where the surface density develops significant inhomogeneity, as in simulations of circumbinary accretion, gap-opening planets, and the thin-disc scenario studied above.
As shown by the tests in this study (with comparisons between \texttt{DISCO} and \texttt{DISCO v2}), the inclusion of the full set viscous fluxes affects the gas morphology close to the secondary. In the cases examined here, the updated version results in a slightly weaker torque onto the binary. 
See \citet{2021ApJ...921...71D} and below for further details. 
Note that the simplified viscosity assumption was also made in some previous studies \citep[e.g.,][]{derdzinski_evolution_2019,derdzinski_evolution_2021}. The comparison in this work demonstrates the dependence of the torque on the full set of viscous flux terms.  

\subsection{\texttt{DISCO v2}}
Many of the settings used in \texttt{DISCO v2} were quite similar to those described above. 
%although 
%the default radial resolution used $N_r=512$ logarithmically-spaced annuli to span the domain, 
%the Courant–Friedrichs–Lewy factor factor was set to 0.5 and the adiabatic index was set to 1.00001. 
However, the viscosity implementations differed significantly between the two codes. Precise expressions for the viscous terms in \texttt{DISCO} and \texttt{DISCO v2} are provided in Appendix A of \citet{2021ApJ...921...71D}, and some of the errors introduced by the approximate viscosity used by \texttt{DISCO} in a circumbinary context are discussed in the appendices of \citet{2021ApJ...921...71D} and \citet{2024ApJ...967...12D}. \texttt{DISCO v2} also includes a number of other upgrades, such as using cell centroids rather than centers where appropriate, which precludes extremely precise head-to-head comparisons. However, for comparison with previous results, \texttt{DISCO v2} implements the approximate viscosity described in \citet{Duffell2016}, which we refer to in the following as ``\texttt{v1} viscosity''

\begin{figure}
    \centering
    \includegraphics[width=.95\textwidth]{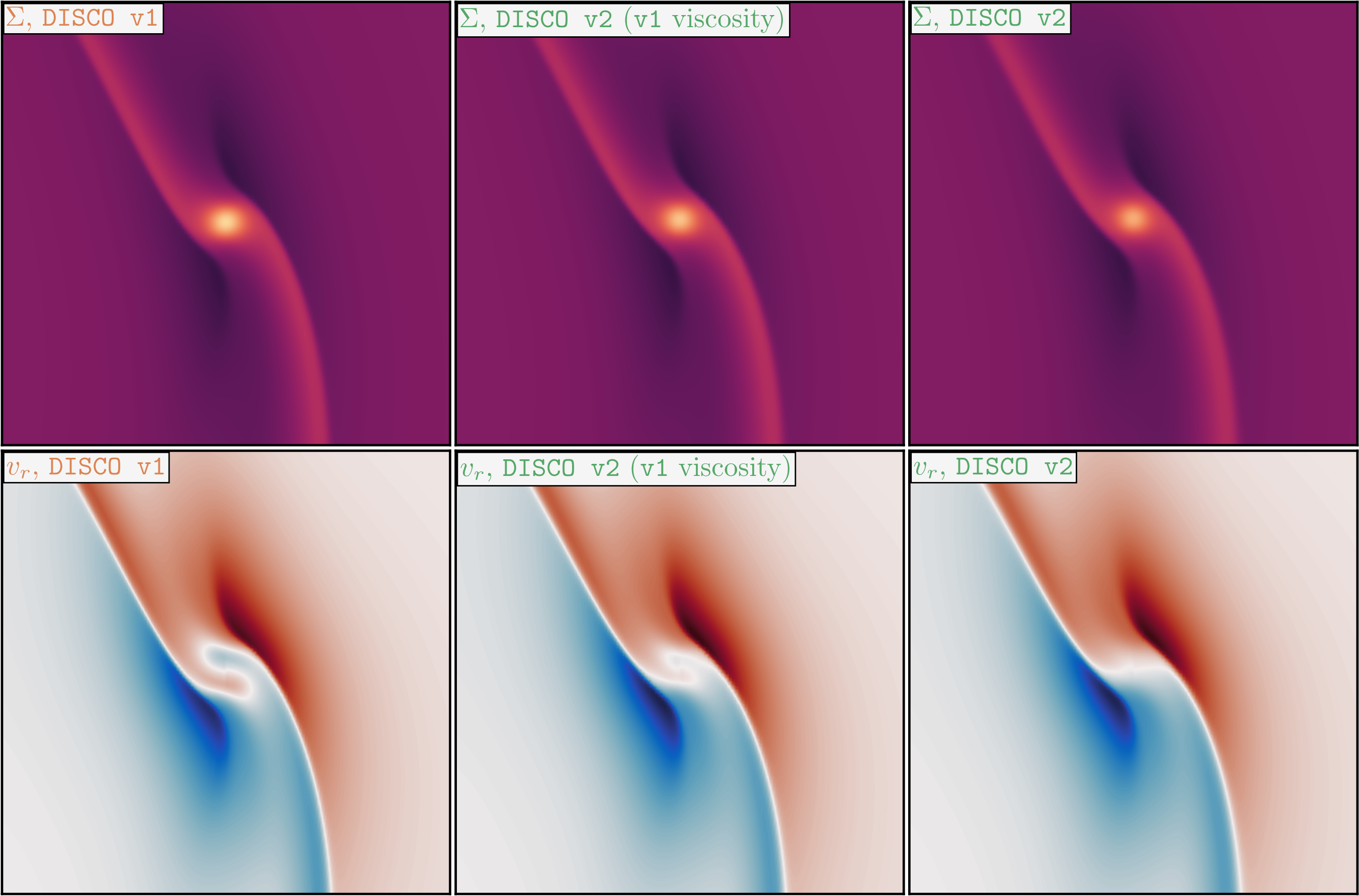}
    \caption{Surface density and radial velocity maps from thin-disc simulations conducted using different version of \texttt{DISCO}, and different viscosity implementations within \texttt{DISCO v2}. The color scale is the same as in Figure \ref{fig:surfacedensitymap_vr_hr003}. }
    \label{fig:disco_viscosity}
\end{figure}

To disentangle the effects of the approximate viscosity of \citet{Duffell2016} from other code changes, we ran an $h=0.03$ simulation using \texttt{DISCO v2} and \texttt{v1} viscosity for comparison. Figure \ref{fig:disco_viscosity} compares the surface density and radial velocity within the Hill sphere for this simulation in addition to the original \texttt{DISCO} and \texttt{DISCO v2} simulations. Qualitatively, the approximations of \citet{Duffell2016} seem responsible for the erroneous circulation patterns within the Hill sphere, and for the slightly higher surface densities near the secondary. The departure of \texttt{DISCO} from the other codes in Figure \ref{fig:torque_h03_M2} can be safely attributed to its incomplete viscosity implementation.

\subsection{\texttt{FARGO3D}}
In addition to the details given in Section~\ref{desc:fargo3d}, we comment that we used orbital advection for all runs, including those for which no gain is expected on the time step -- this happens when the timestep is limited primarily  by the viscosity rather than the velocity (essentially in the thick disc run, at the resolution used in this study). We emphasize that this behavior is specific to simulations of AGN discs, which are much more viscous than protoplanetary discs. For the three dimensional runs, the mesh has a colatitude extent that is three times the aspect ratio. In the thick disc case, the softening length is set to $10^{-2}$ length units (hence one tenth of the pressure length scale), while in the thin disc case it is $6\times 10^{-3}$ length units (hence $20$\% of the pressure length scale). Preliminary runs with a softening length of one-tenth of a pressure length scale, for the thin disc case, yielded very strong periodic variations of the torque. For each 3D simulation, we calculate two flavors of the torque: one for which the material inside the inner Hill sphere is excised, following the prescription of \cite{Crida2009} with $b=0.4$, and one in which all the gas is taken into account, without any specific weighting close to the secondary. Both methods yield comparable results for the thick disc case and very different results for the thin disc case, with a mildly negative torque if the disc is excised, and a strongly positive torque if all the gas is taken into account.

\subsection{\texttt{GASOLINE}}
We implemented physical viscosity in \textsc{Gasoline2} using both the first-derivative and second-derivative methods described in \citet{Price2018}, employing the GDSPH gradient operators. However, for this work, for an easier comparison with other codes, we only applied the second-derivative method. We additionally included artificial viscosity, following a method based on that of \citet{Cullen_Dehnen_2010}, in which we use as shock detector the velocity gradient in the direction of the pressure gradient instead of the velocity divergence \citep[][]{Wadsley_et_al_2017}. Figure \ref{fig:gasoline:taylorgreen} provides an example of a test problem used to verify the viscosity implementation. 
We adopted an $M_4$ cubic spline kernel with 58 neighbors \citep[using the recommended value from][]{Price_2012}.
\begin{figure}
    \centering
    \includegraphics[width=0.5\linewidth]{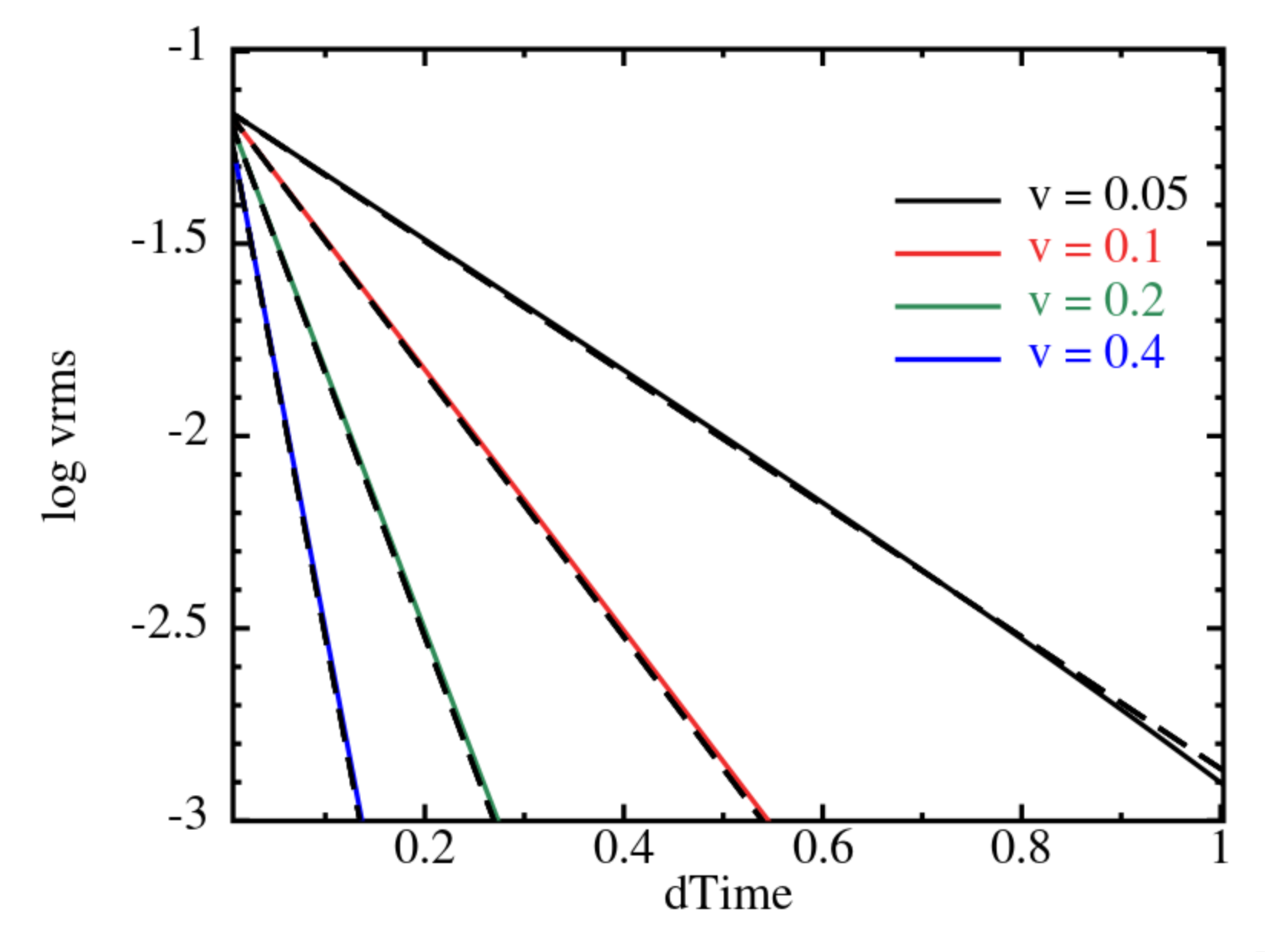}
    \caption{Test of the physical viscosity implementation of Gasoline for the Taylor-Green vortex using the kinematic shear viscosity $\nu=0.05,0.1,0.2,0.4$. The black dashed lines show the analytical solution of the exponential decay rate in the velocity of each case.}
    \label{fig:gasoline:taylorgreen}
\end{figure}

\subsection{\texttt{\texttt{\texttt{GIZMO}}}}

The MFM method is very similar to smoother particle hydrodynamics methods except that a Riemann problem is solved between particles to account for shocks and dissipation rather than assuming an artificial viscosity; the method is Lagrangian, not allowing mass flux between particles.
We adopt a cubic spline kernel and employ 58 neighbors.

The torques are computed by direct summation of the contribution of the particles between $0.5a$ and $3a$, taking into account the different softenings for the two binary components and for the different runs we perform. 

We here include the effect of gas viscosity entering the Navier-Stokes fluid equations as described in \citet{Hopkins2016}, assuming a shear viscosity in the disc $\nu=\alpha c_{\rm s}H$ parametrised using a viscosity parameter $\alpha=0.1$ \citep{shakura73}, and no bulk viscosity. 
We tested the viscosity implementation by simulating a spreading-ring around a single object. We found excellent agreement with the spreading-ring analytical solution \citep{Pringle:1981}.

\subsection{\texttt{PHANTOM}}
The viscosity is implemented following the model by Shakura-Sunyaev, with $\alpha=0.1$. We tested the viscosity implementation by simulating a spreading-ring around a single object and found excellent agreement with the spreading-ring analytical solution \citep{Pringle:1981}.
The primary is modeled using a cubic spline kernel which reduces to a Newtonian potential outside of $r=0.5r_0$, given by 
\begin{equation}
    \Phi(k) = -\frac{M_1}{k\epsilon_\bullet}\left\{\begin{split}
        &\frac{14}{5}k -\frac{16}{3}k^3 +\frac{48}{5}k^5 -\frac{32}{5}k^6 & 0\leq k< 1/2\\
        &-\frac{1}{15}+\frac{16}{5}k-\frac{32}{3}k^3+16k^4-\frac{48}{5}k^5+\frac{32}{15}k^6 & 1/2\leq k<1\\
        &1 & k\geq 1\\
    \end{split}
    \right.
    \label{eq:cubic_spline}
\end{equation}
where $\epsilon_\bullet=0.5$ is the softening kernel extension and $k=|\mathbf{r}|/\epsilon_\bullet$.
The total mass of the binary is set to $G(M_1+M_2)=1$, where $G$ is the gravitational constant.
The secondary BH is placed on a fixed, circular orbit at $r=r_0$ and modeled as a point mass with a smoothed gravitational potential following Equation (\ref{eq:4}).
%\begin{equation}
%    \Phi_2 = -\frac{GM_2}{(r_2^2 + \epsilon^2)^{1/2}}
%\end{equation}
%where the softening is $\epsilon=0.018$. 

\subsection{\texttt{RAMSES}}
\label{app:ramses}

The grid is 2D, Cartesian, with a static (rather than adaptive) mesh refinement centered onto the box center.
Refinement levels range from $9$ to $12$.
%The box size is set to $8 a$, resulting in a finest resolution of $2\times10^{-3} a$. 
The Courant–Friedrichs–Lewy factor was set to $0.8$. %still valid
As mentioned in the main text, the viscosity is not native in \texttt{RAMSES} and has been implemented in the code for the study of binary-disc interaction like the present study.
It incorporates the viscous force in the $j-$th direction

\begin{equation}
F_j^\mathrm{visc} =  (\Delta \cdot T_\mathrm{vis})_j = \partial_k T_\mathrm{vis}^{kj}
\end{equation}

as a source term in the momentum equation (see e.g. \citealt{Duffell2013}) with $T_\mathrm{vis}$ the viscous stress tensor defined as (Eq.~7 of \citealt{tiede_gas-driven_2020})

\begin{equation}
T_\mathrm{vis}^{ij} = \nu \Sigma \left( \frac{\partial v^i}{\partial x^j} + \frac{\partial v^j}{\partial x^i} - \frac{\partial v^k}{\partial x^k} \delta^{ij} \right).
\end{equation}
It will be presented in more details in \cite{Edwin_paper}. %rephrase this paragraph

%Because in our disk model $\Sigma \varpropto r^{-1/2}$, $\alpha=\mathrm{cst}$, $H/r\equiv c_\mathrm{s}/v_\phi =\mathrm{cst}$ so $c_\mathrm{s} \varpropto v_\phi \varpropto r^{-1/2}$ and $H \varpropto r$ , we have $\nu \varpropto r^{1/2} $ and $\Sigma \nu$ is uniform.
%Therefore, the viscous force reduces to
%\begin{equation}
%F_j^\mathrm{visc} = \frac{\Sigma \nu}{2}  \Delta v_j,
%\end{equation}
%where $\Delta v_j$ is the Laplacian operator applied to the velocity vector along direction $j$.
%This assumption breaks when the density profile is perturbed, which happens close to the binary. This is the most probable explanation for the lack of convergence of the torque and the gap observed in the thin disc simulation (see Section \ref{sec:thinDisc}).

We have tested this viscosity implementation in the single-body case, using a cubic spline kernel potential (Eq.~\ref{eq:cubic_spline}). The rest of the parameters are set as in the alignment or thin disc run.
Left panels of Fig.~\ref{fig:RAMSES_visco} show the theoretical -- by theoretical we mean analytical; not to be confused with the steady state --, initial and final azimuthally-averaged density profiles. The deviation with respect to the theoretical value remains below ${\sim}0.4 \%$. 
However, we note that, first, the initial profile slightly deviates from the theoretical one.
We found this deviation to scale as $1/n_r$ with $n_r$ the number of radial bins.
We attribute it partially to the Cartesian grid: closer to the center of the domain, the deviation from spherical symmetry is larger.
The other contribution comes from the initial setup not corresponding exactly to the equilibrium, steady state: the steady state, as set by the combination $\{\Sigma, v_r\}$ (Eq.~5.3 of \citealt{frank_accretion_2002}) ignores the thermal pressure gradient.
The difference with our initial $\Sigma$ and $v_r$ profile is of order $(H/R)^2$, so here we expect ${\sim}1\%$ and ${\sim}0.1\%$ differences here, respectively.
Globally, the density profile retains this deviation overtime, except for an additional source of error at $r=2R_0$: this is where is located the transition between two levels of the AMR grid.
The theoretical, initial and final radial velocity profiles are shown on the right panels of Fig.~\ref{fig:RAMSES_visco}.
This plot shows again how the transition between two AMR levels is a source of error; indeed, while it remains below $0.4\%$ in the thick disc case, it increases to $1\%$ in the thin disc case.

\begin{figure*}
    \centering
    \includegraphics[width=0.48\textwidth]{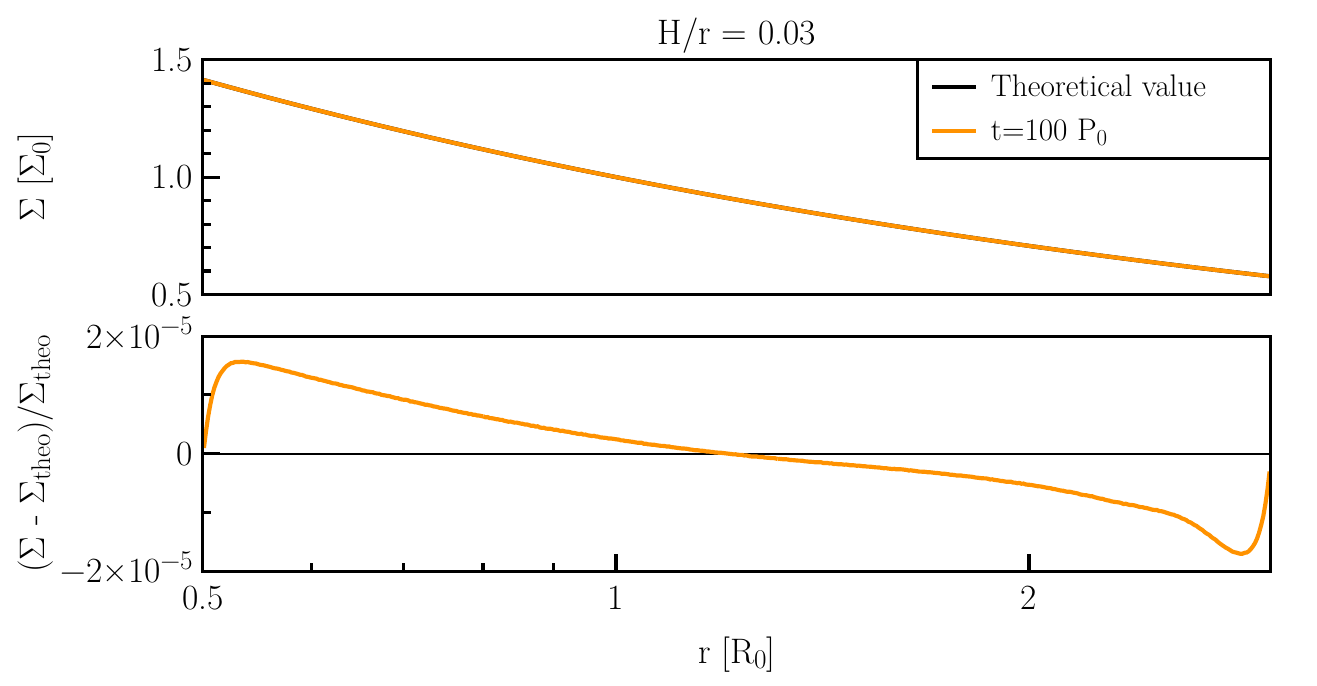}
    \caption{Theoretical and final density profiles in the single-body viscosity test of \texttt{ATHENA++}, for $H/r=0.03$.}
    \label{fig:athenapp_visco}
\end{figure*}

\begin{figure*}
    \centering
    \includegraphics[width=0.48\textwidth]{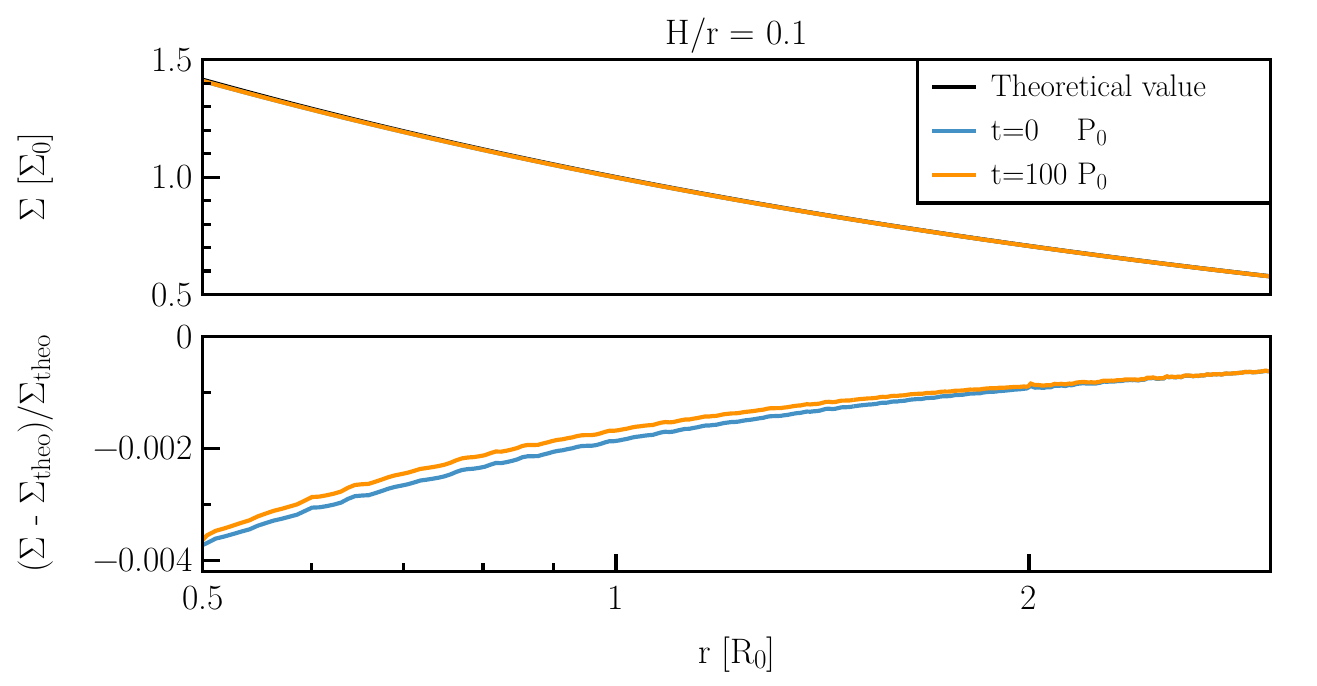}
    \includegraphics[width=0.48\textwidth]{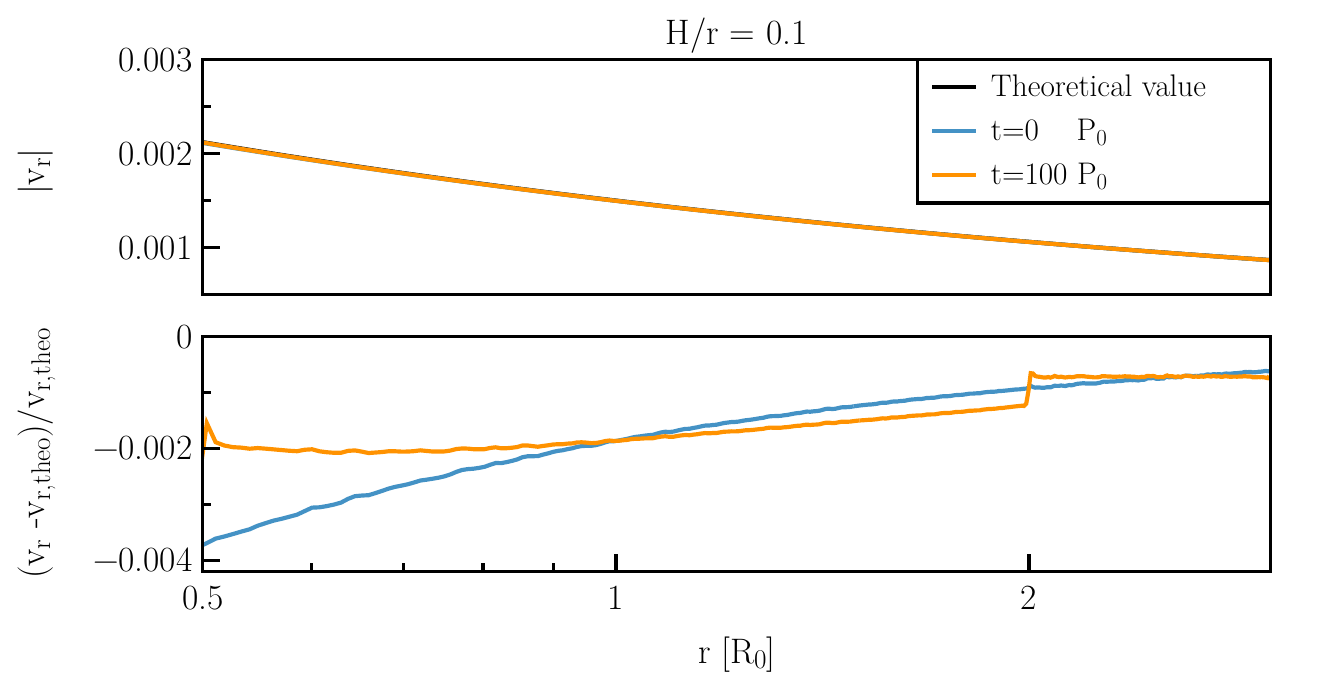}\\
    \includegraphics[width=0.48\textwidth]{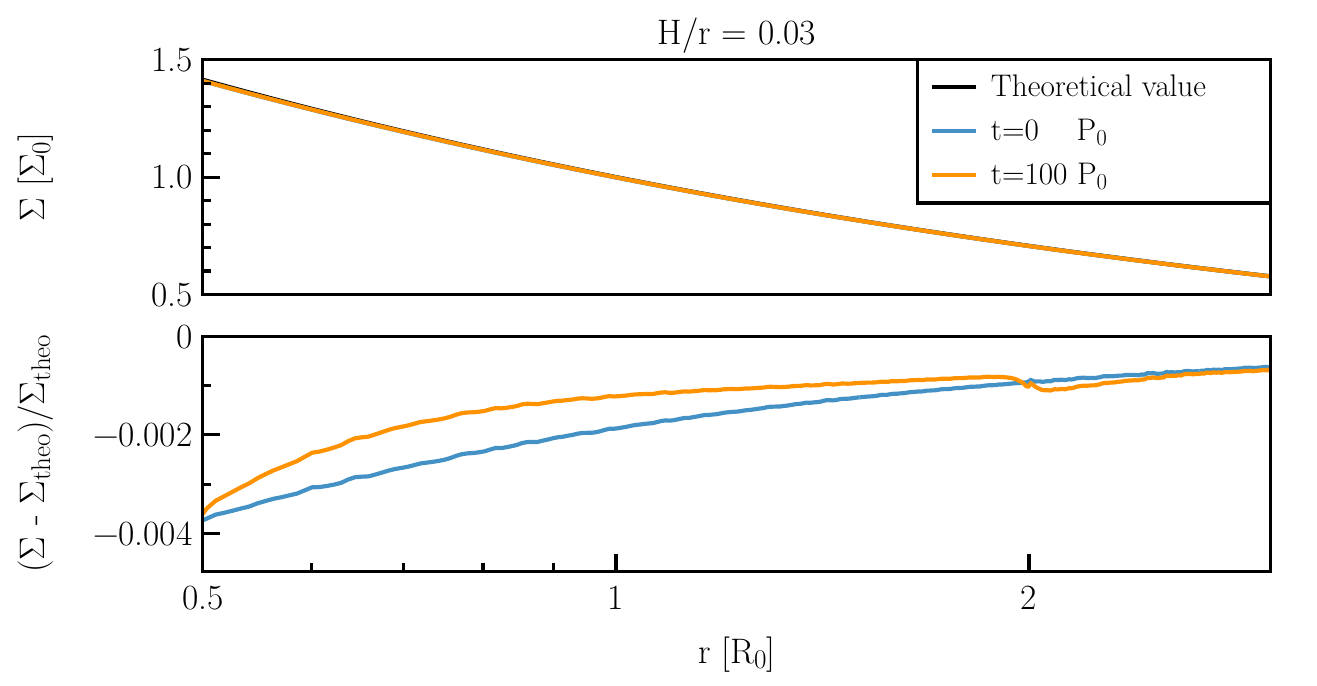}
    \includegraphics[width=0.48\textwidth]{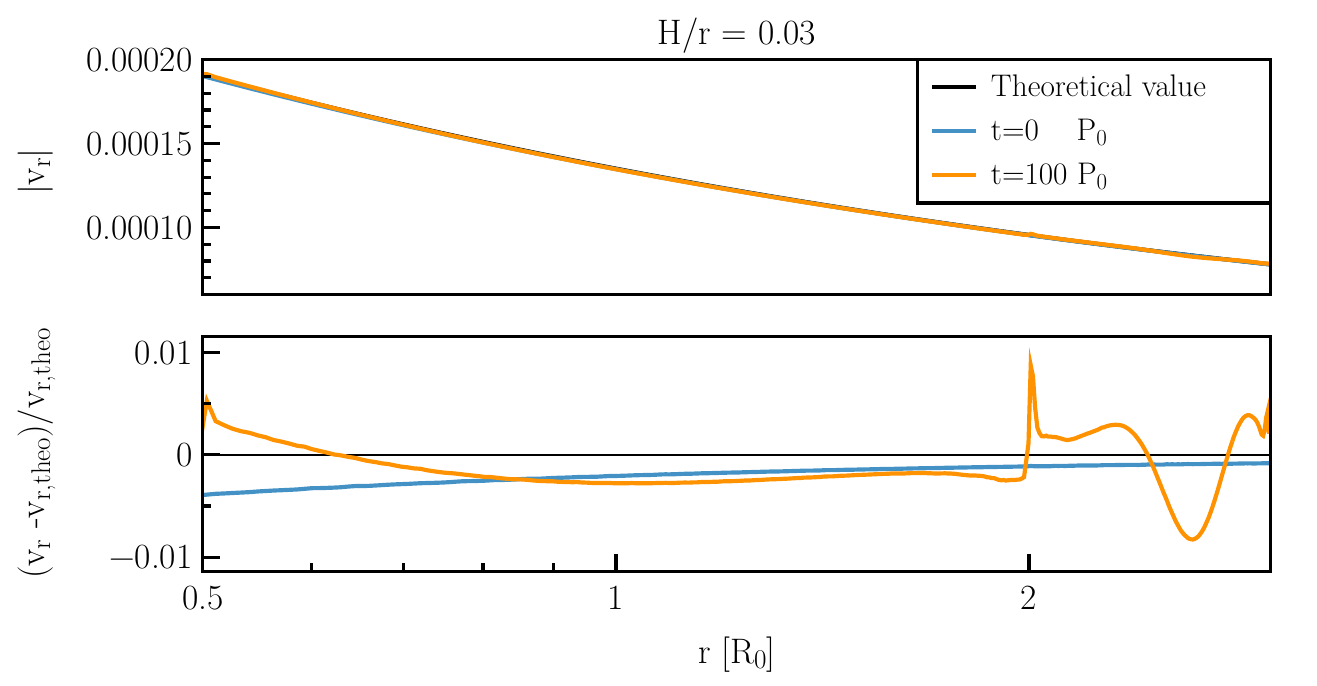}
    \caption{Theoretical, initial and final density profiles (left-hand panels) and radial velocity profiles (right-hand panels) in the single-body viscosity test of \texttt{RAMSES} for $H/r=0.1$ (top panels) and $H/r=0.03$ (bottom panel). We attribute the initial deviation, independent of $H/r$, to the mapping of a disc onto a Cartesian grid (and note that increasing the number of sampling points decreases this error until noise dominates). The transition between two AMR levels, located at $r=2 R_0$, becomes the main source of error. 
    }%these are updated (new, fixed viscosity)
    \label{fig:RAMSES_visco}
\end{figure*}

\section{Additional Morphology Plots}\label{app:extra}
Sometimes it can be useful to view perturbations to accretion discs in $(r,\phi)$ coordinates rather than $(x,y)$. We show such plots here: Figure \ref{fig:surfacedensitymap_hr01_unrolled} in analogy to Figure \ref{fig:surfacedensitymap_vr_hr01}, and Figure \ref{fig:surfacedensitymap_hr003_unrolled} in analogy to Figure \ref{fig:surfacedensitymap_vr_hr003}. 

\subsection{Spiral structure in $(r,\phi)$}
\begin{figure*}
    \centering
    \includegraphics[width=.95\textwidth]{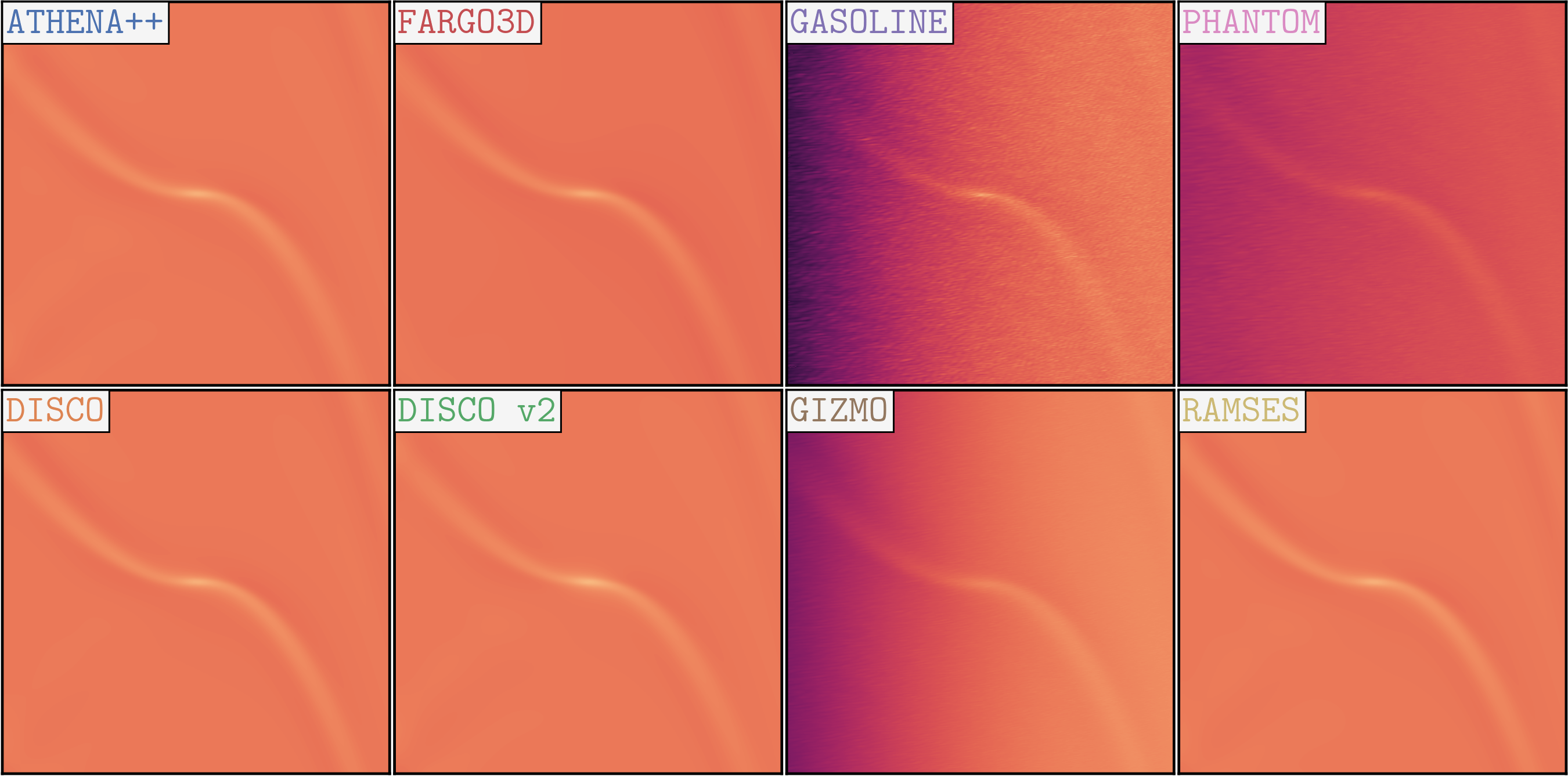}
    \includegraphics[width=.95\textwidth]{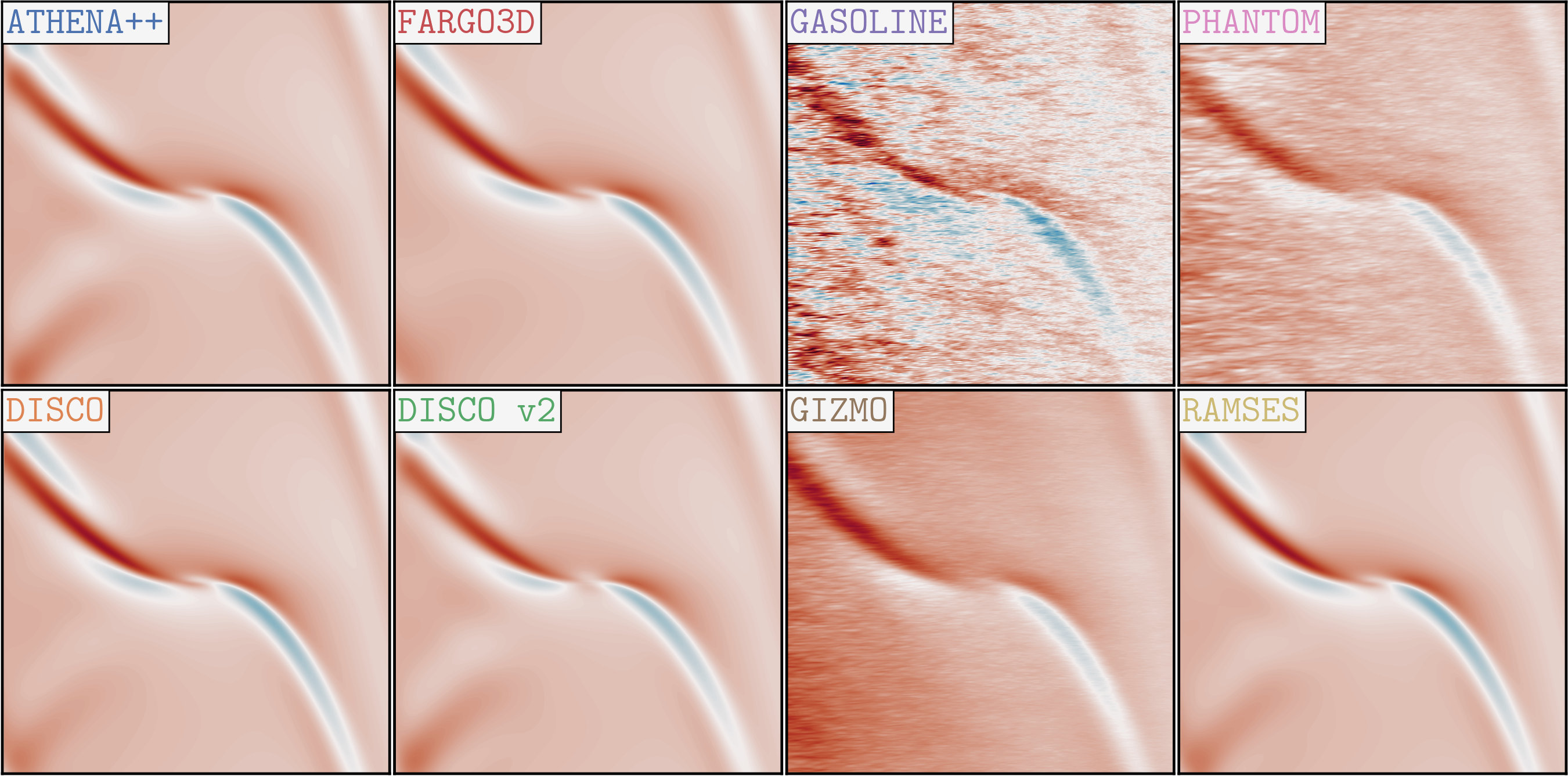}
    \caption{Surface density map (top panel\PC{s}) and radial velocity map (bottom panel\PC{s}) for the alignment run ($h/r=0.1$). Specifically, the top panel\PC{s} plot\sout{s} $\log_{10}{(\Sigma/\Sigma_0)}$ over a range $[-0.2,0.1]$; the bottom panel\PC{s} plot\sout{s} $v_r$ on a scale of $[-0.01,0.01]$, with red indicating negative velocities.}
    \label{fig:surfacedensitymap_hr01_unrolled}
\end{figure*}

\begin{figure*}
    \centering
    \includegraphics[width=.95\textwidth]{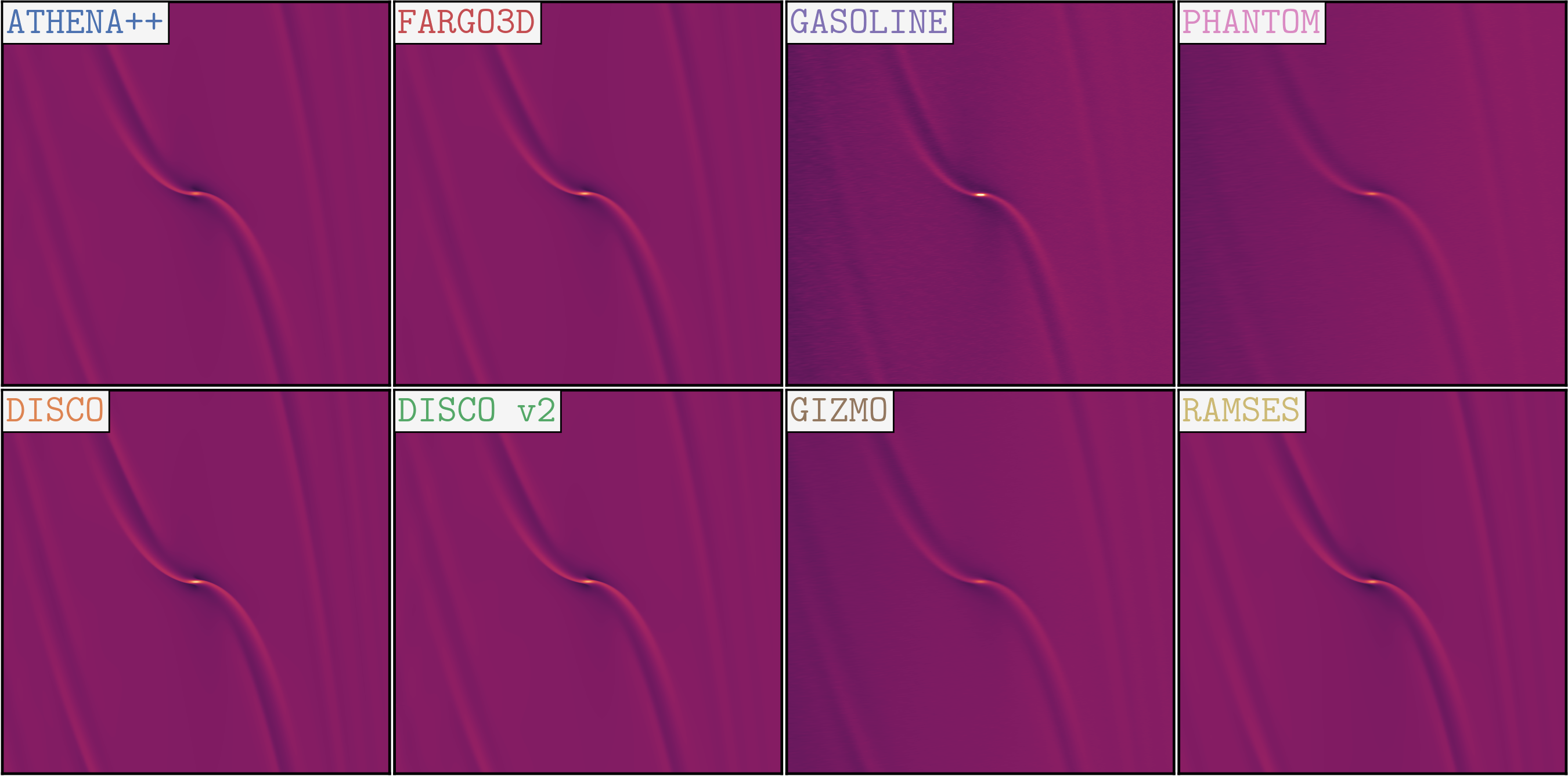}
    \includegraphics[width=.95\textwidth]{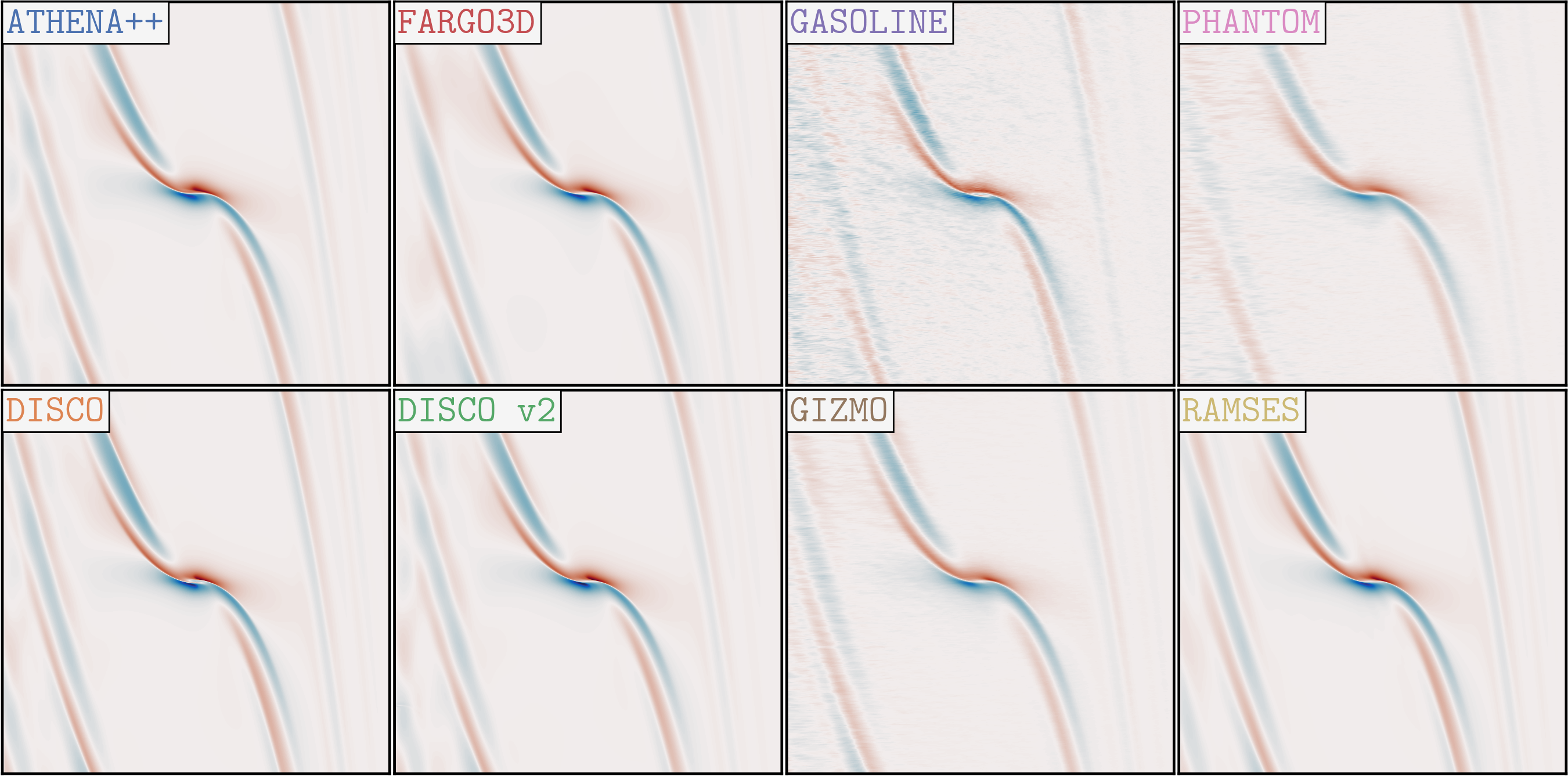}
    \caption{The thin-disc run ($h/r=0.03$). The top panel plots $\log_{10}{(\Sigma/\Sigma_0)}$ over a range $[-0.5,1.5]$; the bottom panel plots $v_r$ on a scale of $[-0.05,0.05]$, with red indicating negative velocities.}
    \label{fig:surfacedensitymap_hr003_unrolled}
\end{figure*}

\end{document}